\documentclass[acmsmall,nonacm=true]{acmart}

\AtBeginDocument{%
  }

\setcopyright{acmcopyright}
\copyrightyear{2023}
\acmYear{2023}
\acmDOI{XXXXXXX.XXXXXXX}

\acmJournal{TACO}
\acmVolume{37}
\acmNumber{4}
\acmMonth{8}

\usepackage{etoolbox}
\newtoggle{releaseStuffAfterDoubleBlind}
\toggletrue{releaseStuffAfterDoubleBlind}
\newtoggle{releaseOnArxivWithoutAckno}
\togglefalse{releaseOnArxivWithoutAckno}
\newtoggle{makeReadyForSubmission}
\toggletrue{makeReadyForSubmission}
\newtoggle{includeAppendix}
\togglefalse{includeAppendix}
\newtoggle{includeResponse}
\togglefalse{includeResponse}

\iftoggle{releaseStuffAfterDoubleBlind}{%
    \toggletrue{makeReadyForSubmission}
}{}

\usepackage{appendix}

\usepackage{units}

\usepackage{tabularx}
\newcolumntype{Z}{>{\centering\let\newline\\\arraybackslash\hspace{0pt}}X}
\newcolumntype{R}{>{\raggedleft\arraybackslash}X}
\newcommand{\tH}[1]{\multicolumn{1}{c}{\textbf{#1}}}
\newcommand{\tR}[1]{\multicolumn{1}{r}{\textbf{#1}}}

\usepackage{textcomp}

\usepackage{multicol}

\usepackage{multirow}

\usepackage[skip=.5\baselineskip]{caption}
\usepackage[skip=.5\baselineskip]{subcaption}
\usepackage{afterpage}

\iftoggle{makeReadyForSubmission}{%
    \usepackage{microtype}
}{}

\newcommand{\TTT}[1]{\textsubscript{#1}}
\newcommand{\TSS}[1]{\textsuperscript{#1}}
\newcommand{\sqmm}{mm\TSS{2}}

\iftoggle{makeReadyForSubmission}{%
    \newcommand{\revision}[1]{}
    \newcommand{\fixme}[1]{}
    \newcommand{\wahib}[1]{}
}{
    \newcommand{\revision}[1]{{\color{green}{#1}}}
    \newcommand{\fixme}[1]{{\color{red}{#1}}}
    \newcommand{\wahib}[1]{{\color{blue}{Wahib}}: {\color{blue}{#1}}}
}
\iftoggle{releaseOnArxivWithoutAckno}{%
    \newcommand{\revised}[1]{#1}
}{
    \newcommand{\revised}[1]{{\color{blue}{#1}}}
}

\newcommand{\proc}[1]{{LARC}}
\newcommand{\procC}[1]{{\proc{}\TTT{C}}}
\newcommand{\procA}[1]{{\proc{}\TSS{A}}}
\usepackage{soul}

\begin{document}

\title{At the Locus of Performance: Quantifying the Effects of Copious 3D-Stacked Cache on HPC Workloads}

\author{Jens Domke}
\authornote{Jens Domke and Emil Vatai are co-first authors.}
\authornote{New Paper, Not an Extension of a Conference Paper.}
\orcid{0000-0002-5343-414X}              
\affiliation{
  \institution{RIKEN Center for Computational Science}            
  \streetaddress{7-1-26 Minatojima-minamimachi, Chuo-ku}
  \city{Kobe}
  \state{Hyogo}
  \country{Japan}                        
  \postcode{650-0047}
}
  
\author{Emil Vatai}
\authornotemark[1]
\orcid{0000-0001-7494-5048}              
\affiliation{
  \institution{RIKEN Center for Computational Science}
  \country{Japan}
}

\author{Balazs Gerofi}
\orcid{0009-0004-8585-6031}             
\affiliation{
  \institution{Intel Corporation}
  \country{USA}
}

\author{Yuetsu Kodama}
\orcid{0000-0001-5787-0363}             
\affiliation{
  \institution{RIKEN Center for Computational Science}
  \country{Japan}
}

\author{Mohamed Wahib}
\orcid{0000-0002-7165-2095}             
\affiliation{
  \institution{RIKEN Center for Computational Science}
  \country{Japan}
}

\author{Artur Podobas}
\orcid{0000-0001-5452-6794}             
\affiliation{
  \institution{KTH Royal Institute of Technology}
  \country{Sweden}
}

\author{Sparsh Mittal}
\orcid{0000-0002-2908-993X}             
\affiliation{
  \institution{Indian Institute Of Technology, Roorkee}
  \country{India}
}

\author{Miquel Peric\`{a}s}
\orcid{0000-0002-7583-6609}             
\affiliation{
  \institution{Chalmers University of Technology}
  \country{Sweden}
}

\author{Lingqi Zhang}
\orcid{0000-0002-2452-1551}             
\affiliation{
  \institution{Tokyo Institute of Technology}
  \country{Japan}
}

\author{Peng Chen}
\orcid{0000-0003-1244-3151}             
\affiliation{
  \institution{National Institute of Advanced Industrial Science and Technology}
  \country{Japan}
}

\author{Aleksandr Drozd}
\orcid{0000-0002-4575-7213}             
\affiliation{
  \institution{RIKEN Center for Computational Science}
  \country{Japan}
}

\author{Satoshi Matsuoka}
\orcid{0000-0003-1910-8532}             
\affiliation{
  \institution{RIKEN Center for Computational Science}
  \country{Japan}
}


\renewcommand{\shortauthors}{Domke and Vatai, et al.}
\acmArticleType{Research}
\acmCodeLink{https://gitlab.com/domke/LARC}
\acmDataLink{https://doi.org/10.5281/zenodo.6420658}


\begin{abstract}
%
Over the last three decades, innovations in the memory subsystem were primarily targeted at overcoming the data movement bottleneck. In this paper, we focus on a specific market trend in memory technology: 3D-stacked memory and caches. We investigate the impact of extending the on-chip memory capabilities in future HPC-focused processors, particularly by 3D-stacked SRAM. First, we propose a method oblivious to the memory subsystem to gauge the upper-bound in performance improvements when data movement costs are eliminated. Then, using the gem5 simulator, we model two variants of a hypothetical LARge Cache processor (LARC), fabricated in \unit[1.5]{nm} and enriched with high-capacity 3D-stacked cache. With a volume of experiments involving a broad set of proxy-applications and benchmarks, we aim to reveal how HPC CPU performance will evolve, and conclude an average boost of 9.56x for cache-sensitive HPC applications, on a per-chip basis.
Additionally, we exhaustively document our methodological exploration to motivate HPC centers to drive their own technological agenda through enhanced co-design.
\end{abstract}




\maketitle

\begin{minipage}{\textwidth}
    \begin{minipage}[tbp]{0.63\textwidth}
        \centering
        \includegraphics[width=\linewidth]{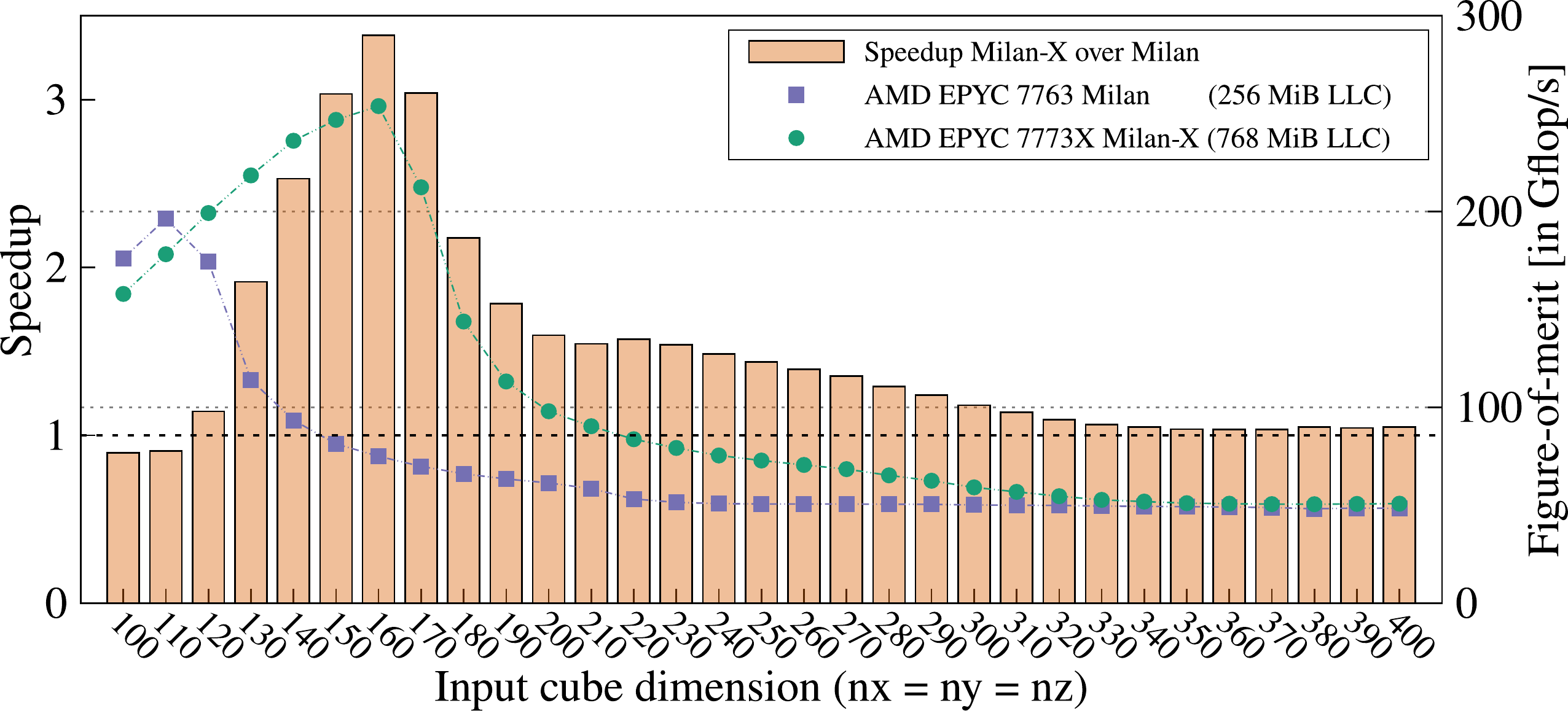}
        \captionof{figure}{MiniFE: relative performance improvement of AMD EPYC 7773X \mbox{Milan-X} over 7763 Milan (for details cf. Table~\ref{tbl:milan-stats}), and Figure of Merit; Input problem scaled from 100$\times$100$\times$100 to 400$\times$400$\times$400; Benchmarks executed with 16 MPI ranks and 8 OpenMP threads} 
        \label{fig:milan-minife}
    \end{minipage}
    \hfill
    \begin{minipage}[tbp]{0.33\linewidth}
        \centering
            {\scriptsize
                \begin{tabularx}{\linewidth}[t]{lRR}
                    \toprule
                    & \tH{7763 Milan} & \tH{7773X Milan-X}            \\
                    \midrule 
                    {Sockets}   &  2               &  2               \\
                    \midrule 
                    & \multicolumn{2}{c}{\textsc{CPU config. per Socket}:} \\
                    {Cores}     & 64               & 64               \\
                    {CCDs}      &  8               &  8               \\
                    {Freq.}     & \unit[2.45]{GHz} & \unit[2.20]{GHz} \\
                    {TDP}       & \unit[280]{W}    & \unit[280]{W}    \\
                    {L3}  & \unit[256]{MiB} & \unit[768]{MiB}         \\
                    \midrule 
                    & \multicolumn{2}{c}{\textsc{Cache per core}:}    \\
                    {L2}     & \unit[512]{KiB} & \unit[512]{KiB}      \\
                    {L1 I+D} & \unit[32+32]{KiB} & \unit[32+32]{KiB}  \\
                    \midrule 
                    {Memory} & \multicolumn{2}{c}{\unit[1]{TiB} DDR4, 16 cha., \unit[409.6]{GB/s}} \\
                    \bottomrule
                \end{tabularx}
            }
        \captionof{table}{Systems configuration for the benchmarked AMD EPYC \mbox{7763 Milan} and \mbox{7773X Milan-X} (for more details: see Zen 3 microarch.)}
        \label{tbl:milan-stats}
    \end{minipage}
\end{minipage}

\section{Introduction}\label{sec:intro} 


Historically, the reliable performance increase of von Neumann-based general-purpose processors (CPUs)
was driven by two technological trends. The first, observed by Gordon E.~Moore~\cite{moore_progress_1975},
is that the number of transistors in an integrated circuit doubles roughly every two years.
The second, called Dennard's scaling~\cite{dennard_design_1974}, postulates that as transistors get
smaller their power density stays constant. These trends synergized well, allowing computer
architectures to continuously improve performance through, for example, aggressive pipelining
and superscalar techniques without running into thermal limitations by, e.g., reducing the
operating voltage. In the early 2000s, Dennard's scaling ended~\cite{hruska_death_2012} and
forced architects to shift their attention from improving instruction-level parallelism
to exploiting on-chip multiple-instruction multiple-data
parallelism~\cite{gottlieb_nyu_1983}. This immediate remedy to the end of Dennard's scaling applies
to this day in the form of processors such as Fujitsu A64FX~\cite{sato_co-design_2020},
AMD Ryzen~\cite{suggs_amd_2020}, or NVIDIA GPUs~\cite{owens_gpu_2008,nickolls_gpu_2010}.

Unfortunately, Moore's law is impending termination~\cite{theis_end_2017-1}, and we are entering
a post-Moore era~\cite{vetter_architectures_2017-1}, home to a diversity of architectures,
such as \mbox{quantum-,} \mbox{neuromorphic-,} or reconfigurable computing~\cite{hemsoth_rogues_2018}.
Many of these prototypes hold promise but are still immature, focus on a niche use case,
or incur long development cycles. However, there is one salient solution
that is growing in maturity and which can facilitate performance improvements in the decades to
come even for the classic von Neumann CPUs we have come to rely upon---3D integrated circuit
(IC) stacking~\cite{black_stacking_2006}.
3D ICs refer to the general technologies of vertically building integrated circuits
and can be done in multiple ways, such as by stacking multiple discrete dies and connecting them
using coarse through-silicon vias (TSVs) or growing the 3D integrated circuit monolithically on the
wafer~\cite{shulaker_monolithic_2015}.

Recent advances in 3D integrated circuits have enabled many times higher capacity for on-chip
memory (caches) than traditional systems (e.g., AMD V-Cache~\cite{evers_amd_2022}).
Intuition tells us that an increased cache size, resulting from 3D-stacking, will help alleviate the performance bottlenecks
of key scientific applications. To demonstrate this, we conduct a pilot study where we execute one
of the important proxy-apps from the DoE ExaScale Computing Project (ECP) suite,
MiniFE~\cite{heroux_improving_2009} (cf.~Section~\ref{ssec:apps}), on AMD EPYC Milan and
\mbox{Milan-X} CPUs---two architecturally similar processors with vastly different L3 cache
sizes~\cite{bonshor_amd_2022}. Figure~\ref{fig:milan-minife} overviews our result of the pilot study, and we see that
for a subset of problem sizes, in particular the 160$\times$160$\times$160 input, the 3-times larger L3 capacity of \mbox{Milan-X} yields up-to~3.4x
improvements over baseline Milan for this memory-bound application, which motivates us to further
research 3D-stacked caches.

3D integrated circuits have various benefits~\cite{hu_stacking_2018}, including
\textbf{(i)}~shorter wire lengths in the interconnect leading to reduced power consumption, 
\textbf{(ii)}~improved memory bandwidth through on-chip integration that can alleviate performance
bottlenecks in memory-bound applications,
\textbf{(iii)}~higher package density yielding more compute and smaller system footprint, and
\textbf{(iv)}~possibly lower fabrication cost due to smaller die size (thus improved yield). 
All these are very desirable benefits in today's exascale (and future) High-Performance Computing (HPC)
systems.  
But how far can 3D ICs (with a focus on increased on-chip cache) take us in HPC?

\textbf{Contributions:} We study our research questions from three different levels of abstraction: \textbf{(i)}~we design a \textbf{novel exploration framework} that allows us
to simulate HPC applications running on a hypothetical processor having infinitely large L1D cache. We use this framework, that is \textbf{orders of magnitude faster} than cycle-accurate simulators, to estimate an upper-bound
for cache-based improvements;
\textbf{(ii)}~we \textbf{model a hypothetical LARge Cache processor (\emph{\proc{}})}, that builds on the design of A64FX, with an LLC (Last Level Caches) designed with eight stacked SRAM dies under \unit[1.5]{nm} manufacturing assumption; \textbf{(iii)}~we complement our study with a plethora of \textbf{simulations of
HPC proxy-applications} and CPU micro-benchmarks; and lastly
\textbf{(iv)}~we find that over half (31 out of~52) of the simulated applications experience a $\geq\,$2x speedup on LARC's Core Memory Group (CMG) that occupies only one fourth the area of the baseline A64FX CMG.
For applications that are responsive to larger cache capacity, this would translate to an average improvement of 9.56x (geometric mean) when we assume ideal scaling and compare at the full chip level.

\begin{figure}[tbp]
    \centering
    \includegraphics[width=.75\linewidth]{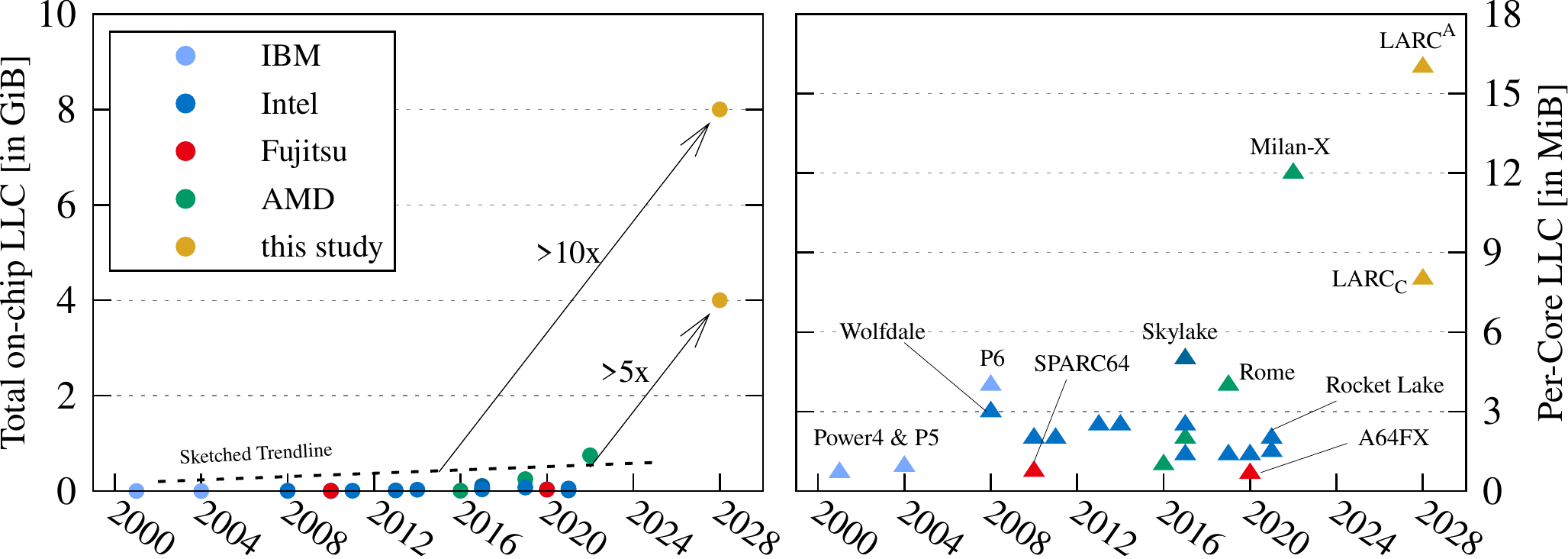}
    \caption{A sample of representative server-grade CPUs of each generational micro-architecture in comparison to our study of \proc{}; Left: total on-chip last-level cache (in~\unit[]{GiB}); Right: per-core last-level cache (in~\unit[]{MiB}) for the same CPUs; The two \proc{} variants will be discussed in detail in Sec.~\ref{ssec:gem5conf}}
    \label{fig:llc-size}
\end{figure}

The novelty in this paper lies in the purpose which \proc{} serves, and not the design of \proc{} itself.
As Figure~\ref{fig:llc-size} shows, the capacity (and bandwidth; not shown) of the LLC have
increased at a moderately gradual slope over the last two decades---with Milan-X being a noticeable
outlier in per-core LLC.
However, we are querying the effect of an LLC, that is an order of magnitude above the trend line
as depicted in Figure~\ref{fig:llc-size}, on HPC applications.
On top of our provided baseline, further application-specific restructuring to utilize large
caches~\cite{ltaief_meeting_2021} will result in even greater benefit. 

\section{CPUs Empowered with High-capacity Cache: the Future of HPC?}\label{sec:projectedArch} 

The memory bandwidth of modern systems has been the bottleneck (the ``memory wall''~\cite{mckee_reflections_2004}) ever since CPU performance started to outgrow the bandwidth of memory
subsystems in the early 1990s~\cite{mccalpin_memory_1995}. Today, this trend continues to shape the performance
optimization landscape in high-performance computing~\cite{or-bach_1000x_2017,oliveira_damov_2021}.
Diverse memory technologies are emerging to overcome said data movement bottleneck, such as
Processing-in-Memory (PIM)~\cite{balasubramonian_near-data_2014}, 3D-stackable High-Bandwidth Memory
(HBM)~\cite{mittal_survey_2016}, deeper (and more complex) memory hierarchies~\cite{warnock_41_2015},
and---the topic of the present paper---novel 3D-stacked
caches~\cite{loh_processor_2007,black_stacking_2006,shiba_96-mb_2021}.
%

In this study, our aspiration is to gauge the far end of processor technology and how it may evolve in six to eight years from now, circa 2028, when processors using~\unit[1.5]{nm} technology are expected to be available according to the IEEE IRDS Roadmap~\cite[Figure ES9]{ieee_irds_international_2021-1}. 
More specifically, as 3D-stacked SRAM memory~\cite{zhang_survey_2014} becomes more common, what are the performance implications for common HPC workloads, and what new challenges lie ahead for the community?
However, before attempting to understand what performance may look like six years from now, we must describe how the processor itself might change.
In this section, we introduce, motivate, and reason about our design choices of what we envision as a hypothetical CPU that capitalizes on large capacity 3D-stacked cache, briefly called \textbf{\proc{}~(LARge Cache processor)}. 
Before looking at \proc{}, we must first set and analyze a baseline processor.

\subsection{\proc{}' Baseline: The A64FX Processor}
We choose to base our future CPU design on the A64FX~\cite{yoshida_fujitsu_2018}.
Fujitsu's Arm-based A64FX is powering Supercomputer Fugaku~\cite{sato_co-design_2020},
leader of the HPCG (TOP500~\cite{strohmaier_top500_2021}; cf.~Section~\ref{ssec:apps}) and Graph500 performance charts.
A64FX is manufactured in~\unit[7]{nm} technology and has a total of 52 Arm cores (with
Scalable Vector Extensions~\cite{stephens_arm_2017}) distributed across four compute
clusters, called Core Memory Groups (CMGs). Twelve cores are available
to the user, and one core is exclusively used for management. Each core has a local~\unit[64]{KiB}
instruction and data-cache, and is capable of delivering~\unit[70.4]{Gflop/s} (IEEE-754 double-precision)
performance---accumulated:~\unit[845]{Gflop/s} per CMG (user cores) or~\unit[3.4]{Tflop/s} for the entire chip.
Each CMG contains a~\unit[8]{MiB}
L2 cache slice, delivering over~\unit[900]{GB/s} bandwidth to the CMG~\cite{yoshida_fujitsu_2018}.
The combined L2 cache, which is the CPU's~\unit[32]{MiB} last level cache (LLC), is kept coherent through a ring interconnect that connects
the four CMGs. Inside the CMG, a crossbar switch is used to connect the cores and the L2 slice.
The L2 cache has 16-way set associativity, a line-size of 256 bytes, and the bus-width between the L1 and L2 cache is set to be
128 bytes (read) and 64 bytes (write).

We emphasize that our aim is not
to propose a successor of A64FX, nor are we particularly restricting our vision by the design
constrains of A64FX (e.g., power budget). However, we build our design on A64FX because:
\textbf{(i)}~as mentioned above, A64FX represents the high-end in performance for commercially
available CPUs, so it is a logical starting point.
\textbf{(ii)}~A64FX is the only commercially-available CPU, currently in continued production, with HBM. The expected bandwidth ratio between future HBM and future 3D-stacked caches is similar
to the ratio between traditional DRAM and LLC bandwidths~\cite{nori_criticality_2018},
which is what applications and performance models are accustomed to.
\textbf{(iii)}~The A64FX LLC cache design (particularly the L2 slices connected by a crossbar switch)
happens to be convenient and thus, requires a minimal effort to extend the design in a
simulated environment.
%
%

In conclusion, while we extend the A64FX architecture, our workflow itself can be generalized to
cover any of the processors supported by CPU simulators (e.g., variants of gem5~\cite{binkert_gem5_2011}
can simulate other architectures, including x86).

\subsection{Floorplan Analysis for Fujitsu A64FX}
\begin{figure}[tbp]
    \centering
    \includegraphics[width=.45\linewidth]{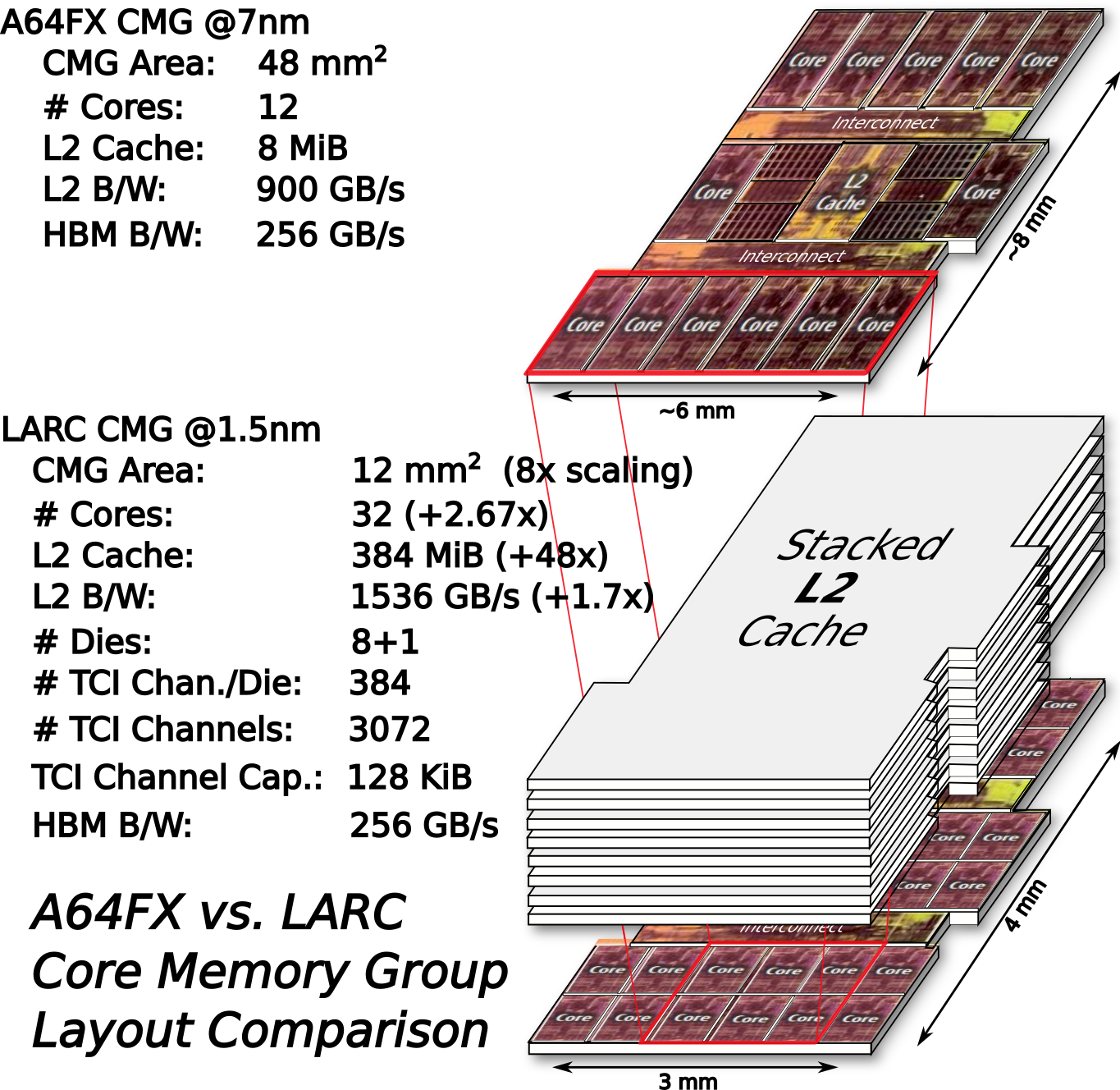}
    \caption{Difference between A64FX's Core Memory Group (CMG) and a \proc{} CMG in various performance-governing parameters; Most notable (for our study) is the~48x increase in per-CMG L2 cache capacity; \textbf{Note:} despite appearing similar in the figure, the \proc{} CMG is, in fact, four times smaller.}
    \label{fig:n64_picture}
\end{figure}

In order to estimate the floorplan of the future \proc{} processor built on~\unit[1.5]{nm} technology,
we first need the floorplan of the current A64FX processor built at~\unit[7]{nm}. We do know that the die size
of A64FX is~${\approx}\,\text{\unit[400]{\sqmm}}$~\cite{sato_co-design_2020}.
With the openly-available die shots including processor core segments
highlighted~\cite{okazaki_supercomputer_2020}, we can estimate most of the A64FX floorplan,
including the size of CMGs and processor cores, as shown in
Figure~\ref{fig:n64_picture}. Overall, each CMG is~${\approx}\,\text{\unit[48]{\sqmm}}$ in area, where
an A64FX core occupies~${\approx}\,\text{\unit[2.25]{\sqmm}}$ area. The remaining parts of the
CMG consist of the L2 cache slice and controller as well as the interconnect for intra-CMG communication.


\subsection{From A64FX's to \proc{}'s CMG Layout}
Knowing the floorplan, we proceed to describe how we envision the CMG design  
with~\unit[1.5]{nm} technology.
We scale the CMG by moving four generations, from~\unit[7]{nm}
to~\unit[1.5]{nm}, and reduce the silicon footprint by around~8x (${\approx}\,\text{1.7x}$ per generation)
for the entire CMG~\cite{esmaeilzadeh_dark_2011}. The new CMG consumes as little
as~\unit[6]{\sqmm} of silicon area. Next, we reclaim the area currently occupied by
the L2 cache and controller and replace it with three additional CPU cores, yielding a
total of 16. Further, inline with the projected year 2019$\rightarrow$2028
growth in the number of cores~\cite[Table SA-1]{ieee_irds_international_2021}, we double the core count of
the CMG to 32, which leads to it occupying~${\approx}\,\text{\unit[12]{\sqmm}}$ of silicon area.
We pessimistically leave the interconnect area unchanged and continue to use it as the
primary means for communication. We call this new variant as \proc{}'s CMG.
Finally, we assume the same die size, and hence, \proc{} would have 16 CMGs,
each with 32 cores, in comparison to A64FX's 4~CMG with 12+1 cores each. For \proc{}, we ignore the management core.
However, our performance analysis will remain on the CMG level, instead of full chip, due to
limitations we detail in Section~\ref{ssec:gem5}. 


\subsection{\proc{}'s Vertically Stacked Cache}\label{ssec:n64fx}
In the above design, we removed the L2 cache and controller from the CMG of \proc{}.
We now assume that the L2 cache can be directly placed vertically on the CMG through 3D
stacking~\cite{loh_processor_2007}.
We build our
estimations based on experiments from Shiba et al.~\cite{shiba_96-mb_2021}, who
demonstrated the feasibility of stacking up-to-eight SRAM dies on top of a processor using a ThruChip Interface (TCI).
The capacity and bandwidth of stacked memory is a function of several parameters: the number of channels available ($N_\text{ch}$),
the per-channel capacity ($N_\text{cap}$ in~\unit[]{KiB}), their width ($W$ in bytes), the number of stacked dies ($N_\text{dies}$), and the operating frequency ($f_\text{clk}$ in~\unit[]{GHz}).
Shiba et al.~\cite{shiba_96-mb_2021} estimated that at a~\unit[10]{nm} process technology, eight stacks would provide~${\approx}\,\text{\unit[512]{MiB}}$ of
aggregated SRAM capacity for a footprint of~${\approx}\,\text{\unit[121]{\sqmm}}$. In their design, each stack has 128 channels of \unit[512]{KiB} capacity.
In our work, we conservatively assume an 8x scaling from~\unit[10]{nm} to~\unit[1.5]{nm}, and thus, at \unit[12]{\sqmm} area (the size of one
\proc{} CMG), $N_\text{ch}$ on each die would be~${\approx}\,\text{102}$ (=128*8/10). 

We approximate $N_\text{ch}$ to a nearby sum of power-of-two number, viz., $N_\text{ch}=\text{96}$.
Thus, with eight stacked dies ($N_\text{dies}=\text{8}$), our 3D SRAM cache has a total storage capacity of $N_\text{dies} \cdot N_\text{ch} \cdot N_\text{cap}=\text{\unit[384]{MiB}}$ per CMG.
We estimate the bandwidth in a similar way. We know from previous studies~\cite{shiba_96-mb_2021}
that 3D-stacked SRAM, built on~\unit[40]{nm} technology, can operate at~\unit[300]{MHz}.
We conservatively expect the same SRAM to operate at ($f_\text{clk}$=)\unit[1]{GHz} when moving
from~\unit[40]{nm}$\rightarrow$\unit[1.5]{nm}.
To account for the increased working set size of future applications, we assume a channel width ($W$) of \unit[16]{byte}, compared to the \unit[4]{byte} width assumed in~\cite{shiba_96-mb_2021}.
With this, the CMG bandwidth becomes:
$N_\text{ch} \cdot f_\text{clk} \cdot W = \text{\unit[1536]{GB/s}}$.  
The read- and
write-latency of their SRAM cache is 3 cycles, including the vertical data movement overhead~\cite{shiba_96-mb_2021}.
While stacked DRAM caches theoretically provide higher capacity than stacked SRAM caches, they have
limitations. For example, the latency of stacked DRAM is only 50\% lower compared to DDR3 DRAM, and hence, they exacerbate miss
latency; they requires refresh operations which consumes energy and reduces availability; and due to their large size, the stacked DRAM caches require special techniques for managing metadata and avoiding
bandwidth bloat~\cite{chou_bear_2015,mittal_survey_2016}. The tag size of a stacked DRAM may exceed
the LLC capacity, and hence, the tags may need to be stored in the DRAM itself which worsens hit latency.
Set-associative designs and serial tag-data accesses further increase hit latency. Proposed architectural
techniques and mitigation strategies, such as Loh-Hill cache~\cite{loh_efficiently_2011}, have yet to solve these problems.
By contrast, 3D SRAM caches do not suffer from any of these issues. In fact, at
iso-capacity, a 3D SRAM cache has even lower access latency than a 2D SRAM cache. Since stacked 3D SRAM caches
have lower capacity than stacked DRAM, its metadata (e.g., tag) can be easily stored in SRAM itself,
further reducing the access latency.

For our cache design, we assume a~\unit[256]{B} cache block
design, which avoids bandwidth bloat. Each tag takes~\unit[6]{B} and as such, the total tag array
size for each CMG becomes~\unit[9]{MiB}. This tag array can be easily placed in the cache itself. We
assume that tag and data accesses happen sequentially. The tags and data of a cache set are
stored on a single die. Hence, on every access, only one die needs to be activated. Since
this takes only few cycles, the overall miss penalty remains small and comparable to that of A64FX' LLC.


To show that our cache projections are realistic, we compare it with AMD's 3D V-cache
design. It uses a single stacked die for the L3 cache, providing~\unit[64]{MiB} capacity (in addition to
the~\unit[32]{MiB} cache in the base die) at~\unit[7]{nm}~\cite{cutress_amd_2021,evers_amd_2022}
and only 3 to 4 cycles of extra latency compared to the non-stacked version~\cite{cheese_amds_2022}. It has~\unit[36]{\sqmm} area
and has a bandwidth of~\unit[2]{TB/s}. When stacking additional dies on top, and assuming an 8x scaling of
the area by going from~\unit[7]{nm} to~\unit[1.5]{nm}, we speculate that the LLC capacity of
this commercial processor could easily exceed that of our proposed \proc{}.


\subsection{\proc{}'s Core Memory Group (CMG)}
At last, we detail our experimental CMG built on a hypothetical~\unit[1.5]{nm}
technology: the \textbf{\proc{}~CMG}. 
An illustration  of this system is shown in
Figure~\ref{fig:n64_picture}. Each CMG consists of 32 A64FX-like cores, which keeps the
L1 instruction- and data-cache to~\unit[64]{KiB} each, yielding a per CMG performance of
$\approx\,$\unit[2.3]{Tflop/s} (IEEE-754 double-precision). A~\unit[384]{MiB} L2 cache is
stacked vertically on the top of the CMG through eight SRAM layers.

We keep the HBM memory bandwidth per CMG to its current A64FX value of~\unit[256]{GB/s}
to be able to quantify performance improvements from the proposed large capacity
3D cache in isolation from any improvements that would come from increased HBM bandwidth.
Furthermore, we make no assumption on the technology scaling of blocks that contain hard-to-scale-down analog components (e.g., TofuD or PCIe IP blocks) and instead focus exclusively on scaling the CMG-part of the System-on-Chip (i.e., processing cores, L1/L2 caches, and intra-chip interconnects).



While our study focuses on evaluating a single CMG, we conclude that a complete, hypothetical \proc{} CPU, with a die size similar to the current A64FX,
would contain 512 processing cores,
\unit[6]{GiB} of stacked L2 cache, a peak L2 bandwidth of~\unit[24.6]{TB/s}, a peak HBM
bandwidth of~\unit[4.1]{TB/s}, and a total of~\unit[36]{Tflop/s} of raw, double-precision, compute.
The A64FX processor has a peak HBM bandwidth of~\unit[1]{TB/s}, whereas our envisioned \proc{} CPU has 4$\times$ more CMGs and hence, a peak HBM bandwidth of~\unit[4.1]{TB/s}. Thus, compared to A64FX, \proc{} has higher \emph{effective bandwidth} of external memory. Further changes to the HBM generation are beyond the scope of this study.



\subsection{\proc{}'s Power and Thermal Considerations}\label{ssec:powerthermal}
To estimate the power consumption of \proc{}, we analyze A64FX's current consumption and
extrapolate to~\unit[1.5]{nm} by leveraging public technology roadmaps. A64FX's peak power,
achieved while running DGEMM, is~\unit[122]{W}~\cite{yamamura_a64fx_2022}; where~\unit[95]{W} correspond
to core power and~\unit[15]{W} correspond to the memory interface (MIF), and hence, we
conclude~\unit[1.98]{W/core} and~\unit[3.75]{W/MIF}. Therefore, a \proc{} CMG with 32 cores
in~\unit[7]{nm} would consume~\unit[67.1]{W}.
TSMC projects that shrinking from~\unit[7]{nm} to~\unit[5]{nm} yields a power reduction of
about~30\%~\cite{shilov_tsmc_2022}, i.e.,~\unit[46.98]{W} for \proc{}'s CMG in~\unit[5]{nm}.
IRDS's roadmap~\cite[Figure ES9]{ieee_irds_international_2021-1} indicates a further
compounded power reduction (at iso frequency) of~42\% when moving from~\unit[5]{nm}
to~\unit[1.5]{nm}, i.e.,~\unit[27.37]{W} for \proc{}'s CMG in~\unit[1.5]{nm}. As the full
\proc{} chip is estimated to include 16 CMGs, we project a total power of~\unit[438]{Watt}
(not including the L2 cache).

Next, we estimate the power consumed by the principal part of this study---the~\unit[384]{MiB} L2 cache.
A~\unit[4]{MiB} SRAM L2 cache in~\unit[7]{nm} consumes~\unit[64]{mW} of static power~\cite{goud_asymmetric_2015}.
Assuming a similar (pessimistic) static power consumption at~\unit[1.5]{nm} and extrapolated
to~\unit[384]{MiB}, we find that our cache would have a static power consumption of~\unit[6.14]{W}.
Scaled to the full 16~CMGs of our hypothetical \proc{}, we arrive at a static power
consumption of~\unit[98.3]{W}. This static power consumption of caches represents
between~90\% and~98\% of the entire power consumption (at~\unit[350]{K} temperature, see 
e.g.,~\cite{chakraborty_analysing_2018,agarwal_single-vsub_2003}), where the remainder is the
dynamic power consumption. If we assume a pessimistic 9:1 ratio between static and dynamic power,
then this yields a total power consumption of~\unit[109.23]{W} for~\unit[6]{GiB} of chip-wide
stacked L2 cache.

To conclude, a \proc{} processor (16 CMG) would have to be designed for a
thermal design power (TDP) of~\textbf{\unit[547]{W}}. While this expected TDP is more
than the current A64FX, it is not entirely unlike emerging architectures, such as NVIDIA's
H100~\cite{nvidia_corporation_nvidia_2022} that consumes up to~\unit[700]{W} or the
AMD Instinct MI250X GPU~\cite{advanced_micro_devices_inc_amd_2021} at \unit[560]{W}.
We stress that our estimate of~\unit[547]{W} is peak power draw achieved only during parallel DGEMM execution.
Adjusting for Stream Triad, based on the breakdown in~\cite{yamamura_a64fx_2022}, we conclude
a realistic, and considerably lower, power consumption of~\unit[420]{W} for
bandwidth-bound applications running on the whole \proc{} chip.
%
%

Finally, while this L2 cache power estimation might appear pessimistic, there are ample
opportunities to further reduce power consumption. To save static energy, all the un-accessed
dies can be changed to data-retentive, low-power (sleep) state. To deal with remaining thermal issues
after stacking the cache layers underneath the cores instead of on top, one can additionally adapt
simple direct-die cooling or advanced techniques~\cite{cao_survey_2019,tavakkoli_analysis_2016},
such as high-$\kappa$ thermal compound~\cite{gomes_81_2020}, microfluid cooling~\cite{wang_3d_2018},
or thermal-aware floorplanning, task-scheduling and data-placement optimizations.
Specifically, microfluid cooling can handle power densities
of~\unit[3.5]{W/mm\TSS{2}} and hot-spot power levels of over~\unit[20]{W/mm\TSS{2}} for
3D-stacked chips~\cite{wesling_heterogeneous_2023}. By contrast, our \proc{} CPU has a
power density of~\unit[2.85]{W/mm\TSS{2}} at~\unit[192]{mm\TSS{2}} if we ignore adjunct components such as I/O die, PCIe,
TofuD interface, etc., and around half the power density at~\unit[400]{mm\TSS{2}} if these components are included.

\begin{figure*}[tbp]
    \centering
    \includegraphics[width=.85\linewidth]{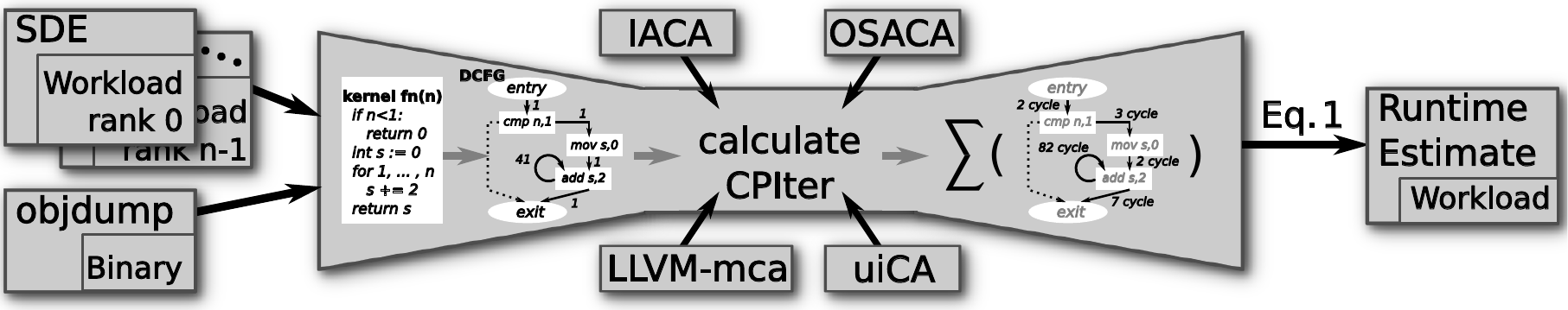}
    \caption{Illustration of our runtime estimation pipeline with the MCA-based tool for an accumulative kernel executed with $n=\text{42}$; Dotted line: branch not taken; Solid line: kernel execution as recorded by SDE; Edges in directed CFG annotated by number of jumps between basic blocks; Details in Section~\ref{ssec:L1sim}}
    \label{fig:mcatool}
\end{figure*}

\section{Projecting Performance Improvement in Simulated Environments}\label{sec:methods} 

Analyzing \proc{}'s feasibility is only the first step, and hence we have to
demonstrate the effects of the proposed changes on real workloads to allow a meaningful cost-benefit analysis
by CPU vendors. This section details two simulation approaches (one novel; one established) and discusses the
HPC applications, which we evaluate extensively in Sections~\ref{sec:MCAeval} and~\ref{sec:GEMeval}.


\subsection{Simulating Unrestricted Locality with MCA}\label{ssec:L1sim}

Designing and executing even initial studies (i.e. no complex memory models, etc.) with cycle-level gem5 simulations for realistic workloads takes substantial time with unknown outcome. Therefore, one would
want to have a first-order
approximation of a very large and fast cache.
Regrettably, and to the best of our knowledge, existing approaches for fast
first-order approximations do generally not support complex HPC applications, i.e.,
the existing tools neither handle multi-threading correctly nor do they have support
for MPI applications~\cite{akram_survey_2019}.
Hence, we design a simulation approach, using Machine Code Analyzers 
(MCA), which can estimate the speedup for a given application orders-of-magnitude faster than gem5 (typically hours instead of months; cf.~next
section). This upper bound in expected performance improvement allows us to:
\textbf{(i)}~get a perspective on the best possible performance improvement if all read/writes can be satisfied from the cache; and
\textbf{(ii)}~justify more accurate simulations and classify their results with respect to the baseline and the upper bound.

Machine Code Analyzers, such as llvm-mca~\cite{llvm_project_llvm-mca_2022}, have been designed to study
microarchitectures, improve compilers, and investigate resource pressure for application kernels. Usually,
the input for these tools is a short Assembly sequence and they output, among other things, an expected
throughput for a given CPU when the sequence is executed many times and all data is available in L1 data
cache. For most real applications, the latter assumption is obviously incorrect, however, it is ideal to
gauge an upper bound on performance when all the memory-bottlenecks disappear.

Unfortunately, it is neither feasible to record all executed instructions in one long sequence, nor to analyze
a full program sequence with llvm-mca. Hence, we break the program execution into basic blocks (at most tens or
hundreds of instructions) and evaluate their throughput individually. For a given combination of a program
and input (called \textit{workload} hereafter), the basic blocks and their dependencies create a directed Control Flow
Graph (CFG)~\cite{intel_corporation_dynamic_2020} with one source (program start) and one sink (program
termination). All intermediate nodes (representing basic blocks) of the graph can have multiple parent- and
dependent-nodes, as well as self-references (e.g. basic blocks of \texttt{for}-loops). Knowing the ``runtime''
of each basic block and the number of invocations per basic block, we can estimate the runtime of the entire
workload by summation of the parts.

We utilize the Software Development Emulator (SDE)~\cite{intel_corporation_intel_2021} from Intel to record the basic
blocks and their caller/callee dependencies for a workload with modest runtime overhead (typically in order of
1000x slowdown).
SDE also notes down the number of invocations per CFG edge for a workload, i.e., how often the program counter (PC)
jumped from one specific basic block to another specific block. We developed a program which parses the
output of Intel SDE and establishes an internal representation of the Control Flow Graph. The internal
CFG nodes are then amended with Assembly extracted from the program's binary, since SDE's Assembly output is
not compatible with Machine Code Analyzers. Our program subsequently executes a Machine Code Analyzer for
each basic block, getting in return an estimated cycles-per-iteration metric (CPIter). We record the per-block CPIter
at the directed CFG edge from caller to callee, which already holds the number of invocations of this edge,
effectively creating a ``weighted'' graph. Figure~\ref{fig:mcatool} showcases the result and it is
easy to see that the summation of all edges in the CFG is equivalent to the estimated runtime of the entire workload (assuming
all data is inside the L1 data cache).

The above outlined approach works for both sequential and  parallel programs. Intel SDE can record the instruction execution and caller/callee dependencies for thread-parallel
programs, e.g. pthreads, OpenMP, or TBB. Furthermore, we can attach SDE to individual MPI ranks to get the
data for it. Therefore, we are able to estimate the runtime for MPI+X parallelized HPC applications by the
following equation:
\begin{equation}\textstyle
    \text{t}_\text{app} := \frac{ \max\limits_{r\,\in\,\text{ranks}} \big( \max\limits_{t\,\in\,\text{threads}_{r}} (\,\sum\limits_{\text{edges}\,e\,\in\,\text{CFG}_{t,r}} \text{CPIter}_{e} \cdot \#\text{calls}_{e}\,)\big) }{ \text{processor frequency in Hz} }
    \label{eq:mca}
\end{equation}
under the assumption that MPI ranks and threads do not share computational resources\footnote{Resource over-subscription
is outside the scope of this study and our tool.}, where we sum up the number of cycles required for each block
(i.e., CFG edges) considering only the ``slowest'' thread and rank, and divide by the CPU frequency to convert the
total cycles into runtime.

The self-imposed restriction of Machine Code Analyzers is the limited accuracy compared to cycle-accurate
simulators, due to their distinct design goal. To improve our CPIter estimate, we rely on four different MCAs,
namely llvm-mca~\cite{llvm_project_llvm-mca_2022}, Intel ACA (IACA)~\cite{intel_corporation_intel_2012},
uiCA~\cite{abel_parametric_2021}, and OSACA~\cite{laukemann_automated_2018}, and take the median of the results.
Another shortcoming of MCA tools is that most of them estimate the throughput of basic blocks in isolation while
assuming looping behavior of the assembly block (PC jumps from last back to first instruction).
Neither ``block looping'' nor an empty instruction pipeline (single iteration of the block) are realistic for
some blocks.
Hence, for non-looping basic blocks, we estimate the CPIter by feeding the MCA tool with the blocks of caller and
callee, and the callee's CPIter is calculated by subtracting the cycle of retirement of its last instruction from
the caller's last instruction retirement (instead of when the callee's first instructions are decoded, which
can overlap with execution of caller instructions). Further, we correct some cycle estimates for specific
instructions within our tool in post-processing, since we encountered a few unsupported or grossly
mis-estimated instructions while validating our tool against benchmarks.
We refer the reader to Section~\ref{ssec:L1simVali} for more details.

\subsection{Cycle-level Accuracy: CPUs Simulated in gem5}\label{ssec:gem5}

While the MCAs can give a first-order approximation, we
still require highly accurate predictions for our 3D-stacked, cache-rich CPU.
Hence, we employ an open-source system architecture simulator, called gem5~\cite{binkert_gem5_2011}. 
It supports Arm, x86, and RISC-V CPUs to varying degrees of accuracy, and can be
extended with memory models for higher simulation fidelity of the memory
subsystem. We use gem5's ``syscall emulation'' mode to executes applications directly without booting a Linux kernel.

Fortunately, RIKEN released their gem5 version
which was specially tailored for A64FX's co-design to support SVE, HBM2, and other advanced
features~\cite{riken-rccs_riken_simulator_2020}. Hence it is well suited to simulate our
\proc{} proposal in Section~\ref{ssec:n64fx}. 
This version of gem5 has been validated
for A64FX~\cite{kodama_accuracy_2020}, and can be used with production compilers from Fujitsu.
Albeit, while evaluating RIKEN's gem5, we noticed a few drawbacks, such as the lack of support for:
\textbf{(i)}~dynamically linked binaries;
\textbf{(ii)}~adequate memory management (freeing memory after application's \texttt{free()} calls);
\textbf{(iii)}~simulating more than 16 CPU cores due to limits in the cache coherence protocol;
\textbf{(iv)}~multi-rank MPI-based programs; and
\textbf{(v)}~simulating more than one A64FX CMG.

We modify gem5 to remedy the first three problems. However, the last
two problems remain intractable without major changes to the simulator's codebase, 
and hence we limit ourselves to single-CMG simulations (with one MPI rank).
Relying on the assumption that most HPC codes are weak scaled across multiple NUMA domains
and compute nodes, we believe the single-rank approach still serves as a solid foundation for
future performance projection.
However, even single-rank MPI binaries require numerous unsupported system calls.
To circumvent this problem, we extend and deploy a MPI stub library~\cite{sorby_mpi_2017}.

\subsection{Relevant HPC (Proxy-)Apps and Benchmarks}\label{ssec:apps}
Instead of relying on a narrow set of cherry-picked applications, we attempt to cover a broad spectrum of typical
scientific/HPC workloads. 
We customize and extend a publicly
available benchmarking framework\footnote{Exact benchmark versions, git commits, inputs, etc.~are provided in our artifacts which are referenced in Section~\ref{sec:aead}.}~\cite{domke_matrix_2021,domke_matrix_2021-1} with a few additional benchmarks
and necessary features to perform the MCA- and gem5-based simulations.
The benchmark complexity ranges from simple kernels to large code bases (O(100,000s) lines-of-code) which are used by vendors for architecture comparisons and used by HPC centers for hardware procurements~\cite{exascale_computing_project_ecp_2018}.
Hereafter, we detail the list of
127 included workloads, summed up across all benchmark suites, which are sized to fit within a single node and which could be simulated with gem5 in a
reasonable time ($\leq\,$six months).


\subsubsection*{Polyhedral Benchmark Suite}
The PolyBench/C suite contains 30 single-threaded, scientific kernels which can be parameterized in
memory occupancy ($\in[\text{\unit[16]{KiB}},\text{\unit[120]{MiB}}]$)~\cite{pouchet_polybenchc_2016}. Unless stated otherwise,
we use the largest configuration.

\subsubsection*{TOP500, STREAM, and Deep Learning Benchmarks}\mbox{}\\
High Performance Linpack (HPL)~\cite{dongarra_linpack_1988} solves a dense system of linear equations $Ax = b$ of size
36,864 in our case. High Performance Conjugate Gradients (HPCG)~\cite{dongarra_hpcg_2015}  applies  a conjugate gradient
solver to a system of linear equation (with sparse matrix $A$). We choose $120^3$ for
HPCG's global problem size. BabelStream~\cite{deakin_gpu-stream_2016} evaluates the memory subsystem of CPUs and
accelerators, and we configure~\unit[2]{GiB} input vectors. Moreover, we implement a micro-benchmark, DLproxy, to isolate
the single-precision GEMM operation ($m=\text{1577088}; n=\text{27}; k=\text{32}$) which is commonly found in 2D deep
convolutional neural networks, such as 224$\times$224 ImageNet classification workloads~\cite{vasudevan_parallel_2017}.

\subsubsection*{NASA Advanced Supercomputing Parallel Benchmarks}
The NAS Parallel Benchmarks (NPB)~\cite{bailey_nas_1991,van_der_wijngaart_nas_2002} consists of nine kernels and proxy-apps
which are common in computational fluid dynamics (CFD). The original MPI-only set has been expanded with ten OpenMP-only
benchmarks~\cite{jin_openmp_1999} and we select the \texttt{class B} input size for all of them.

\subsubsection*{RIKEN's Fiber Mini-Apps and TAPP Kernels}
To aid the co-design of Supercomputer Fugaku, RIKEN developed the Fiber proxy-application set~\cite{riken_aics_fiber_2015},
a benchmark suite representing the scientific priority areas of Japan. Additionally, RIKEN released
scaled-down TAPP kernels~\cite{riken_center_for_computational_science_kernel_2021} of their priority applications
which are tailored for fast simulations with gem5~\cite{kodama_accuracy_2020}. 
Our workloads are as follows:
\emph{FFB}~\cite{guo_basic_2006} with the 3D-flow problem discretized into 50$\times$50$\times$50 sub-regions; 
\emph{FFVC}~\cite{ono_ffv-c_2016} using 144$\times$144$\times$144 cuboids;
\emph{MODYLAS}~\cite{andoh_modylas:_2013} with the \texttt{wat222} workload;
\emph{mVMC}~\cite{misawa_mvmc--open-source_2018} with the strong-scaling test reduced to 1/8th of the samples and 1/3rd of the lattice size;
\emph{NICAM}~\cite{tomita_new_2004} with a single (not 11) simulated day;
\emph{NTChem}~\cite{nakajima_ntchem:_2014} with the H\textsubscript{2}O workload;
\emph{QCD}~\cite{boku_multi-block/multi-core_2012} with the \texttt{class 2} input.

\subsubsection*{Exascale Computing Project Proxy-Applications}
The US-based supercomputing centers curated a co-design benchmarking suite
for their recent exascale efforts~\cite{exascale_computing_project_ecp_2018}. 
We select eleven applications of the aforementioned benchmarking framework with the following workloads.
\emph{AMG}~\cite{park_high-performance_2015} with the \texttt{problem~1} workload;
\emph{CoMD}~\cite{mohd-yusof_co-design_2013} with the 256,000-atom strong-scaling test;
\emph{Laghos}~\cite{dobrev_high-order_2012} modelling a 3D Sedov blast but with 1/6th of the timesteps;
\mbox{\emph{MACSio}}~\cite{dickson_replicating_2016} with an $\approx\,$\unit[1.14]{GiB} data dump distributed across many \texttt{JSON} files;
\emph{MiniAMR}~\cite{heroux_improving_2009} simulating a sphere moving diagonally through 3D space;
\emph{MiniFE}~\cite{heroux_improving_2009} with 128$\times$128$\times$128 grid size;
\emph{MiniTri}~\cite{wolf_task-based_2015} testing triangle- and largest clique-detection on \texttt{BCSSTK30} (MatrixMarket~\cite{boisvert_matrix_1997});
\emph{Nekbone}~\cite{argonne_national_laboratory_nek5000_2022} with 8,640 elements and polynomial order of~8;
\emph{SW4lite}~\cite{petersson_users_2017} simulating a \texttt{pointsource};
\emph{SWFFT}~\cite{habib_hacc:_2016} with 32 forward and backward tests for a 128$\times$128$\times$128
grid;
\emph{XSBench}~\cite{tramm_xsbench_2014} with the \texttt{small} problem and 15 million particle lookups.

\subsubsection{SPEC CPU \& SPEC OMP Benchmarks}
The Standard Performance Evaluation Corporation~\cite{standard_performance_evaluation_corporation_specs_2020} offers,
among others, two HPC-focused benchmark suits: SPEC CPU\textsuperscript{\textregistered} 2017[speed] (ten integer-heavy, single-threaded; ten
OpenMP-parallelized, floating-point benchmarks) and SPEC OMP\textsuperscript{\textregistered} 2012 (14 OpenMP-parallelized benchmarks). All SPEC tests hereafter
are based on non-compliant runs with the \texttt{train} input configuration.

\section{MCA-based Simulation Results}\label{sec:MCAeval} 



Sections~\ref{ssec:L1simVali} and~\ref{ssec:L1simResults} are dedicated to our MCA-based estimation of the upper bound on performance improvement with abundant L1 cache. 
First, we evaluate the accuracy of this approach, and then apply the novel methodology to our benchmarking sets.



\subsection{MCA-based Simulator Validation}\label{ssec:L1simVali}

During the development of our MCA-based simulator, we implemented numerous micro-benchmarks to fine-tune the
CPI estimation capabilities while comparing the results to an \mbox{Intel}\textsuperscript{\textregistered} Xeon\textsuperscript{\textregistered} processor \mbox{E5-2650v4} (formerly code named Broadwell). Our
micro-benchmarks comprise MPI-/OpenMP-only, MPI+OpenMP, and single-threaded tests (exercising
recursive functions, floating-point- or integer-intensive operations, L1-localised, or stream-like operation).

\begin{figure}[tbp]
    \centering
    \includegraphics[width=.75\linewidth]{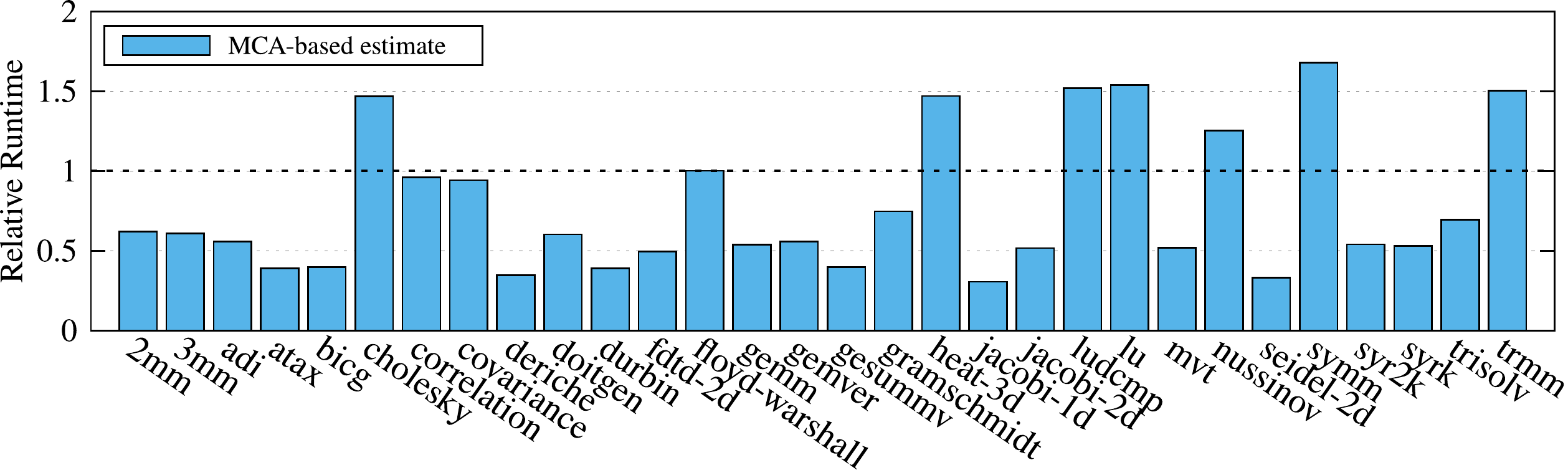}
    \caption{Validation of MCA-based runtime predictions against PolyBench/C \texttt{MINI} with inputs fitting into L1D; Relative runtime shown (vs.~Intel E5-2650v4 measurements); Values $\leq\,$1 show prediction of faster execution}
    \label{fig:l1simvali}
\end{figure}

Needless to say, applying MCA-based simulations to full workloads or complex application kernels is still
error-prone, since these tools are designed to analyze small Assembly sequences without guarantee for
accurate absolute performance numbers. Regardless, we validate the current status of our tool
using PolyBench/C with \texttt{MINI} inputs. In theory, these input sizes ($\approx\,$\unit[16]{KiB}) should
all fit into the \unit[32]{KiB} L1D cache of the Broadwell. Hence, measuring the kernel execution time for these
PolyBench tests should yield numbers close to MCA-based runtime estimates. For the baseline measurements,
we set all cores of the Broadwell to \unit[2.2]{GHz}, set the uncore to~\unit[2.7]{GHz}, and disable turbo boost; compile each workload with Intel's
Parallel Studio XE\footnote{For details of flags, tools, versions, and executions
environments, please refer to Section~\ref{sec:aead}.}, and execute every test for 100 times (since
many only run for a few \unit[]{ms}) to determine the fastest possible execution time.
The difference between the real baseline results and our MCA-based estimates is visualized in
Figure~\ref{fig:l1simvali} as projected relative runtime difference.

The data shows that on average our MCA-based method slightly overestimates: MCA approach predicts faster execution times then it should. Only seven out of 30 workloads are expected to run slower than what we observe on the real Broadwell (i.e., y-value $\leq$1).
For eight of the PolyBench tests, our tool estimates the runtime to be over 2x faster than our measurements.
Hence, we can conclude that for 73\% of the micro-benchmarks, the MCA-based method is reasonably accurate: within
2x slower-to-2x faster. While a 2x discrepancy might appear high, we have to point out that our
cross-validations using SST~\cite{rodrigues_improvements_2012,voskuilen_analyzing_2016} and third-party
gem5 models~\cite{akram_validation_2019} for Intel CPUs yield similar inaccuracies\footnote{A large-scale
survey of academic simulators in realistic scenarios, beyond carefully selected and tuned micro-kernels, is---in our humble opinion---consequential, and yet outside the scope of this paper. Although, reference~\cite{akram_survey_2019} provides a data point.}, but our MCA-based method is substantially faster.

Another indicator for the accuracy of our MCA-approach can be drawn from DGEMM (double precision \texttt{gemm} benchmark in Figure~\ref{fig:l1simvali}).
Theoretically, DGEMM
performs close to peak and is not memory-bound for large matrices, and hence the measured runtime and MCA-based
estimates are expected to match. Unfortunately, PolyBench's \unit[]{Gflop/s} rate for \texttt{gemm} is far
from peak (due to its hand-coded loop-nest), and therefore we replace it with an Intel MKL-based implementation
of equal matrix dimensions. For the PolyBench input sizes $\texttt{MINI},\ldots,\texttt{EXTRALARGE}$ in our MKL-based
implementation, our MCA tool estimates a faster runtime by 6.4x, 75\%, 11\%, 1.9\%, and 1.5\%, respectively.
This closely matches the achievable single-core \unit[]{Gflop/s} of the \mbox{E5-2650v4}: for \texttt{MINI}
and the MKL-based runs, we measure only \unit[2]{Gflop/s}, while for \texttt{EXTRALARGE} we peak out at the expected
\unit[32]{Gflop/s}. The low \unit[]{Gflop/s} measurements for \texttt{MINI} (and \texttt{SMALL}) demonstrate
that MKL is not yet compute-bound, and hence causes the 6.4x (and 75\%) misprediction.

\subsection{Speedup-potential with Unrestricted Locality}
\label{ssec:L1simResults}

In this Section, we take on the entire benchmark suite from Section~\ref{ssec:apps} with the MCA-based approach and evaluate their speedup potential when all data fits into L1.

\begin{figure*}[tbp]
    \includegraphics[width=\linewidth]{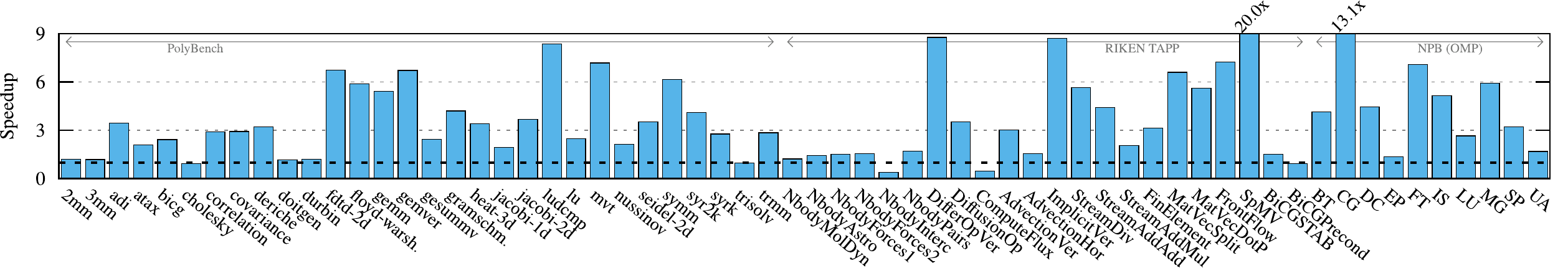}
    \hfil
    \includegraphics[width=\linewidth]{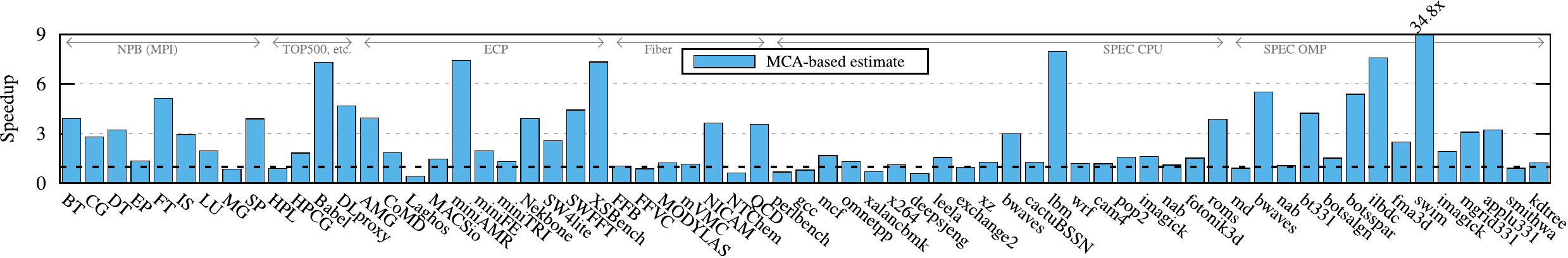}
    \caption{Projected speedup against a baseline dual-socket Intel Broadwell E5-2650v4 system while assuming all data fits into L1D with ``optimistic'' load-to-use latency; Top row, left to right: PolyBench, RIKEN TAPP kernels, NPB~(OMP); Bottom row, left to right: NPB (MPI), TOP500 etc., ECP proxies, RIKEN Fiber apps, SPEC CPU[int/single] and CPU[float/OMP], SPEC OMP}
    \label{fig:l1simresults}
\end{figure*}

The baseline measurements for the speedup estimates are conducted on a dual-socket Intel Broadwell E5-2650v4 system with 48 cores (2-way hyper-threading enabled, cores are set to~\unit[2.2]{GHz}, turbo boost disabled).
For all listed benchmarks, excluding SPEC CPU and OMP, we focus on the solver times only, i.e., we ignore data initialization and
post-processing phases. Since most proxy-apps are parallelized with MPI and/or OpenMP, we
perform an initial sweep of possible configurations of ranks and threads to determine the fastest
time-to-solution (TTS) for our strong-scaling benchmarks, and the highest figure-of-merit (as reported by the
benchmarks) for weak-scaling workloads. The highest performing configurations is executed ten times to determine
the TTS of the kernel as our reference point in Figure~\ref{fig:l1simresults}.

The same MPI/OMP configurations are then used for our MCA-based estimate. Under the assumption
that some MPI-parallized benchmarks experience imbalances, we randomly sample up to nine ranks
(in addition to rank~0)\footnote{Sampling at most ten out of all MPI ranks should not substantially alter
the result but saves resources, since we have to execute SDE once per rank.}, execute the selected rank with
Intel SDE (and the remaining ranks normally), and calculate the estimated runtime using Equation~\eqref{eq:mca}
and the~\unit[2.2]{GHz} processor frequency. The resulting runtime estimate is divided by the measured runtime to
determine the upper-bound speedup potential per application when all its data would fit into L1D,
see Figure~\ref{fig:l1simresults}.

For PolyBench/C workloads, we see similar speedup trends as for its smallest inputs which we used in \fixme{Figure~\ref{fig:l1simvali}}, although the expected speedup for \texttt{EXTRALARGE} increases to a peak of~8.4x for the ludcomp kernel.
Only four kernels show no performance increase, presumably by being compute-bound and not bandwidth-bound: 2mm, 3mm, doitgen, and trisolv.
Overall, the MCA-based approach estimates a geometric mean~($GM$) speedup of 2.9x from fitting all data into L1D.
RIKEN's TAPP kernels benefit the most from unrestricted locality. Especially kernel~20 (SpMV), which represents
one core function of the FFB application, shows a speedup of~20x. Altogether, we see a projection of
($GM$=)2.6x increased performance, but also two cases (kernels 5 and 9) where the MCA tool estimates
a $\approx\,$50\% slowdown. These two are from GENESIS~\cite{jung_genesis_2015} and NICAM,
respectively, but as detailed in Section~\ref{ssec:L1simVali}, some inaccuracy is expected as the trade-off for the faster simulation time.

NPB's OpenMP version of a conjugate gradient (CG) solver is another workload with a large theoretical performance gain of~13.1x. 
In total, we expect a ($GM$=)3x gain for all NAS Parallel Benchmarks; specifically, ($GM$=)4x for the OpenMP versions and ($GM$=)2.3x for the MPI versions. The potential gain
for CG is not surprising, since these solvers are predominantly bound by memory bandwidth and are sensitive to memory latency~\cite{dongarra_new_2016}.
High Performance Linpack is unsurprisingly not expected to gain any performance by placing all
its data into L1 cache, as this benchmarks is compute-bound. In fact, our MCA tool expected
a small runtime decrease of~11\%. By contrast, DLproxy, which uses MKL's SGEMM, would benefit
from a large L1, since MKL cannot achieve peak~\unit[]{Gflop/s} for the tall/skinny matrix in this
workload (cf.~Section~\ref{ssec:apps}).
XSBench and miniAMR show the highest gains for ECP's and RIKEN's proxy-apps, with a value of 7.3x and 7.4x,
respectively. This appears to be in line with expectation from the roofline characteristics of the benchmarks
when measured on a similar compute node~\cite{domke_double-precision_2019}.

A deeper look at roofline analysis in~\cite{domke_double-precision_2019} reveal that there is no strong correlation
between the position of an application on the roofline model and the expected performance gain from solely running out of L1D cache.
We speculate that other, hidden bottlenecks are exposed by our MCA approach, such as data
dependencies and lack of concurrency in the applications, which limit the expected speedup.
Apart from noticeable outliers in the expected speedup, such as lbm, ilbdc, and especially swim,
the potential from enlarged L1D is rather slim for SPEC, and only ($GM$=)1.9x runtime reduction
can be expected across all 34 workloads.

\section{gem5-based Simulation Results}\label{sec:GEMeval} 
In Section~\ref{ssec:gem5conf}, we detail our choice for the simulated architectures in gem5.
Similarly structured to the MCA-based simulations, Sections~\ref{ssec:gem5Vali} and~\ref{ssec:gem5Results} highlight our validation of gem5 for our proposed CPU architectures and evaluate numerous benchmarks and proxy applications on said architecture, and we summarize the results in Section~\ref{ssec:gem5Sum}.

\subsection{\proc{} CMG Models in gem5 and A64FX\TTT{S} Baseline}\label{ssec:gem5conf}

\begin{table}[btp]
    \caption{Chip area and simulator configurations for gem5}
    \label{tbl:gem5params}
    \centering
    {\scriptsize
        \begin{tabularx}{0.75\linewidth}{l@{\extracolsep{\fill}}ZZZZ}
            \toprule
            & \tH{A64FX\TTT{S}} & \tH{A64FX\TSS{32}} & \tH{\procC{}} & \tH{\procA{}} \\
            \midrule 
            {Cores}     & 12    & 32    & 32    & 32    \\
            {CMGs}      &  4    &  4    & 16    & 16    \\
            \addlinespace[.3em]
            {Core config.} & \multicolumn{4}{c}{Arm v8.2 + SVE, \unit[512]{bit} SIMD, \unit[2.2]{GHz}, OoO 128 ROB entries, dispatch width 4} \\
            {Branch pred.} & \multicolumn{4}{c}{Bi-mode: \unit[16]{K} global predictor, \unit[16]{K} choice predictor} \\
            {Per-core L1D} & \multicolumn{4}{c}{\unit[64]{KiB}  4-way set-assoc, 3 cycles, adjacent line prefetcher} \\
            \midrule 
            & \multicolumn{4}{c}{\textsc{L2 cache per CMG}:} \\
            L2 size &  \multicolumn{2}{c}{\makebox[0pt]{\unit[8]{MiB}}}  & \unit[256]{MiB} & \unit[512]{MiB} \\
            {BW} & \multicolumn{2}{c}{\makebox[0pt]{$\sim\,$\unit[800]{GB/s}}}  & $\sim\,$\unit[800]{GB/s} & $\sim\,$\unit[1600]{GB/s} \\
            \addlinespace[.3em]
            & \multicolumn{4}{c}{\textsc{L2 Cache Aggregated}:} \\
            L2 size &  \multicolumn{2}{c}{\makebox[0pt]{\unit[32]{MiB}}}  & \unit[4096]{MiB} & \unit[8192]{MiB} \\
            {BW} &  \multicolumn{2}{c}{\makebox[0pt]{$\sim\,$\unit[3.2]{TB/s}}}  & $\sim\,$\unit[12.8]{TB/s} & $\sim\,$\unit[25.6]{TB/s} \\
            \addlinespace[.3em]
            {L2 config.} & \multicolumn{4}{c}{16-way set-associative, 37 cycles, inclusive, \unit[256]{B} block} \\
            \midrule 
            {Main Memory} & \multicolumn{4}{c}{\unit[32]{GiB} HBM2, 4 channels, \unit[256]{GB/s}} \\
            \bottomrule
        \end{tabularx}
    }
\end{table}

As we discussed in Section~\ref{ssec:n64fx}, we envision one \proc{} CMG to have 32 cores,~\unit[384]{MiB} L2 cache, and~\unit[1.6]{TB/s}
L2 bandwidth. 
Regretfully, gem5 (at least RIKEN's version) can only be configured with L2 cache sizes that are 2\TSS{X}, and therefore
we either have to scale up or down \proc{}'s L2 cache size. Hence, we explore both as distinct options, one conservative and one
technologically aggressive configuration. The conservative option, called \procC{}, is limited to~\unit[256]{MiB} L2 cache at~$\sim\,$\unit[800]{GB/s},
while the aggressive version, \procA{}, doubles both values, to~\unit[512]{MiB} and~$\sim\,$\unit[1.6]{TB/s}, respectively.

Starting at a baseline, i.e., a simulated version of A64FX which we label as A64FX\TTT{S}, and in order to materialize
the properties of the \proc{} CMG (cf.~Section~\ref{ssec:n64fx}), we modify three parameters in our gem5 model.
We modify:
\textbf{(i)} the number of cores in the system to match 32 (up from A64FX\TTT{S}' baseline of 12);
\textbf{(ii)} the size of the total L2 cache to match the capacity of the eight stacked layers (\unit[256/512]{MiB}, up from A64FX\TTT{S}'~L2 size of~\unit[8]{MiB} per CMG); and
\textbf{(iii)} we adjust the number of L2 banks in \procA{} to control the bandwidth.

We introduce
a fourth gem5 configuration, called A64FX\TSS{32}, which simulates one baseline A64FX\TTT{S} CMG but with 32 cores. These four configurations
A64FX\TTT{S} $\to$A64FX\TSS{32} $\to$\procC{} $\to$\procA{} should allow us to determine the speedup gains from the larger core count
and larger L2 cache, individually. The core frequency is universally set to~\unit[2.2]{GHz}.
Table~\ref{tbl:gem5params} summarizes the four gem5 configurations.



\subsection{gem5-based Simulation and Configuration Validation}\label{ssec:gem5Vali}

\begin{figure}[tbp]
    \centering
    \begin{subfigure}[b]{0.485\textwidth}
    \includegraphics[width=\linewidth]{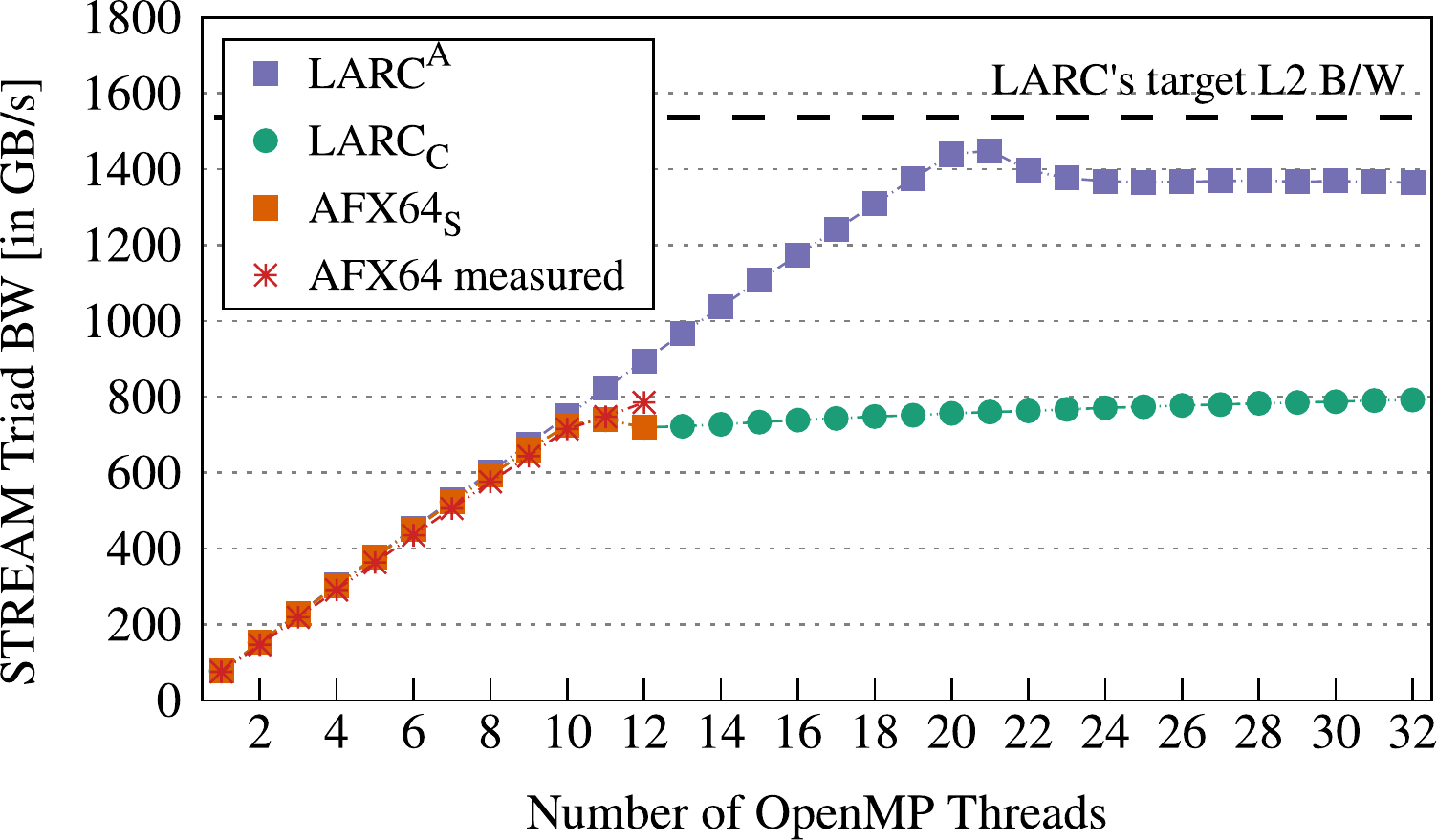}
    \caption{Validation using fixed \unit[128]{KiB} vectors per core}
    \label{fig:gem5valid-cores}
    \end{subfigure}
    \hfill
    \begin{subfigure}[b]{0.485\textwidth}
    \includegraphics[width=\linewidth]{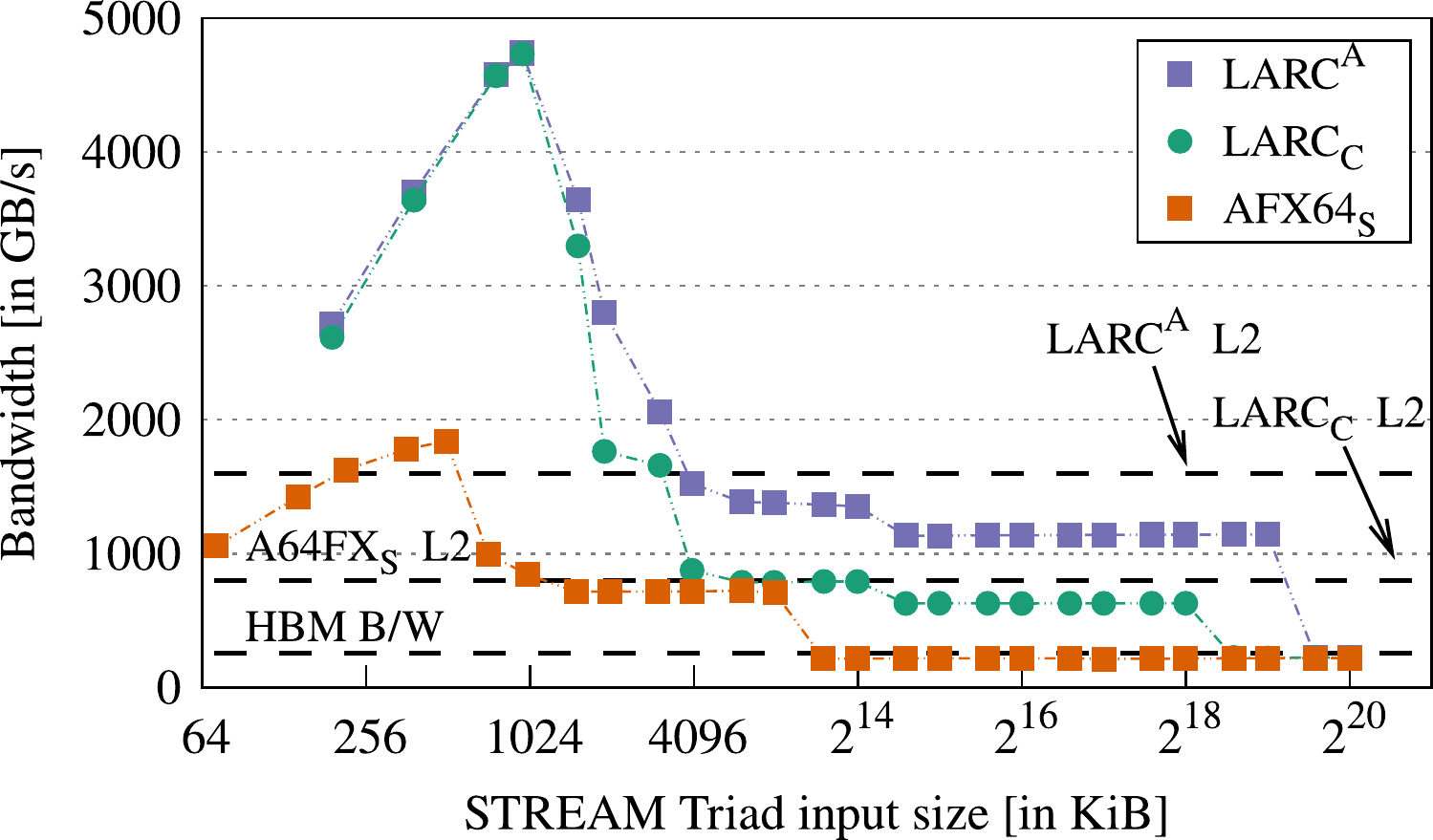}
    \caption{Validation using input range from few \unit[]{KiB} to \unit[1]{GiB}}
    \label{fig:gem5valid-size}
    \end{subfigure}
    \caption{Validation with simulated STREAM Triad; Both LARC configurations with 32 cores; A64FX\TTT{S} scaled to 12 cores; Real A64FX measurements on 1 CMG for reference; Dashed lines highlight trend (not measured)}
    \label{fig:gem5valid-both}
\end{figure}

We perform OpenMP tests to verify our gem5 simulator for up to 32 cores. For the L2 cache size and bandwidth
changes, we employ a STREAM Triad benchmark, parameterized to avoid cache line conflicts among participating
threads. Splitting the A64FX\TTT{S} CMG L2 cache into 12 chunks (one per thread) yields a working size of~\unit[683]{KiB}.
Hence, the three~\unit[128]{KiB} vectors of the Triad operation will fit into the L2 cache. We increase the total vector
size in proportion to the number of threads and test the achievable L2 bandwidth for \procC{} and \procA{}.
Additionally, Figure~\ref{fig:gem5valid-cores} includes the baseline A64FX\TTT{S} CMG scaled to 12 cores.
The simulation shows that \procC{}'s L2 bandwidth peaks out at~\unit[792]{GB/s} and \procA{}'s bandwidth
goes up to~\unit[1450]{GB/s} for this particular test case, which is, respectively, 1\% and 9\% lower than our estimates shown above.
The baseline A64FX\TTT{S} closely matches the bandwidth of the real A64FX CPU executing
this test.

Another validation test we perform is setting the number of cores to the maximum (12 and 32, respectively)
and scale the vector size from~\unit[2]{KiB} per core to a total of~\unit[1]{GiB} for the three vectors.
Figure~\ref{fig:gem5valid-size} shows the results for this simulation. In the memory range of tens to hundreds
of KiB, the Triad operation can be done from L1 cache, for which \procC{} and \procA{} show higher bandwidth.
Their 2.7x higher core count results in 2.6x higher aggregated L1 bandwidth. For the Triad, for the memory sizes that fit into L2 cache, we see a behavior similar to Figure~\ref{fig:gem5valid-cores}. Past~\unit[8]{MiB}, the A64FX\TTT{S}
configuration shows the expected bandwidth drop to HBM2 level, while for \procC{} and \procA{}, the expected
L2 cache bandwidth is maintained until~\unit[256]{MiB} and~\unit[512]{MiB}, respectively. This validates that
our gem5 settings yield the expected LLC characteristics.

\begin{figure}[tbp]
    \centering
    \includegraphics[width=\linewidth]{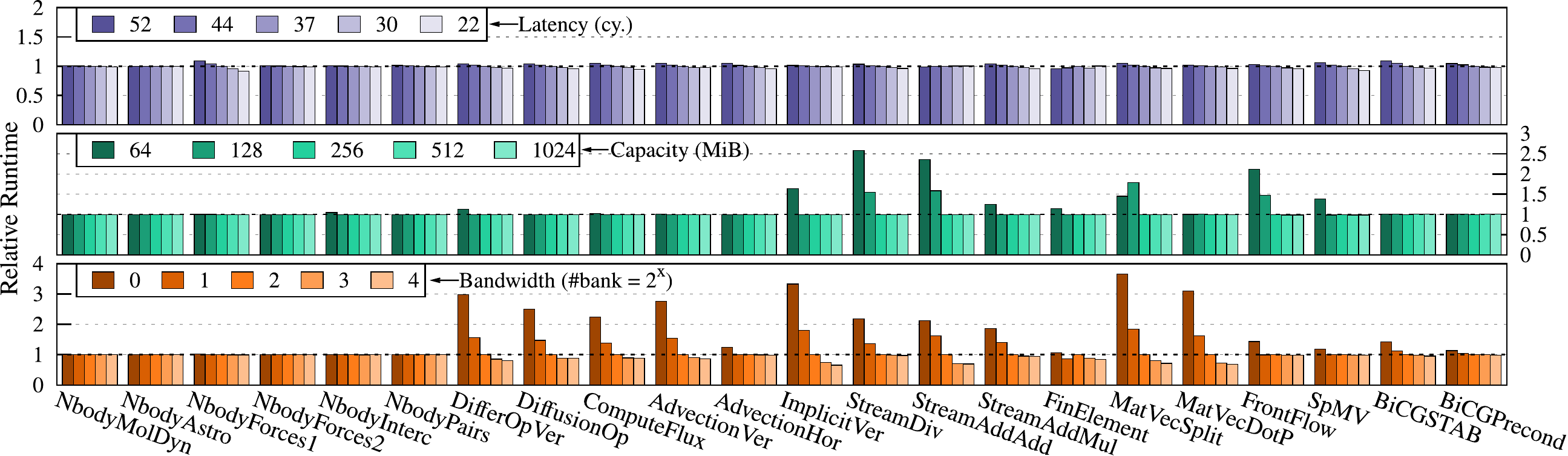}
    \caption{Sensitivity study of cache parameters using RIKEN's TAPP~\cite{riken_center_for_computational_science_kernel_2021} kernels; Relative runtime compared to \procC{} baseline (37 cycle latency, \unit[256]{MiB}, 2 bankbits; middle bar among the five) is shown; Top row: L2 latency modified; Middle row: L2 capacity; Bottom row: adjusting L2 bandwidth via bankbits (\#banks = $2^x$)}
    \label{fig:sensitivity}
\end{figure}

Lastly, to validate the \proc{} configuration and to see the changes applied to more complex science kernels, we perform a sensitivity analysis of
cache parameters with the RIKEN TAPP kernels. In Figure~\ref{fig:sensitivity}, we vary L2 cache access latency, size, and bandwidth in ranges beyond
our \procC{} and \procA{} target architectures. This analysis will help us in adjusting our expectations when future \proc{}-like architectures
deviate from our design parameters, e.g., by stacking less SRAM layers or having higher L2 access latency. In this parameter sweep,
\procC{} will be the baseline and we vary one parameter while keeping the others fixed. The top row of Figure~\ref{fig:sensitivity} shows the
latency sweep, where we choose 22 cycles as best latency (which is 2$\times$ the data load latency from L1 for SVE instructions in A64FX). The worst case
of 52 cycles is equidistant to our baseline in the opposite direction, and two additional latencies are selected in between. Similarly, we adjust
the L2 size (middle row; simulating more or less SRAM stacks or a larger or smaller semiconductor process nodes) and L2 bank bits in gem5, see
bottom row of Figure~\ref{fig:sensitivity}. The latter indirectly controls the L2 bandwidth of the simulated architectures.
The latency change has minimal impact, since HPC applications are typically not latency bound. However, the L2 cache capacity and bandwidth can
have a significant impact on performance, as expected, since they determine the amount of data that can be stored and accessed quickly. For some of
the TAPP kernels, though, the performance is unaffected by these parameters\footnote{The MatVecSplit oddity (runtime increases for
\unit[128]{MiB}) needs further investigation. It shows an enlarged counter of \texttt{LoadLockedRequests}---this artifact could be attributed to software (such as barrier implementation in the OpenMP runtime).}, since these kernels are actually shrunk-down versions specifically
designed for cycle-level architecture simulations, and therefore have low memory footprint.

\subsection{Speedup-potential with Restricted Locality}\label{ssec:gem5Results}

To further refine our projections gained by abundant cache, we proceed with the
cycle-level simulations of the proxy-applications and benchmarks listed in Section~\ref{ssec:apps}.

We compile all benchmarks with Fujitsu's Software
Technical Computing Suite (v4.6.1) targeting the real A64FX, and simulate the single-rank workloads in gem5 for
our four configurations.
Unfortunately, three of our MPI-based benchmarks require multi-rank MPI: \mbox{MODYLAS},
NICAM, and NTChem, and hence
we omit them. Furthermore, we skip the MPI-only versions of NPB.
%
Hereafter, we only report proxy applications and benchmarks which ran to completion within gem5 (i.e., gem5-crashes or simulated application-crashes are excluded when infeasible to patch, and simulations exceeding the 6-months time limit are ignored).
%

\begin{figure}[tbp]
    \centering
    \includegraphics[width=\linewidth]{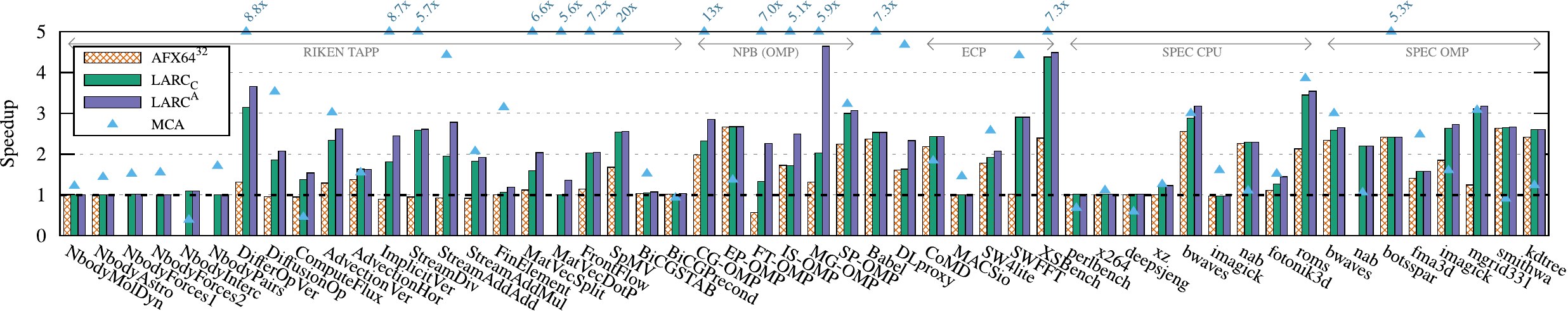}
    \caption{gem5-based, simulated speedups of A64FX\TSS{32}, \procC{} and \procA{} in comparison to baseline A64FX\TTT{S}; Left to right: RIKEN TAPP kernels, NPB (OMP), TOP500 etc., ECP proxies, SPEC CPU[int/single] and CPU[float/OMP], SPEC OMP; Added MCA-based estimations from Fig.~\ref{fig:l1simresults} for reference; TAPP kernels \mbox{3--6} (multiple Nbody kernels) and 18 (MatVecDotP) are limited to 12 threads, hence we omit A64FX\TSS{32}; Missing benchmarks (cf.~Fig.~\ref{fig:l1simresults}) primarily due to gem5 issues or exceeding simulation time limit. PolyBench results (single core) are also omitted due to limited speedup across all of them and no noteworthy outliers.}
    \label{fig:gem5results}
\end{figure}

The per-configuration speedup is given relative to the baseline A64FX\TTT{S} configuration.
We exclude initialization and post-processing times, and measure only the main kernel runtime, except for the SPEC benchmarks as described in Section~\ref{ssec:L1simResults}.
These results are presented in Figure~\ref{fig:gem5results} and show the effects of the gradual expansion of simulated resources.
The average (single CMG) speedups from \procC{} and \procA{} are $\approx\,$1.9x and $\approx\,$2.1x, respectively, with some applications reaching $\approx\,$4.4x for \procC{} and $\approx\,$4.6x for \procA{}.

As expected, most benchmarks benefit from the additional cores and cache capacity, most prominently MG-OMP which gains a small speedup of $\approx\,$1.3x from the extra cores, $\approx\,$2x speedup from the extra cache, and with~\unit[512]{MiB} cache and higher bandwidth reaches $\approx\,$4.6x speedup.
Comparable incremental improvements with all three architecture steps are observable in other workloads, such as TAPP kernel~7 (DifferOpVer) and~17 (MatVecSplit), showing good scaling on multiple cores and being memory-bound since they benefit from the additional cores and cache capacity.
TAPP kernels 19 and 20, XSBench, roms, and \mbox{imagick~(SPEC OMP)} show similar gain in runtime, but the difference between \procC{} and \procA{} is smaller, implying that the problem size either fits into the \unit[256]{MiB} L2 (e.g., XSBench) or the workload arrives at a point of diminishing returns from the~2x larger cache.
TAPP kernels 8, 9, 12--15, and FT-OMP suffer a slowdown from cache contention in A64FX\TSS{32}.
\procC{} and \procA{} avoid the cache contention, resulting in speedups similar to the benchmarks discussed earlier.
EP-OMP, CoMD, and other compute-bound benchmarks benefit only from the higher core count, with both \proc{}s providing  similar speedup as A64FX\TSS{32}.

\begin{table}[tbp]
    \caption{L2 cache-miss rate [in \%] of representative proxies}
    \label{tbl:gem5missrate}
    \centering
    \setlength{\tabcolsep}{4pt}
    {\scriptsize
        \begin{tabularx}{.75\columnwidth}{rRRRR}
            \toprule
            \tH{Proxy-App}  & \tR{A64FX\TTT{S}} & \tR{A64FX\TSS{32}}    & \tR{\procC{}} & \tR{\procA{}} \\
            \midrule
            NICAM's ImplicitVer (kernel 12)	& 36.6	& 47.6	& 10.5	& 9.1   \\
            ADVENTURE's MatVecSplit (kernel 17)   & 46.7  & 49.5  & 48.7  & 34.8  \\
            FFB's FrontFlow (kernel 19)	& 73.8	& 69.6	& 49.1	& 48.9  \\\addlinespace[.2em]
            FT-OMP	    & 11.6	& 48.2	&  6.4	& 3.8   \\
            MG-OMP      & 59.8	& 70.9	& 29.4	& 0.4   \\\addlinespace[.2em]
            XSBench	    & 32.1	& 36.4	&  0.1	& 0.1   \\
            \bottomrule
        \end{tabularx}
    }
\end{table}

Expectedly, single-threaded workloads (all of PolyBench's benchmarks) show little to no improvements over A64FX\TTT{S}, i.e.,\ they do not benefit from more cores.
However, these
benchmarks also do not show a performance gain from a larger 3D-stacked L2 cache, albeit their working set size exceeding A64FX\TTT{S}' \unit[8]{MiB} L2 yet fitting into \proc{}' larger cache. We only see a limited speedup of ($GM$=)4.3\% across
all of them and no noteworthy outliers, and hence omit them in Figure~\ref{fig:gem5results}.
%
%
We attribute other outliers, such as the slowdown of imagick (SPEC-CPU), to similar intrinsic property of the benchmark:
our testing on a real A64FX reveals that imagick has a sweetspot at 8 OpenMP threads, and scales negatively thereafter; and
the TAPP kernels 3--6 and 18 were customized for the 12-core A64FX CMG and cannot run effectively on 32 threads without a rewrite. 
Hence, we limit gem5 to 12 cores for these TAPP kernels, and we see that only the MatVecDotP kernel of the ADVENTURE application~\cite{adventure_project_2019} benefits from a larger L2.
Further proxy-applications and benchmarks missing from Figure~\ref{fig:gem5results}, yet appearing in Figure~\ref{fig:l1simresults},
are the unfavorable result of persistent, repeatable simulator errors---sometimes occurring after months of simulation.

We should note that in some cases the benchmarks' implementation and the quality of the compiler may skew the results,
for instance, BabelStream measuring memory bandwidth on a~\unit[2]{GiB} buffer.
Being unoptimized for A64FX, BabelStream's baseline underperforms in terms of per-core bandwidth
(compared to STREAM Triad tests in Figure~\ref{fig:gem5valid-cores} and~\ref{fig:gem5valid-size})
which in turn results in performance gain when the number of cores increases
to~32.

Overall, the speedup on A64FX\TSS{32} can originate from the following reasons:
\textbf{(i)}~the program is compute-bound (a valid result);
\textbf{(ii)}~the workload exhibits both compute-bound and memory-bound tendencies in different components of a proxy-application (a valid result);
\textbf{(iii)}~the program is highly latency-bound, and hence the speedup can be the result of the larger aggregate L1 cache (a valid result); or
\textbf{(iv)}~a poor baseline resulting in a slightly misleading result.
%

We confirm the validity of attributing improvement to the high capacity L2 by inspecting the L2 cache-miss rates of our gem5 simulations (with the miss rate of some selected examples listed in Table~\ref{tbl:gem5missrate}). The reduction in cache-miss rates reported in the table is consistent with  the performance improvements we observe in Figure~\ref{fig:gem5results}.

\subsection{Summary of the Results}\label{ssec:gem5Sum}
Our gem5 simulations indicate that more than half (31 out of 52) of the applications experienced a larger than two times speedup on \procA{} compared to the baseline A64FX\TTT{S} CMG. For over two-thirds (24 out of 31) of these applications, the performance gains are directly attributed to the larger (3D-stacked) cache, i.e., with at least 10\% gain by either of the two \proc{} configurations over the A64FX\TSS{32} variant. Most notably, out of all the RIKEN TAPP kernels that experienced meaningful speedup on \proc{}, a majority benefited from the expanded cache, rather than the increase in number of cores. This carries particular importance as these kernels are highly tuned for A64FX.
%

%
%
\section{Discussion and Limitations}\label{sec:discussion} 

In this study, we simulated a single \proc{} CMG in gem5, and its potential future effect on common HPC workloads.

\subsection{The Prospect of \proc{}}
In reality, if a \proc{} processor were incepted in 2028, it would contain 16 \proc{} CMGs, which correspond to the same silicon area as the current A64FX CPU, and it is important to understand what impact such a processor would have on the HPC community and its applications. Unfortunately, it is hard to give a conclusive answer to such a forward looking question today. However, if we do ideal scaling of both A64FX and \proc{} CMGs and compare at the full chip level, then a \proc{} system in 2028 could give between 4.91x~(\texttt{xz}; SPEC CPU) and 18.57x~(\texttt{MG-OMP}; NPB) performance improvements over the current A64FX processor with an average improvement of ($GM$=)9.56x for applications that are responsive to larger cache capacity. For applications that do not yet benefit from a larger cache, future studies should (continue to) consider algorithmic improvements~\cite{ltaief_meeting_2021}, as well as investigate the potential of allocating parts of the cache to vary compute capabilities, for example, processing-in-memory~\cite{balasubramonian_near-data_2014} or alternative compute modules, e.g., CGRAs~\cite{podobas_survey_2020-1}.

\subsection{Considerations and Limitations}
Our MCA-based estimation framework only gives a first-order approximation for a hypothetical
CPU with sufficiently large L1 cache to host the entire data structures of a specific workload.
This approach has some advantages and disadvantages and should be used with caution, but it also has
capabilities which we have not yet detailed, such
as estimating the runtime of the same binary/workload for different (ISA-compatible) x86 systems
by simply replacing the MCA target architecture and altering the CPU clock frequency.

We emphasize that we run applications as they are, i.e., without any algorithmic optimizations
to the larger last level cache, in our MCA- and gem5-based simulators. This is also true to our
motivating experiment shown in Figure~\ref{fig:milan-minife}. While the cache capacity of AMD's \mbox{Milan-X} CPU
is about three times that of Milan, it is far from what we envision in 2028.
Hence, our \mbox{Milan-X} results serve as a first-order indication of what SRAM---in its current available SoTA---can offer. 

Another notable aspect, which is outside the main scope of this extrapolation study, is the heat dissipation of CPU cores
in face of the 3D-stacked cache. It has been reported that AMD's \mbox{Milan-X} carefully stacks caches
above areas of the chip that are not used for compute, i.e., mostly above caches~\cite{morgan_milan-x_2022}. Our assumption is that, by 2028, manufacturing technologies will have advanced enough to overcome this limitation.
Yet, for interested readers we provide further details on thermal and power estimates for our hypothetical \proc{} CPU in Section~\ref{ssec:powerthermal}.
\vspace{-5pt}
\section{Related Work}\label{sec:relatedWork} 

\textit{Stacked Memory and Caches:}
The size of LLC has increased for the last 25 years~\cite{iyer_advances_2021}, a trend anticipated to continue
into the future. Yet, 2D IC becomes hard to exploit for additional performance, despite recent attempts
by IBM~\cite{cutress_did_2021,jacobi_real-time_2021}. However, 3D-stacking is becoming a
promising alternative~\cite{hu_stacking_2018}, as demonstrated by AMD's 3D V-Cache~\cite{evers_amd_2022}, Samsung's proposed 3D SRAM stacking solution~\cite{lau_3d_2021} based on~\unit[7]{nm} TSVs, or the most recent study of~\unit[7]{nm} TCI-based 2- and 4-layer SRAM stacks by Shiba et al.~\cite{shiba_7-nm_2022}.
Moreover, academics explored 3D-stacked DRAM cache~\cite{young_accord_2018,hameed_improving_2021}, but these incur
much higher latency and power consumption~\cite{shiba_96-mb_2021,mittal_survey_2016}.
Non-Volatile Memory
is considered as LLC alternative, yet it suffers similar latency issues~\cite{korgaonkar_density_2018}.
Lastly, NVIDIA applied for a patent of an 8-layer memory stack fused with a processor die~\cite{dally_memory_nodate}, theorizing a 50x improvement in bytes-to-flop ratio.
However, what differs our work from the work of our peers is: \textbf{(i)} we focus on the real-world impact of future
caches, several magnitudes larger than those found today.

\textit{Performance modeling tools and methodologies:}
Computer architecture research is often based on simulators, such as the Structural Simulation Toolkit
(SST)~\cite{rodrigues_improvements_2012} or CODES~\cite{cope_codes:_2011}, for efficiently evaluating
and optimizing HPC architectures and applications.
The gem5 simulator, by Binkert et al.~\cite{binkert_gem5_2011}, is widely used by academia and vendors for micro-
and full-system architecture emulation and simulation. It supports validated models for x86~\cite{akram_validation_2019}
and Arm~\cite{kodama_accuracy_2020}. We refer the interested reader to \url{www.gem5.org/publications/} for an
comprehensive library of gem5-based research and derivative works.
However, what differs our work from the work of our peers is:
\textbf{(ii)}: unlike prior work that utilizes (relatively) small kernels, our work operates on large-scale
MPI/OpenMP-parallelized proxy-applications in order to quantify the impact of caches on realistic workloads.
To our knowledge of reported research-driven gem5 simulations, this is the largest scale of cycle-accurate
simulations conducted in terms of the aggregate number of instructions simulated ($6.08 \times 10^{13}$). 

Other methods such as MUSA by Grass et al.~\cite{grass_musa_2016}
are closer to our MCA-based approach, since MUSA uses PIN which is the basis for Intel SDE (used in this study),
but focus on MPI analysis and multi-node workloads. We are not the first to utilize Machine Code Analyzers, 
see~\cite{abel_parametric_2021,laukemann_automated_2018} and derivative works such
as~\cite{chen_bhive_2019,renda_difftune_2020,alappat_execution-cache-memory_2021,mendis_ithemal_2019,corda_memory_2019,caparros_cabezas_parallelism_2011}.
However, what differs our work from the work of our peers is:
\textbf{(iii)}: instead of estimating accurate performance of existing system architectures, our MCA-based
approach tries to gauge the upper-bound in obtainable performance, and exposes bottlenecks better than
the roofline approach, for common HPC applications.

\section{Conclusion}
\label{sec:conclusion}
We aspire to understand the performance implications of emerging SRAM-based die-stacking on future HPC processors. We first designed a methodology to project the upper bound that an infinitely large cache would have on relevant HPC applications. We find that several well-known HPC applications and benchmarks have ample opportunities to exploit an increased cache capacity.

We further expand our study by proposing a hypothetical processor (called \proc{}) in~\unit[1.5]{nm} technology. This processor would have nearly~\unit[6]{GiB} L2 cache memory; compared to our baseline A64FX\TTT{S} CPU architecture with~\unit[32]{MiB} L2 cache. Next, we exercise a single \proc{} CMG using a plethora of HPC applications and benchmarks using the gem5 simulator and contrast the observed performance against the existing A64FX\TTT{S} CMG. We find that the \proc{} CMG would (on average) be 1.9x faster than the corresponding A64FX\TTT{S} CMG, albeit consuming $\frac{1}{4}$th of the area. When area-normalized to the real A64FX CMG (by assuming optimistic ideal scaling), we can expect to see an average boost of 9.56x for cache-sensitive HPC applications by the end of this decade.

Finally, we expect that the larger caches will motivate and facilitate algorithmic advances that in combination with the abundant cache can potentially yield an order of magnitude gain in performance, as demonstrated by the tile low-rank (TLR) approximations~\cite{ltaief_meeting_2021}. These approaches however require a minimum size of the cache to reach their fullest potential. We firmly believe that the combination of high-bandwidth, large, 3D-stacked caches, and algorithmic advances, is the path forward for the next generation of HPC processors when attempting to break the ``memory wall''.
\section{FAIR Commitment by the Authors}\label{sec:aead}

We developed a framework of scripts and git submodules to manage the R\&D of LARC,
to set up the benchmarking infrastructure, and to perform the simulations.
After cloning our repository \url{https://gitlab.com/domke/LARC} (or downloading
the artifacts from \url{https://doi.org/10.5281/zenodo.6420658}),
one has access to all benchmarks (see Section~\ref{ssec:apps}), patches,
scripts, and our collected data. Only minor modifications to the configuration
files should be necessary, such as changing host names, paths to compilers, or
downloading licensed third-party software, before testing on another system.
If users deviate from our OS version (CentOS Linux release 7.9.2009, and
\verb|intel_pstate=disable| kernel parameter) then some additional changes might
be required.\\

\iftoggle{releaseStuffAfterDoubleBlind}{
    \section*{Acknowledgment}

This work was supported by
PRESTO Grant Number JPMJPR20MA, Japan;
the New Energy and Industrial Technology Development Organization (NEDO);
and the AIST/TokyoTech Real-world Big-Data Computation Open Innovation Laboratory (RWBC-OIL).
Furthermore, we thank Masazumi Nakamura from AMD for providing us early access to the Milan-X platform.

JD, EV, BG, and MW designed the study; JD and EV conducted the experiments; BG fixed gem5 issues;
JD, EV, MW, AP, MP, LZ and PC analyzed the results; SM and AP developed the stacked cache model,
and all authors participated in brainstorming and writing the manuscript.
}


\bibliographystyle{ACM-Reference-Format}
\bibliography{main}


\begin{thebibliography}{120}


\ifx \showCODEN    \undefined \def \showCODEN     #1{\unskip}     \fi
\ifx \showDOI      \undefined \def \showDOI       #1{#1}\fi
\ifx \showISBNx    \undefined \def \showISBNx     #1{\unskip}     \fi
\ifx \showISBNxiii \undefined \def \showISBNxiii  #1{\unskip}     \fi
\ifx \showISSN     \undefined \def \showISSN      #1{\unskip}     \fi
\ifx \showLCCN     \undefined \def \showLCCN      #1{\unskip}     \fi
\ifx \shownote     \undefined \def \shownote      #1{#1}          \fi
\ifx \showarticletitle \undefined \def \showarticletitle #1{#1}   \fi
\ifx \showURL      \undefined \def \showURL       {\relax}        \fi
\providecommand\bibfield[2]{#2}
\providecommand\bibinfo[2]{#2}
\providecommand\natexlab[1]{#1}
\providecommand\showeprint[2][]{arXiv:#2}

\bibitem[wes(2023)]%
        {wesling_heterogeneous_2023}
 \bibinfo{year}{2023}\natexlab{}.
\newblock \bibinfo{booktitle}{\emph{Heterogeneous {{Integration Roadmap}} 2023
  {{Edition}} - {{Chapter}} 20: {{Thermal}}}}.
\newblock \bibinfo{type}{{T}echnical {R}eport}. \bibinfo{institution}{{IEEE
  Electronics Packaging Society}}. \bibinfo{pages}{1--39} pages.
\newblock
\newblock
\shownote{\url{https://eps.ieee.org/images/files/HIR_2023/ch20_thermalfinal.pdf}}.


\bibitem[Abel and Reineke(2021)]%
        {abel_parametric_2021}
\bibfield{author}{\bibinfo{person}{Andreas Abel} {and} \bibinfo{person}{Jan
  Reineke}.} \bibinfo{year}{2021}\natexlab{}.
\newblock \bibinfo{title}{A {{Parametric Microarchitecture Model}} for
  {{Accurate Basic Block Throughput Prediction}} on {{Recent Intel CPUs}}}.
\newblock \bibinfo{howpublished}{\url{https://arxiv.org/pdf/2107.14210.pdf}}.
\newblock


\bibitem[{Advanced Micro Devices, Inc}(2021)]%
        {advanced_micro_devices_inc_amd_2021}
\bibfield{author}{\bibinfo{person}{{Advanced Micro Devices, Inc}}.}
  \bibinfo{year}{2021}\natexlab{}.
\newblock \bibinfo{title}{{{AMD Instinct}}\texttrademark{} {{MI250X
  Accelerator}}}.
\newblock
  \bibinfo{howpublished}{\url{https://www.amd.com/en/products/server-accelerators/instinct-mi250x}}.
\newblock


\bibitem[{ADVENTURE Project}(2019)]%
        {adventure_project_2019}
\bibfield{author}{\bibinfo{person}{{ADVENTURE Project}}.}
  \bibinfo{year}{2019}\natexlab{}.
\newblock \bibinfo{title}{Development of Computational Mechanics System for
  Large Scale Analysis and Design --- ADVENTURE Project}.
\newblock \bibinfo{howpublished}{\url{https://adventure.sys.t.u-tokyo.ac.jp/}}.
\newblock


\bibitem[Agarwal et~al\mbox{.}(2003)]%
        {agarwal_single-vsub_2003}
\bibfield{author}{\bibinfo{person}{A. Agarwal}, \bibinfo{person}{Hai Li}, {and}
  \bibinfo{person}{K. Roy}.} \bibinfo{year}{2003}\natexlab{}.
\newblock \showarticletitle{A Single-{{V}}/Sub t/ Low-Leakage Gated-Ground
  Cache for Deep Submicron}.
\newblock \bibinfo{journal}{\emph{IEEE Journal of Solid-State Circuits}}
  \bibinfo{volume}{38}, \bibinfo{number}{2} (\bibinfo{year}{2003}),
  \bibinfo{pages}{319--328}.
\newblock
\urldef\tempurl%
\url{https://doi.org/10.1109/JSSC.2002.807414}
\showDOI{\tempurl}


\bibitem[Akram and Sawalha(2019a)]%
        {akram_survey_2019}
\bibfield{author}{\bibinfo{person}{Ayaz Akram} {and} \bibinfo{person}{Lina
  Sawalha}.} \bibinfo{year}{2019}\natexlab{a}.
\newblock \showarticletitle{A {{Survey}} of {{Computer Architecture Simulation
  Techniques}} and {{Tools}}}.
\newblock \bibinfo{journal}{\emph{IEEE Access}}  \bibinfo{volume}{7}
  (\bibinfo{year}{2019}), \bibinfo{pages}{78120--78145}.
\newblock
\urldef\tempurl%
\url{https://doi.org/10.1109/ACCESS.2019.2917698}
\showDOI{\tempurl}


\bibitem[Akram and Sawalha(2019b)]%
        {akram_validation_2019}
\bibfield{author}{\bibinfo{person}{Ayaz Akram} {and} \bibinfo{person}{Lina
  Sawalha}.} \bibinfo{year}{2019}\natexlab{b}.
\newblock \showarticletitle{Validation of the Gem5 {{Simulator}} for X86
  {{Architectures}}}. In \bibinfo{booktitle}{\emph{2019 {{IEEE}}/{{ACM
  Performance Modeling}}, {{Benchmarking}} and {{Simulation}} of {{High
  Performance Computer Systems}} ({{PMBS}})}}. \bibinfo{publisher}{{IEEE}},
  \bibinfo{address}{{Denver, CO, USA}}, \bibinfo{pages}{53--58}.
\newblock
\urldef\tempurl%
\url{https://doi.org/10.1109/PMBS49563.2019.00012}
\showDOI{\tempurl}


\bibitem[Alappat et~al\mbox{.}(2021)]%
        {alappat_execution-cache-memory_2021}
\bibfield{author}{\bibinfo{person}{Christie Alappat}, \bibinfo{person}{Nils
  Meyer}, \bibinfo{person}{Jan Laukemann}, \bibinfo{person}{Thomas Gruber},
  \bibinfo{person}{Georg Hager}, \bibinfo{person}{Gerhard Wellein}, {and}
  \bibinfo{person}{Tilo Wettig}.} \bibinfo{year}{2021}\natexlab{}.
\newblock \showarticletitle{Execution-{{Cache-Memory}} Modeling and Performance
  Tuning of Sparse Matrix-Vector Multiplication and {{Lattice}} Quantum
  Chromodynamics on {{A64FX}}}.
\newblock \bibinfo{journal}{\emph{Concurrency and Computation: Practice and
  Experience}} (\bibinfo{date}{Aug.} \bibinfo{year}{2021}),
  \bibinfo{pages}{30}.
\newblock
\urldef\tempurl%
\url{https://doi.org/10.1002/cpe.6512}
\showDOI{\tempurl}


\bibitem[Andoh et~al\mbox{.}(2013)]%
        {andoh_modylas:_2013}
\bibfield{author}{\bibinfo{person}{Yoshimichi Andoh}, \bibinfo{person}{Noriyuki
  Yoshii}, \bibinfo{person}{Kazushi Fujimoto}, \bibinfo{person}{Keisuke
  Mizutani}, \bibinfo{person}{Hidekazu Kojima}, \bibinfo{person}{Atsushi
  Yamada}, \bibinfo{person}{Susumu Okazaki}, \bibinfo{person}{Kazutomo
  Kawaguchi}, \bibinfo{person}{Hidemi Nagao}, \bibinfo{person}{Kensuke
  Iwahashi}, \bibinfo{person}{Fumiyasu Mizutani}, \bibinfo{person}{Kazuo
  Minami}, \bibinfo{person}{Shin-ichi Ichikawa}, \bibinfo{person}{Hidemi
  Komatsu}, \bibinfo{person}{Shigeru Ishizuki}, \bibinfo{person}{Yasuhiro
  Takeda}, {and} \bibinfo{person}{Masao Fukushima}.}
  \bibinfo{year}{2013}\natexlab{}.
\newblock \showarticletitle{{{MODYLAS}}: {{A Highly Parallelized
  General-Purpose Molecular Dynamics Simulation Program}} for {{Large-Scale
  Systems}} with {{Long-Range Forces Calculated}} by {{Fast Multipole Method}}
  ({{FMM}}) and {{Highly Scalable Fine-Grained New Parallel Processing
  Algorithms}}}.
\newblock \bibinfo{journal}{\emph{Journal of Chemical Theory and Computation}}
  \bibinfo{volume}{9}, \bibinfo{number}{7} (\bibinfo{year}{2013}),
  \bibinfo{pages}{3201--3209}.
\newblock
\urldef\tempurl%
\url{https://doi.org/10.1021/ct400203a}
\showDOI{\tempurl}


\bibitem[{Argonne National Laboratory}(2022)]%
        {argonne_national_laboratory_nek5000_2022}
\bibfield{author}{\bibinfo{person}{{Argonne National Laboratory}}.}
  \bibinfo{year}{2022}\natexlab{}.
\newblock \bibinfo{title}{{{NEK5000}}}.
\newblock \bibinfo{howpublished}{\url{http://nek5000.mcs.anl.gov}}.
\newblock


\bibitem[Bailey et~al\mbox{.}(1991)]%
        {bailey_nas_1991}
\bibfield{author}{\bibinfo{person}{D.~H. Bailey}, \bibinfo{person}{E. Barszcz},
  \bibinfo{person}{J.~T. Barton}, \bibinfo{person}{D.~S. Browning},
  \bibinfo{person}{R.~L. Carter}, \bibinfo{person}{L. Dagum},
  \bibinfo{person}{R.~A. Fatoohi}, \bibinfo{person}{P.~O. Frederickson},
  \bibinfo{person}{T.~A. Lasinski}, \bibinfo{person}{R.~S. Schreiber},
  \bibinfo{person}{H.~D. Simon}, \bibinfo{person}{V. Venkatakrishnan}, {and}
  \bibinfo{person}{S.~K. Weeratunga}.} \bibinfo{year}{1991}\natexlab{}.
\newblock \showarticletitle{The {{NAS}} Parallel Benchmarks \textendash{}
  Summary and Preliminary Results}. In \bibinfo{booktitle}{\emph{Proceedings of
  the 1991 {{ACM}}/{{IEEE Conference}} on {{Supercomputing}}}}
  \emph{(\bibinfo{series}{{{SC}} '91})}. \bibinfo{publisher}{{ACM}},
  \bibinfo{address}{{New York, NY, USA}}, \bibinfo{pages}{158--165}.
\newblock
\showISBNx{0-89791-459-7}
\urldef\tempurl%
\url{https://doi.org/10.1145/125826.125925}
\showDOI{\tempurl}


\bibitem[Balasubramonian et~al\mbox{.}(2014)]%
        {balasubramonian_near-data_2014}
\bibfield{author}{\bibinfo{person}{Rajeev Balasubramonian},
  \bibinfo{person}{Jichuan Chang}, \bibinfo{person}{Troy Manning},
  \bibinfo{person}{Jaime~H. Moreno}, \bibinfo{person}{Richard Murphy},
  \bibinfo{person}{Ravi Nair}, {and} \bibinfo{person}{Steven Swanson}.}
  \bibinfo{year}{2014}\natexlab{}.
\newblock \showarticletitle{Near-{{Data Processing}}: {{Insights}} from a
  {{MICRO-46 Workshop}}}.
\newblock \bibinfo{journal}{\emph{IEEE Micro}} \bibinfo{volume}{34},
  \bibinfo{number}{4} (\bibinfo{year}{2014}), \bibinfo{pages}{36--42}.
\newblock
\urldef\tempurl%
\url{https://doi.org/10.1109/MM.2014.55}
\showDOI{\tempurl}


\bibitem[Binkert et~al\mbox{.}(2011)]%
        {binkert_gem5_2011}
\bibfield{author}{\bibinfo{person}{Nathan Binkert}, \bibinfo{person}{Bradford
  Beckmann}, \bibinfo{person}{Gabriel Black}, \bibinfo{person}{Steven~K.
  Reinhardt}, \bibinfo{person}{Ali Saidi}, \bibinfo{person}{Arkaprava Basu},
  \bibinfo{person}{Joel Hestness}, \bibinfo{person}{Derek~R. Hower},
  \bibinfo{person}{Tushar Krishna}, \bibinfo{person}{Somayeh Sardashti},
  \bibinfo{person}{Rathijit Sen}, \bibinfo{person}{Korey Sewell},
  \bibinfo{person}{Muhammad Shoaib}, \bibinfo{person}{Nilay Vaish},
  \bibinfo{person}{Mark~D. Hill}, {and} \bibinfo{person}{David~A. Wood}.}
  \bibinfo{year}{2011}\natexlab{}.
\newblock \showarticletitle{The {{Gem5 Simulator}}}.
\newblock \bibinfo{journal}{\emph{ACM SIGARCH Computer Architecture News}}
  \bibinfo{volume}{39}, \bibinfo{number}{2} (\bibinfo{date}{Aug.}
  \bibinfo{year}{2011}), \bibinfo{pages}{1--7}.
\newblock
\showISSN{0163-5964}
\urldef\tempurl%
\url{https://doi.org/10.1145/2024716.2024718}
\showDOI{\tempurl}


\bibitem[Black et~al\mbox{.}(2006)]%
        {black_stacking_2006}
\bibfield{author}{\bibinfo{person}{Bryan Black}, \bibinfo{person}{Murali
  Annavaram}, \bibinfo{person}{Ned Brekelbaum}, \bibinfo{person}{John DeVale},
  \bibinfo{person}{Lei Jiang}, \bibinfo{person}{Gabriel~H. Loh},
  \bibinfo{person}{Don McCaule}, \bibinfo{person}{Pat Morrow},
  \bibinfo{person}{Donald~W. Nelson}, \bibinfo{person}{Daniel Pantuso},
  \bibinfo{person}{Paul Reed}, \bibinfo{person}{Jeff Rupley},
  \bibinfo{person}{Sadasivan Shankar}, \bibinfo{person}{John Shen}, {and}
  \bibinfo{person}{Clair Webb}.} \bibinfo{year}{2006}\natexlab{}.
\newblock \showarticletitle{Die {{Stacking}} ({{3D}}) {{Microarchitecture}}}.
  In \bibinfo{booktitle}{\emph{Proceedings of the 39th {{Annual IEEE}}/{{ACM
  International Symposium}} on {{Microarchitecture}}}}
  \emph{(\bibinfo{series}{{{MICRO}} 39})}. \bibinfo{publisher}{{IEEE Computer
  Society}}, \bibinfo{address}{{USA}}, \bibinfo{pages}{469--479}.
\newblock
\showISBNx{0-7695-2732-9}
\urldef\tempurl%
\url{https://doi.org/10.1109/MICRO.2006.18}
\showDOI{\tempurl}


\bibitem[Boisvert et~al\mbox{.}(1997)]%
        {boisvert_matrix_1997}
\bibfield{author}{\bibinfo{person}{Ronald~F. Boisvert}, \bibinfo{person}{Roldan
  Pozo}, \bibinfo{person}{Karin Remington}, \bibinfo{person}{Richard~F.
  Barrett}, {and} \bibinfo{person}{Jack~J. Dongarra}.}
  \bibinfo{year}{1997}\natexlab{}.
\newblock \showarticletitle{Matrix {{Market}}: {{A Web Resource}} for {{Test
  Matrix Collections}}}. In \bibinfo{booktitle}{\emph{Proceedings of the {{IFIP
  TC2}}/{{WG2}}.5 {{Working Conference}} on {{Quality}} of {{Numerical
  Software}}: {{Assessment}} and {{Enhancement}}}}.
  \bibinfo{publisher}{{Chapman \& Hall, Ltd.}}, \bibinfo{address}{{London, UK,
  UK}}, \bibinfo{pages}{125--137}.
\newblock
\showISBNx{0-412-80530-8}
\newblock
\shownote{\url{http://dl.acm.org/citation.cfm?id=265834.265854}}.


\bibitem[Boku et~al\mbox{.}(2012)]%
        {boku_multi-block/multi-core_2012}
\bibfield{author}{\bibinfo{person}{T. Boku}, \bibinfo{person}{K.~I. Ishikawa},
  \bibinfo{person}{Y. Kuramashi}, \bibinfo{person}{K. Minami},
  \bibinfo{person}{Y. Nakamura}, \bibinfo{person}{F. Shoji},
  \bibinfo{person}{D. Takahashi}, \bibinfo{person}{M. Terai},
  \bibinfo{person}{A. Ukawa}, {and} \bibinfo{person}{T. Yoshie}.}
  \bibinfo{year}{2012}\natexlab{}.
\newblock \showarticletitle{Multi-Block/Multi-Core {{SSOR}} Preconditioner for
  the {{QCD}} Quark Solver for {{K}} Computer}.
\newblock \bibinfo{journal}{\emph{Proceedings, 30th International Symposium on
  Lattice Field Theory (Lattice 2012): Cairns, Australia, June 24-29, 2012}}
  \bibinfo{volume}{LATTICE2012} (\bibinfo{year}{2012}), \bibinfo{pages}{188}.
\newblock
\urldef\tempurl%
\url{https://doi.org/10.22323/1.164.0188}
\showDOI{\tempurl}


\bibitem[Bonshor(2022)]%
        {bonshor_amd_2022}
\bibfield{author}{\bibinfo{person}{Gavin Bonshor}.}
  \bibinfo{year}{2022}\natexlab{}.
\newblock \bibinfo{title}{{{AMD Releases Milan-X CPUs With 3D V-Cache}}:
  {{EPYC}} 7003 {{Up}} to 64 {{Cores}} and 768 {{MB L3 Cache}}}.
\newblock
  \bibinfo{howpublished}{\url{https://www.anandtech.com/show/17323/amd-releases-milan-x-cpus-with-3d-vcache-epyc-7003}}.
\newblock


\bibitem[Cao et~al\mbox{.}(2019)]%
        {cao_survey_2019}
\bibfield{author}{\bibinfo{person}{Kun Cao}, \bibinfo{person}{Junlong Zhou},
  \bibinfo{person}{Tongquan Wei}, \bibinfo{person}{Mingsong Chen},
  \bibinfo{person}{Shiyan Hu}, {and} \bibinfo{person}{Keqin Li}.}
  \bibinfo{year}{2019}\natexlab{}.
\newblock \showarticletitle{A {{Survey}} of {{Optimization Techniques}} for
  {{Thermal-Aware 3D Processors}}}.
\newblock \bibinfo{journal}{\emph{Journal of Systems Architecture}}
  \bibinfo{volume}{97}, \bibinfo{number}{C} (\bibinfo{date}{Aug.}
  \bibinfo{year}{2019}), \bibinfo{pages}{397--415}.
\newblock
\showISSN{1383-7621}
\urldef\tempurl%
\url{https://doi.org/10.1016/j.sysarc.2019.01.003}
\showDOI{\tempurl}


\bibitem[Caparr{\'o}s~Cabezas and {Stanley-Marbell}(2011)]%
        {caparros_cabezas_parallelism_2011}
\bibfield{author}{\bibinfo{person}{Victoria Caparr{\'o}s~Cabezas} {and}
  \bibinfo{person}{Phillip {Stanley-Marbell}}.}
  \bibinfo{year}{2011}\natexlab{}.
\newblock \showarticletitle{Parallelism and {{Data Movement Characterization}}
  of {{Contemporary Application Classes}}}. In
  \bibinfo{booktitle}{\emph{Proceedings of the {{Twenty-Third Annual ACM
  Symposium}} on {{Parallelism}} in {{Algorithms}} and {{Architectures}}}}
  \emph{(\bibinfo{series}{{{SPAA}} '11})}. \bibinfo{publisher}{{Association for
  Computing Machinery}}, \bibinfo{address}{{New York, NY, USA}},
  \bibinfo{pages}{95--104}.
\newblock
\showISBNx{978-1-4503-0743-7}
\urldef\tempurl%
\url{https://doi.org/10.1145/1989493.1989506}
\showDOI{\tempurl}


\bibitem[Chakraborty and Kapoor(2018)]%
        {chakraborty_analysing_2018}
\bibfield{author}{\bibinfo{person}{Shounak Chakraborty} {and}
  \bibinfo{person}{Hemangee~K. Kapoor}.} \bibinfo{year}{2018}\natexlab{}.
\newblock \showarticletitle{Analysing the {{Role}} of {{Last Level Caches}} in
  {{Controlling Chip Temperature}}}.
\newblock \bibinfo{journal}{\emph{IEEE Transactions on Sustainable Computing}}
  \bibinfo{volume}{3}, \bibinfo{number}{4} (\bibinfo{year}{2018}),
  \bibinfo{pages}{289--305}.
\newblock


\bibitem[{Cheese}(2022)]%
        {cheese_amds_2022}
\bibfield{author}{\bibinfo{person}{{Cheese}}.} \bibinfo{year}{2022}\natexlab{}.
\newblock \bibinfo{title}{{{AMD}}'s {{V-Cache Tested}}: {{The Latency
  Teaser}}}.
\newblock
  \bibinfo{howpublished}{\url{https://chipsandcheese.com/2022/01/14/amds-v-cache-tested-the-latency-teaser/}}.
\newblock


\bibitem[Chen et~al\mbox{.}(2019)]%
        {chen_bhive_2019}
\bibfield{author}{\bibinfo{person}{Yishen Chen}, \bibinfo{person}{Ajay
  Brahmakshatriya}, \bibinfo{person}{Charith Mendis}, \bibinfo{person}{Alex
  Renda}, \bibinfo{person}{Eric Atkinson}, \bibinfo{person}{Ond{\v r}ej
  S{\'y}kora}, \bibinfo{person}{Saman Amarasinghe}, {and}
  \bibinfo{person}{Michael Carbin}.} \bibinfo{year}{2019}\natexlab{}.
\newblock \showarticletitle{{{BHive}}: {{A Benchmark Suite}} and {{Measurement
  Framework}} for {{Validating}} X86-64 {{Basic Block Performance Models}}}. In
  \bibinfo{booktitle}{\emph{2019 {{IEEE International Symposium}} on {{Workload
  Characterization}} ({{IISWC}})}}. \bibinfo{publisher}{{IEEE Press}},
  \bibinfo{address}{{Orlando, FL, USA}}, \bibinfo{pages}{167--177}.
\newblock
\urldef\tempurl%
\url{https://doi.org/10.1109/IISWC47752.2019.9042166}
\showDOI{\tempurl}


\bibitem[Chou et~al\mbox{.}(2015)]%
        {chou_bear_2015}
\bibfield{author}{\bibinfo{person}{Chiachen Chou}, \bibinfo{person}{Aamer
  Jaleel}, {and} \bibinfo{person}{Moinuddin~K. Qureshi}.}
  \bibinfo{year}{2015}\natexlab{}.
\newblock \showarticletitle{{{BEAR}}: {{Techniques}} for {{Mitigating Bandwidth
  Bloat}} in {{Gigascale DRAM Caches}}}. In
  \bibinfo{booktitle}{\emph{Proceedings of the 42nd {{Annual International
  Symposium}} on {{Computer Architecture}}}} \emph{(\bibinfo{series}{{{ISCA}}
  '15})}. \bibinfo{publisher}{{Association for Computing Machinery}},
  \bibinfo{address}{{New York, NY, USA}}, \bibinfo{pages}{198--210}.
\newblock
\showISBNx{978-1-4503-3402-0}
\urldef\tempurl%
\url{https://doi.org/10.1145/2749469.2750387}
\showDOI{\tempurl}


\bibitem[Cope et~al\mbox{.}(2011)]%
        {cope_codes:_2011}
\bibfield{author}{\bibinfo{person}{J. Cope}, \bibinfo{person}{N. Liu},
  \bibinfo{person}{Samuel Lang}, \bibinfo{person}{C.~D. Carothers}, {and}
  \bibinfo{person}{Robert~B. Ross}.} \bibinfo{year}{2011}\natexlab{}.
\newblock \showarticletitle{{{CODES}}: {{Enabling Co-Design}} of {{Multi-Layer
  Exascale Storage Architectures}}}. In \bibinfo{booktitle}{\emph{Workshop on
  {{Emerging Supercomputing Technologies}} 2011 ({{WEST}} 2011)}}.
  \bibinfo{publisher}{{OSTI.GOV}}, \bibinfo{address}{{Tuscon, Arizona, USA}},
  \bibinfo{pages}{1--6}.
\newblock
\urldef\tempurl%
\url{https://doi.org/10.2172/1311761}
\showDOI{\tempurl}


\bibitem[Corda et~al\mbox{.}(2019)]%
        {corda_memory_2019}
\bibfield{author}{\bibinfo{person}{Stefano Corda}, \bibinfo{person}{Gagandeep
  Singh}, \bibinfo{person}{Ahsan~Javed Awan}, \bibinfo{person}{Roel Jordans},
  {and} \bibinfo{person}{Henk Corporaal}.} \bibinfo{year}{2019}\natexlab{}.
\newblock \showarticletitle{Memory and {{Parallelism Analysis Using}} a
  {{Platform-Independent Approach}}}. In \bibinfo{booktitle}{\emph{Proceedings
  of the 22nd {{International Workshop}} on {{Software}} and {{Compilers}} for
  {{Embedded Systems}}}} \emph{(\bibinfo{series}{{{SCOPES}} '19})}.
  \bibinfo{publisher}{{Association for Computing Machinery}},
  \bibinfo{address}{{New York, NY, USA}}, \bibinfo{pages}{23--26}.
\newblock
\showISBNx{978-1-4503-6762-2}
\urldef\tempurl%
\url{https://doi.org/10.1145/3323439.3323988}
\showDOI{\tempurl}


\bibitem[Cutress(2021a)]%
        {cutress_amd_2021}
\bibfield{author}{\bibinfo{person}{Ian Cutress}.}
  \bibinfo{year}{2021}\natexlab{a}.
\newblock \bibinfo{title}{{{AMD Demonstrates Stacked 3D V-Cache Technology}}:
  192 {{MB}} at 2 {{TB}}/Sec}.
\newblock
  \bibinfo{howpublished}{\url{https://www.anandtech.com/show/16725/amd-demonstrates-stacked-vcache-technology-2-tbsec-for-15-gaming}}.
\newblock


\bibitem[Cutress(2021b)]%
        {cutress_did_2021}
\bibfield{author}{\bibinfo{person}{Ian Cutress}.}
  \bibinfo{year}{2021}\natexlab{b}.
\newblock \bibinfo{title}{Did {{IBM Just Preview The Future}} of {{Caches}}?}
\newblock
  \bibinfo{howpublished}{\url{https://www.anandtech.com/show/16924/did-ibm-just-preview-the-future-of-caches}}.
\newblock


\bibitem[Dally et~al\mbox{.}({[n.\,d.]})]%
        {dally_memory_nodate}
\bibfield{author}{\bibinfo{person}{William~James Dally},
  \bibinfo{person}{Carl~Thomas Gray}, \bibinfo{person}{Stephen~W. Keckler},
  {and} \bibinfo{person}{James~Michael O'connor}.}
  \bibinfo{year}{[n.\,d.]}\natexlab{}.
\newblock \bibinfo{title}{{{MEMORY STACKED ON PROCESSOR FOR HIGH BANDWIDTH}}}.
\newblock
\newblock
\newblock
\shownote{\url{https://patents.justia.com/patent/20230275068}}.


\bibitem[Deakin et~al\mbox{.}(2016)]%
        {deakin_gpu-stream_2016}
\bibfield{author}{\bibinfo{person}{Tom Deakin}, \bibinfo{person}{James Price},
  \bibinfo{person}{Matt Martineau}, {and} \bibinfo{person}{Simon
  {McIntosh-Smith}}.} \bibinfo{year}{2016}\natexlab{}.
\newblock \showarticletitle{{{GPU-STREAM}} v2.0: {{Benchmarking}} the
  {{Achievable Memory Bandwidth}} of {{Many-Core Processors Across Diverse
  Parallel Programming Models}}}. In \bibinfo{booktitle}{\emph{High
  {{Performance Computing}}}}, \bibfield{editor}{\bibinfo{person}{Michela
  Taufer}, \bibinfo{person}{Bernd Mohr}, {and} \bibinfo{person}{Julian~M.
  Kunkel}} (Eds.). \bibinfo{publisher}{{Springer}}, \bibinfo{address}{{Cham}},
  \bibinfo{pages}{489--507}.
\newblock
\showISBNx{978-3-319-46079-6}


\bibitem[Dennard et~al\mbox{.}(1974)]%
        {dennard_design_1974}
\bibfield{author}{\bibinfo{person}{Robert~H. Dennard},
  \bibinfo{person}{Fritz~H. Gaensslen}, \bibinfo{person}{Hwa-Nien Yu},
  \bibinfo{person}{V.~Leo Rideout}, \bibinfo{person}{Ernest Bassous}, {and}
  \bibinfo{person}{Andre~R. LeBlanc}.} \bibinfo{year}{1974}\natexlab{}.
\newblock \showarticletitle{Design of Ion-Implanted {{MOSFET}}'s with Very
  Small Physical Dimensions}.
\newblock \bibinfo{journal}{\emph{IEEE Journal of Solid-State Circuits}}
  \bibinfo{volume}{9}, \bibinfo{number}{5} (\bibinfo{year}{1974}),
  \bibinfo{pages}{256--268}.
\newblock
\urldef\tempurl%
\url{https://doi.org/10.1109/JSSC.1974.1050511}
\showDOI{\tempurl}


\bibitem[Dickson et~al\mbox{.}(2016)]%
        {dickson_replicating_2016}
\bibfield{author}{\bibinfo{person}{James Dickson}, \bibinfo{person}{Steven
  Wright}, \bibinfo{person}{Satheesh Maheswaran}, \bibinfo{person}{Andy
  Herdmant}, \bibinfo{person}{Mark~C. Miller}, {and} \bibinfo{person}{Stephen
  Jarvis}.} \bibinfo{year}{2016}\natexlab{}.
\newblock \showarticletitle{Replicating {{HPC I}}/{{O Workloads}} with {{Proxy
  Applications}}}. In \bibinfo{booktitle}{\emph{Proceedings of the 1st {{Joint
  International Workshop}} on {{Parallel Data Storage}} \& {{Data Intensive
  Scalable Computing Systems}}}} \emph{(\bibinfo{series}{{{PDSW-DISCS}} '16})}.
  \bibinfo{publisher}{{IEEE Press}}, \bibinfo{address}{{Piscataway, NJ, USA}},
  \bibinfo{pages}{13--18}.
\newblock
\showISBNx{978-1-5090-5216-5}
\urldef\tempurl%
\url{https://doi.org/10.1109/PDSW-DISCS.2016.6}
\showDOI{\tempurl}


\bibitem[Dobrev et~al\mbox{.}(2012)]%
        {dobrev_high-order_2012}
\bibfield{author}{\bibinfo{person}{V. Dobrev}, \bibinfo{person}{T. Kolev},
  {and} \bibinfo{person}{R. Rieben}.} \bibinfo{year}{2012}\natexlab{}.
\newblock \showarticletitle{High-{{Order Curvilinear Finite Element Methods}}
  for {{Lagrangian Hydrodynamics}}}.
\newblock \bibinfo{journal}{\emph{SIAM Journal on Scientific Computing}}
  \bibinfo{volume}{34}, \bibinfo{number}{5} (\bibinfo{year}{2012}),
  \bibinfo{pages}{B606--B641}.
\newblock
\urldef\tempurl%
\url{https://doi.org/10.1137/120864672}
\showDOI{\tempurl}


\bibitem[Domke et~al\mbox{.}( may)]%
        {domke_double-precision_2019}
\bibfield{author}{\bibinfo{person}{Jens Domke}, \bibinfo{person}{Kazuaki
  Matsumura}, \bibinfo{person}{Mohamed Wahib}, \bibinfo{person}{Haoyu Zhang},
  \bibinfo{person}{Keita Yashima}, \bibinfo{person}{Toshiki Tsuchikawa},
  \bibinfo{person}{Yohei Tsuji}, \bibinfo{person}{Artur Podobas}, {and}
  \bibinfo{person}{Satoshi Matsuoka}.} \bibinfo{year}{2019, may}\natexlab{}.
\newblock \showarticletitle{Double-Precision {{FPUs}} in {{High-Performance
  Computing}}: An {{Embarrassment}} of {{Riches}}?}. In
  \bibinfo{booktitle}{\emph{2019 {{IEEE International Parallel}} and
  {{Distributed Processing Symposium}}, {{IPDPS}} 2019, {{Rio}} de {{Janeiro}},
  {{Brazil}}, {{May}} 20-24, 2019}}. \bibinfo{publisher}{{IEEE Press}},
  \bibinfo{address}{{Rio de Janeiro, Brazil}}, \bibinfo{pages}{78--88}.
\newblock
\urldef\tempurl%
\url{https://doi.org/10.1109/IPDPS.2019.00019}
\showDOI{\tempurl}


\bibitem[Domke and Vatai(2021)]%
        {domke_matrix_2021-1}
\bibfield{author}{\bibinfo{person}{Jens Domke} {and} \bibinfo{person}{Emil
  Vatai}.} \bibinfo{year}{2021}\natexlab{}.
\newblock \bibinfo{title}{Matrix {{Engine Study}}}.
\newblock \bibinfo{howpublished}{\url{https://gitlab.com/domke/MEstudy}}.
\newblock


\bibitem[Domke et~al\mbox{.}(2021)]%
        {domke_matrix_2021}
\bibfield{author}{\bibinfo{person}{Jens Domke}, \bibinfo{person}{Emil Vatai},
  \bibinfo{person}{Aleksandr Drozd}, \bibinfo{person}{Chen Peng},
  \bibinfo{person}{Yosuke Oyama}, \bibinfo{person}{Lingqi Zhang},
  \bibinfo{person}{Shweta Salaria}, \bibinfo{person}{Daichi Mukunoki},
  \bibinfo{person}{Artur Podobas}, \bibinfo{person}{Mohamed Wahib}, {and}
  \bibinfo{person}{Satoshi Matsuoka}.} \bibinfo{year}{2021}\natexlab{}.
\newblock \showarticletitle{Matrix {{Engines}} for {{High Performance
  Computing}}: {{A Paragon}} of {{Performance}} or {{Grasping}} at
  {{Straws}}?}. In \bibinfo{booktitle}{\emph{2021 {{IEEE International
  Parallel}} and {{Distributed Processing Symposium}}, {{IPDPS}} 2021,
  {{Portland}}, {{Oregon}}, {{USA}}, {{May}} 17-21, 2021}}.
  \bibinfo{publisher}{{IEEE Press}}, \bibinfo{address}{{Portland, Oregon,
  USA}}, \bibinfo{pages}{1056--1065}.
\newblock


\bibitem[Dongarra(1988)]%
        {dongarra_linpack_1988}
\bibfield{author}{\bibinfo{person}{Jack Dongarra}.}
  \bibinfo{year}{1988}\natexlab{}.
\newblock \showarticletitle{The {{LINPACK Benchmark}}: {{An Explanation}}}. In
  \bibinfo{booktitle}{\emph{Proceedings of the 1st {{International Conference}}
  on {{Supercomputing}}}}. \bibinfo{publisher}{{Springer-Verlag}},
  \bibinfo{address}{{London, UK, UK}}, \bibinfo{pages}{456--474}.
\newblock
\showISBNx{3-540-18991-2}
\newblock
\shownote{\url{http://dl.acm.org/citation.cfm?id=647970.742568}}.


\bibitem[Dongarra et~al\mbox{.}(2015)]%
        {dongarra_hpcg_2015}
\bibfield{author}{\bibinfo{person}{Jack Dongarra}, \bibinfo{person}{Michael
  Heroux}, {and} \bibinfo{person}{Piotr Luszczek}.}
  \bibinfo{year}{2015}\natexlab{}.
\newblock \bibinfo{booktitle}{\emph{{{HPCG Benchmark}}: A {{New Metric}} for
  {{Ranking High Performance Computing Systems}}}}.
\newblock \bibinfo{type}{{T}echnical {R}eport} ut-eecs-15-736.
  \bibinfo{institution}{{University of Tennessee}}.
\newblock
\newblock
\shownote{\url{https://library.eecs.utk.edu/pub/594}}.


\bibitem[Dongarra et~al\mbox{.}(2016)]%
        {dongarra_new_2016}
\bibfield{author}{\bibinfo{person}{Jack Dongarra}, \bibinfo{person}{Michael~A.
  Heroux}, {and} \bibinfo{person}{Piotr Luszczek}.}
  \bibinfo{year}{2016}\natexlab{}.
\newblock \showarticletitle{A New Metric for Ranking High-Performance Computing
  Systems}.
\newblock \bibinfo{journal}{\emph{National Science Review}}
  \bibinfo{volume}{3}, \bibinfo{number}{1} (\bibinfo{year}{2016}),
  \bibinfo{pages}{30--35}.
\newblock
\urldef\tempurl%
\url{https://doi.org/10.1093/nsr/nwv084}
\showDOI{\tempurl}


\bibitem[Esmaeilzadeh et~al\mbox{.}(2011)]%
        {esmaeilzadeh_dark_2011}
\bibfield{author}{\bibinfo{person}{Hadi Esmaeilzadeh}, \bibinfo{person}{Emily
  Blem}, \bibinfo{person}{Renee St.~Amant}, \bibinfo{person}{Karthikeyan
  Sankaralingam}, {and} \bibinfo{person}{Doug Burger}.}
  \bibinfo{year}{2011}\natexlab{}.
\newblock \showarticletitle{Dark {{Silicon}} and the {{End}} of {{Multicore
  Scaling}}}. In \bibinfo{booktitle}{\emph{Proceedings of the 38th {{Annual
  International Symposium}} on {{Computer Architecture}}}}
  \emph{(\bibinfo{series}{{{ISCA}} '11})}. \bibinfo{publisher}{{Association for
  Computing Machinery}}, \bibinfo{address}{{New York, NY, USA}},
  \bibinfo{pages}{365--376}.
\newblock
\showISBNx{978-1-4503-0472-6}
\urldef\tempurl%
\url{https://doi.org/10.1145/2000064.2000108}
\showDOI{\tempurl}


\bibitem[Evers et~al\mbox{.}(2022)]%
        {evers_amd_2022}
\bibfield{author}{\bibinfo{person}{Mark Evers}, \bibinfo{person}{Leslie
  Barnes}, {and} \bibinfo{person}{Mike Clark}.}
  \bibinfo{year}{2022}\natexlab{}.
\newblock \showarticletitle{The {{AMD Next Generation Zen}} 3 {{Core}}}.
\newblock \bibinfo{journal}{\emph{IEEE Micro}} \bibinfo{volume}{42},
  \bibinfo{number}{3} (\bibinfo{year}{2022}), \bibinfo{pages}{7--12}.
\newblock
\urldef\tempurl%
\url{https://doi.org/10.1109/MM.2022.3152788}
\showDOI{\tempurl}


\bibitem[{Exascale Computing Project}(2018)]%
        {exascale_computing_project_ecp_2018}
\bibfield{author}{\bibinfo{person}{{Exascale Computing Project}}.}
  \bibinfo{year}{2018}\natexlab{}.
\newblock \bibinfo{title}{{{ECP Proxy Apps Suite}}}.
\newblock
  \bibinfo{howpublished}{\url{https://proxyapps.exascaleproject.org/ecp-proxy-apps-suite/}}.
\newblock


\bibitem[Gomes et~al\mbox{.}(2020)]%
        {gomes_81_2020}
\bibfield{author}{\bibinfo{person}{Wilfred Gomes}, \bibinfo{person}{Sanjeev
  Khushu}, \bibinfo{person}{Doug~B. Ingerly}, \bibinfo{person}{Patrick~N.
  Stover}, \bibinfo{person}{Nasirul~I. Chowdhury}, \bibinfo{person}{Frank
  O'Mahony}, \bibinfo{person}{Ajay Balankutty}, \bibinfo{person}{Noam Dolev},
  \bibinfo{person}{Martin~G. Dixon}, \bibinfo{person}{Lei Jiang},
  \bibinfo{person}{Surya Prekke}, \bibinfo{person}{Biswajit Patra},
  \bibinfo{person}{Pavel~V. Rott}, {and} \bibinfo{person}{Rajesh Kumar}.}
  \bibinfo{year}{2020}\natexlab{}.
\newblock \showarticletitle{8.1 {{Lakefield}} and {{Mobility Compute}}: {{A 3D
  Stacked}} 10nm and {{22FFL Hybrid Processor System}} in 12\texttimes
  12mm{\textsuperscript{2}}, 1mm {{Package-on-Package}}}. In
  \bibinfo{booktitle}{\emph{2020 {{IEEE International Solid- State Circuits
  Conference}} - ({{ISSCC}})}}. \bibinfo{publisher}{{IEEE Press}},
  \bibinfo{address}{{San Francisco, CA, USA}}, \bibinfo{pages}{144--146}.
\newblock
\urldef\tempurl%
\url{https://doi.org/10.1109/ISSCC19947.2020.9062957}
\showDOI{\tempurl}


\bibitem[{Gottlieb} et~al\mbox{.}(1983)]%
        {gottlieb_nyu_1983}
\bibfield{author}{\bibinfo{person}{{Gottlieb}}, \bibinfo{person}{{Grishman}},
  \bibinfo{person}{{Kruskal}}, \bibinfo{person}{{McAuliffe}},
  \bibinfo{person}{{Rudolph}}, {and} \bibinfo{person}{{Snir}}.}
  \bibinfo{year}{1983}\natexlab{}.
\newblock \showarticletitle{The {{NYU Ultracomputer}}\textemdash{{Designing}}
  an {{MIMD Shared Memory Parallel Computer}}}.
\newblock \bibinfo{journal}{\emph{IEEE Trans. Comput.}} \bibinfo{volume}{C-32},
  \bibinfo{number}{2} (\bibinfo{year}{1983}), \bibinfo{pages}{175--189}.
\newblock
\urldef\tempurl%
\url{https://doi.org/10.1109/TC.1983.1676201}
\showDOI{\tempurl}


\bibitem[Goud et~al\mbox{.}(2015)]%
        {goud_asymmetric_2015}
\bibfield{author}{\bibinfo{person}{A.~Arun Goud}, \bibinfo{person}{Rangharajan
  Venkatesan}, \bibinfo{person}{Anand Raghunathan}, {and}
  \bibinfo{person}{Kaushik Roy}.} \bibinfo{year}{2015}\natexlab{}.
\newblock \showarticletitle{Asymmetric {{Underlapped FinFET Based Robust SRAM
  Design}} at 7nm {{Node}}}. In \bibinfo{booktitle}{\emph{Proceedings of the
  2015 {{Design}}, {{Automation}} \&amp; {{Test}} in {{Europe Conference}}
  \&amp; {{Exhibition}}}} \emph{(\bibinfo{series}{{{DATE}} '15})}.
  \bibinfo{publisher}{{EDA Consortium}}, \bibinfo{address}{{San Jose, CA,
  USA}}, \bibinfo{pages}{659--664}.
\newblock
\showISBNx{978-3-9815370-4-8}


\bibitem[Grass et~al\mbox{.}(2016)]%
        {grass_musa_2016}
\bibfield{author}{\bibinfo{person}{Thomas Grass}, \bibinfo{person}{C{\'e}sar
  Allande}, \bibinfo{person}{Adri{\`a} Armejach}, \bibinfo{person}{Alejandro
  Rico}, \bibinfo{person}{Eduard Ayguad{\'e}}, \bibinfo{person}{Jesus Labarta},
  \bibinfo{person}{Mateo Valero}, \bibinfo{person}{Marc Casas}, {and}
  \bibinfo{person}{Miquel Moreto}.} \bibinfo{year}{2016}\natexlab{}.
\newblock \showarticletitle{{{MUSA}}: {{A Multi-Level Simulation Approach}} for
  next-{{Generation HPC Machines}}}. In \bibinfo{booktitle}{\emph{Proceedings
  of the {{International Conference}} for {{High Performance Computing}},
  {{Networking}}, {{Storage}} and {{Analysis}}}} \emph{(\bibinfo{series}{{{SC}}
  '16})}. \bibinfo{publisher}{{IEEE Press}}, \bibinfo{address}{{Salt Lake City,
  UT, USA}}, \bibinfo{pages}{526--537}.
\newblock
\showISBNx{978-1-4673-8815-3}
\urldef\tempurl%
\url{https://doi.org/10.1109/SC.2016.44}
\showDOI{\tempurl}


\bibitem[GUO et~al\mbox{.}(2006)]%
        {guo_basic_2006}
\bibfield{author}{\bibinfo{person}{Yang GUO}, \bibinfo{person}{Chisachi KATO},
  {and} \bibinfo{person}{Yoshinobu YAMADE}.} \bibinfo{year}{2006}\natexlab{}.
\newblock \showarticletitle{Basic {{Features}} of the {{Fluid Dynamics
  Simulation Software}} ``{{FrontFlow}}/{{Blue}}''}.
\newblock \bibinfo{journal}{\emph{SEISAN KENKYU}} \bibinfo{volume}{58},
  \bibinfo{number}{1} (\bibinfo{year}{2006}), \bibinfo{pages}{11--15}.
\newblock
\urldef\tempurl%
\url{https://doi.org/10.11188/seisankenkyu.58.11}
\showDOI{\tempurl}


\bibitem[Habib et~al\mbox{.}(2016)]%
        {habib_hacc:_2016}
\bibfield{author}{\bibinfo{person}{Salman Habib}, \bibinfo{person}{Vitali
  Morozov}, \bibinfo{person}{Nicholas Frontiere}, \bibinfo{person}{Hal Finkel},
  \bibinfo{person}{Adrian Pope}, \bibinfo{person}{Katrin Heitmann},
  \bibinfo{person}{Kalyan Kumaran}, \bibinfo{person}{Venkatram Vishwanath},
  \bibinfo{person}{Tom Peterka}, \bibinfo{person}{Joe Insley},
  \bibinfo{person}{David Daniel}, \bibinfo{person}{Patricia Fasel}, {and}
  \bibinfo{person}{Zarija Luki{\'c}}.} \bibinfo{year}{2016}\natexlab{}.
\newblock \showarticletitle{{{HACC}}: {{Extreme Scaling}} and {{Performance
  Across Diverse Architectures}}}.
\newblock \bibinfo{journal}{\emph{Commun. ACM}} \bibinfo{volume}{60},
  \bibinfo{number}{1} (\bibinfo{date}{Dec.} \bibinfo{year}{2016}),
  \bibinfo{pages}{97--104}.
\newblock
\showISSN{0001-0782}
\urldef\tempurl%
\url{https://doi.org/10.1145/3015569}
\showDOI{\tempurl}


\bibitem[Hameed et~al\mbox{.}(2021)]%
        {hameed_improving_2021}
\bibfield{author}{\bibinfo{person}{Fazal Hameed}, \bibinfo{person}{Asif~Ali
  Khan}, {and} \bibinfo{person}{Jeronimo Castrillon}.}
  \bibinfo{year}{2021}\natexlab{}.
\newblock \showarticletitle{Improving the {{Performance}} of {{Block-based DRAM
  Caches Via Tag-Data Decoupling}}}.
\newblock \bibinfo{journal}{\emph{IEEE Trans. Comput.}} \bibinfo{volume}{70},
  \bibinfo{number}{11} (\bibinfo{year}{2021}), \bibinfo{pages}{1914--1927}.
\newblock
\urldef\tempurl%
\url{https://doi.org/10.1109/TC.2020.3029615}
\showDOI{\tempurl}


\bibitem[Hemsoth(2018)]%
        {hemsoth_rogues_2018}
\bibfield{author}{\bibinfo{person}{Nicole Hemsoth}.}
  \bibinfo{year}{2018}\natexlab{}.
\newblock \bibinfo{title}{A {{Rogues Gallery}} of {{Post-Moore}}'s {{Law
  Options}}}.
\newblock
  \bibinfo{howpublished}{\url{https://www.nextplatform.com/2018/08/27/a-rogues-gallery-of-post-moores-law-options/}}.
\newblock


\bibitem[Heroux et~al\mbox{.}(2009)]%
        {heroux_improving_2009}
\bibfield{author}{\bibinfo{person}{Michael~A Heroux},
  \bibinfo{person}{Douglas~W Doerfler}, \bibinfo{person}{Paul~S Crozier},
  \bibinfo{person}{James~M Willenbring}, \bibinfo{person}{H~Carter Edwards},
  \bibinfo{person}{Alan Williams}, \bibinfo{person}{Mahesh Rajan},
  \bibinfo{person}{Eric~R Keiter}, \bibinfo{person}{Heidi~K Thornquist}, {and}
  \bibinfo{person}{Robert~W Numrich}.} \bibinfo{year}{2009}\natexlab{}.
\newblock \bibinfo{booktitle}{\emph{Improving {{Performance}} via
  {{Mini-applications}}}}.
\newblock \bibinfo{type}{{T}echnical {R}eport} SAND2009-5574.
  \bibinfo{institution}{{Sandia National Laboratories}}.
\newblock


\bibitem[Hruska(2012)]%
        {hruska_death_2012}
\bibfield{author}{\bibinfo{person}{Joel Hruska}.}
  \bibinfo{year}{2012}\natexlab{}.
\newblock \bibinfo{title}{The Death of {{CPU}} Scaling: {{From}} One Core to
  Many -- and Why We're Still Stuck}.
\newblock
  \bibinfo{howpublished}{\url{https://www.extremetech.com/computing/116561-the-death-of-cpu-scaling-from-one-core-to-many-and-why-were-still-stuck}}.
\newblock


\bibitem[Hu et~al\mbox{.}(2018)]%
        {hu_stacking_2018}
\bibfield{author}{\bibinfo{person}{Xing Hu}, \bibinfo{person}{Dylan~C. Stow},
  {and} \bibinfo{person}{Yuan Xie}.} \bibinfo{year}{2018}\natexlab{}.
\newblock \showarticletitle{Die {{Stacking Is Happening}}}.
\newblock \bibinfo{journal}{\emph{IEEE Micro}} \bibinfo{volume}{38},
  \bibinfo{number}{1} (\bibinfo{year}{2018}), \bibinfo{pages}{22--28}.
\newblock
\urldef\tempurl%
\url{https://doi.org/10.1109/MM.2018.011441561}
\showDOI{\tempurl}


\bibitem[{IEEE IRDS\texttrademark}(2021a)]%
        {ieee_irds_international_2021-1}
\bibfield{author}{\bibinfo{person}{{IEEE IRDS\texttrademark}}.}
  \bibinfo{year}{2021}\natexlab{a}.
\newblock \bibinfo{booktitle}{\emph{International {{Roadmap}} for {{Devices}}
  and {{Systems}} ({{IRDS}}\texttrademark ) 2021 {{Edition}} -- {{Executive
  Summary}}}}.
\newblock \bibinfo{type}{{{IEEE IRDS}}\texttrademark{} {{Roadmap}}}.
  \bibinfo{institution}{{IEEE}}. \bibinfo{pages}{64} pages.
\newblock
\newblock
\shownote{\url{https://irds.ieee.org/images/files/pdf/2021/2021IRDS_ES.pdf}}.


\bibitem[{IEEE IRDS\texttrademark}(2021b)]%
        {ieee_irds_international_2021}
\bibfield{author}{\bibinfo{person}{{IEEE IRDS\texttrademark}}.}
  \bibinfo{year}{2021}\natexlab{b}.
\newblock \bibinfo{booktitle}{\emph{International {{Roadmap}} for {{Devices}}
  and {{Systems}} ({{IRDS}}\texttrademark ) 2021 {{Edition}} -- {{Systems}} and
  {{Architectures}}}}.
\newblock \bibinfo{type}{{{IEEE IRDS}}\texttrademark{} {{Roadmap}}}.
  \bibinfo{institution}{{IEEE}}. \bibinfo{pages}{23} pages.
\newblock
\newblock
\shownote{\url{https://irds.ieee.org/images/files/pdf/2021/2021IRDS_SA.pdf}}.


\bibitem[{Intel Corporation}(2012)]%
        {intel_corporation_intel_2012}
\bibfield{author}{\bibinfo{person}{{Intel Corporation}}.}
  \bibinfo{year}{2012}\natexlab{}.
\newblock \bibinfo{title}{{{Intel}}\textregistered{} {{Architecture Code
  Analyzer}} -- {{User}}'s {{Guide}}}.
\newblock
  \bibinfo{howpublished}{\url{https://www.intel.com/content/dam/develop/external/us/en/documents/intel-architecture-code-analyzer-2-0-users-guide-157548.pdf}}.
\newblock


\bibitem[{Intel Corporation}(2020)]%
        {intel_corporation_dynamic_2020}
\bibfield{author}{\bibinfo{person}{{Intel Corporation}}.}
  \bibinfo{year}{2020}\natexlab{}.
\newblock \bibinfo{title}{Dynamic {{Control- Flow Graph}} ({{DCFG}}) and
  {{DCFG-Trace Format Specifications}} -- {{For Format Version}} 1.00}.
\newblock
  \bibinfo{howpublished}{\url{https://www.intel.com/content/dam/develop/external/us/en/documents/dcfg-format-548994.pdf}}.
\newblock


\bibitem[{Intel Corporation}(2021)]%
        {intel_corporation_intel_2021}
\bibfield{author}{\bibinfo{person}{{Intel Corporation}}.}
  \bibinfo{year}{2021}\natexlab{}.
\newblock \bibinfo{title}{{{Intel}}\textregistered{} {{Software Development
  Emulator}}}.
\newblock
  \bibinfo{howpublished}{\url{https://www.intel.com/content/www/us/en/developer/articles/tool/software-development-emulator.html}}.
\newblock


\bibitem[Iyer et~al\mbox{.}(2021)]%
        {iyer_advances_2021}
\bibfield{author}{\bibinfo{person}{Ravi~R. Iyer}, \bibinfo{person}{Vivek De},
  \bibinfo{person}{Ramesh Illikkal}, \bibinfo{person}{David~A. Koufaty},
  \bibinfo{person}{Bhushan Chitlur}, \bibinfo{person}{Andrew Herdrich},
  \bibinfo{person}{Muhammad~M. Khellah}, \bibinfo{person}{Fatih Hamzaoglu},
  {and} \bibinfo{person}{Eric Karl}.} \bibinfo{year}{2021}\natexlab{}.
\newblock \showarticletitle{Advances in {{Microprocessor Cache Architectures
  Over}} the {{Last}} 25 {{Years}}}.
\newblock \bibinfo{journal}{\emph{IEEE Micro}} \bibinfo{volume}{41},
  \bibinfo{number}{6} (\bibinfo{year}{2021}), \bibinfo{pages}{78--88}.
\newblock
\urldef\tempurl%
\url{https://doi.org/10.1109/MM.2021.3114903}
\showDOI{\tempurl}


\bibitem[Jacobi(2021)]%
        {jacobi_real-time_2021}
\bibfield{author}{\bibinfo{person}{Christian Jacobi}.}
  \bibinfo{year}{2021}\natexlab{}.
\newblock \showarticletitle{Real-Time {{AI}} for {{Enterprise Workloads}}: The
  {{IBM Telum Processor}}}. In \bibinfo{booktitle}{\emph{2021 {{IEEE Hot
  Chips}} 33 {{Symposium}} ({{HCS}})}}. \bibinfo{publisher}{{IEEE Computer
  Society}}, \bibinfo{address}{{Palo Alto, CA, USA}}, \bibinfo{pages}{22}.
\newblock
\urldef\tempurl%
\url{https://doi.org/10.1109/HCS52781.2021.9567422}
\showDOI{\tempurl}
\newblock
\shownote{\url{https://hc33.hotchips.org/assets/program/conference/day1/HC2021.C1.3
  IBM Cristian Jacobi Final.pdf}}.


\bibitem[Jin et~al\mbox{.}(1999)]%
        {jin_openmp_1999}
\bibfield{author}{\bibinfo{person}{H. Jin}, \bibinfo{person}{M. Frumkin}, {and}
  \bibinfo{person}{J. Yan}.} \bibinfo{year}{1999}\natexlab{}.
\newblock \bibinfo{booktitle}{\emph{The {{OpenMP Implementation}} of {{NAS
  Parallel Benchmarks}} and {{Its Performance}}}}.
\newblock \bibinfo{type}{{T}echnical {R}eport} NAS-99-011.
  \bibinfo{institution}{{NASA Ames Research Center}}. \bibinfo{pages}{26}
  pages.
\newblock
\newblock
\shownote{\url{https://www.nas.nasa.gov/assets/pdf/techreports/1999/nas-99-011.pdf}}.


\bibitem[Jung et~al\mbox{.}(2015)]%
        {jung_genesis_2015}
\bibfield{author}{\bibinfo{person}{Jaewoon Jung}, \bibinfo{person}{Takaharu
  Mori}, \bibinfo{person}{Chigusa Kobayashi}, \bibinfo{person}{Yasuhiro
  Matsunaga}, \bibinfo{person}{Takao Yoda}, \bibinfo{person}{Michael Feig},
  {and} \bibinfo{person}{Yuji Sugita}.} \bibinfo{year}{2015}\natexlab{}.
\newblock \showarticletitle{{{GENESIS}}: A Hybrid-Parallel and Multi-Scale
  Molecular Dynamics Simulator with Enhanced Sampling Algorithms for
  Biomolecular and Cellular Simulations}.
\newblock \bibinfo{journal}{\emph{WIREs Computational Molecular Science}}
  \bibinfo{volume}{5}, \bibinfo{number}{4} (\bibinfo{year}{2015}),
  \bibinfo{pages}{310--323}.
\newblock
\urldef\tempurl%
\url{https://doi.org/10.1002/wcms.1220}
\showDOI{\tempurl}


\bibitem[Kodama et~al\mbox{.}(2020)]%
        {kodama_accuracy_2020}
\bibfield{author}{\bibinfo{person}{Yuetsu Kodama}, \bibinfo{person}{Tetsuya
  Odajima}, \bibinfo{person}{Akira Asato}, {and} \bibinfo{person}{Mitsuhisa
  Sato}.} \bibinfo{year}{2020}\natexlab{}.
\newblock \showarticletitle{Accuracy {{Improvement}} of {{Memory System
  Simulation}} for {{Modern Shared Memory Processor}}}. In
  \bibinfo{booktitle}{\emph{Proceedings of the {{International Conference}} on
  {{High Performance Computing}} in {{Asia-Pacific Region}}}}
  \emph{(\bibinfo{series}{{{HPCAsia2020}}})}. \bibinfo{publisher}{{Association
  for Computing Machinery}}, \bibinfo{address}{{New York, NY, USA}},
  \bibinfo{pages}{142--149}.
\newblock
\showISBNx{978-1-4503-7236-7}
\urldef\tempurl%
\url{https://doi.org/10.1145/3368474.3368483}
\showDOI{\tempurl}


\bibitem[Korgaonkar et~al\mbox{.}(2018)]%
        {korgaonkar_density_2018}
\bibfield{author}{\bibinfo{person}{Kunal Korgaonkar}, \bibinfo{person}{Ishwar
  Bhati}, \bibinfo{person}{Huichu Liu}, \bibinfo{person}{Jayesh Gaur},
  \bibinfo{person}{Sasikanth Manipatruni}, \bibinfo{person}{Sreenivas
  Subramoney}, \bibinfo{person}{Tanay Karnik}, \bibinfo{person}{Steven
  Swanson}, \bibinfo{person}{Ian Young}, {and} \bibinfo{person}{Hong Wang}.}
  \bibinfo{year}{2018}\natexlab{}.
\newblock \showarticletitle{Density {{Tradeoffs}} of {{Non-Volatile Memory}} as
  a {{Replacement}} for {{SRAM Based Last Level Cache}}}. In
  \bibinfo{booktitle}{\emph{Proceedings of the 45th {{Annual International
  Symposium}} on {{Computer Architecture}}}} \emph{(\bibinfo{series}{{{ISCA}}
  '18})}. \bibinfo{publisher}{{IEEE Press}}, \bibinfo{address}{{Los Angeles,
  CA, USA}}, \bibinfo{pages}{315--327}.
\newblock
\showISBNx{978-1-5386-5984-7}
\urldef\tempurl%
\url{https://doi.org/10.1109/ISCA.2018.00035}
\showDOI{\tempurl}


\bibitem[Lau(2021)]%
        {lau_3d_2021}
\bibfield{author}{\bibinfo{person}{John~H. Lau}.}
  \bibinfo{year}{2021}\natexlab{}.
\newblock \showarticletitle{{{3D IC Integration}} and {{3D IC Packaging}}}.
\newblock In \bibinfo{booktitle}{\emph{Semiconductor {{Advanced Packaging}}}}.
  \bibinfo{publisher}{{Springer}}, \bibinfo{address}{{Singapore}},
  \bibinfo{pages}{343--378}.
\newblock
\showISBNx{978-981-16-1376-0}
\newblock
\shownote{\url{https://doi.org/10.1007/978-981-16-1376-0_7}}.


\bibitem[Laukemann et~al\mbox{.}(2018)]%
        {laukemann_automated_2018}
\bibfield{author}{\bibinfo{person}{Jan Laukemann}, \bibinfo{person}{Julian
  Hammer}, \bibinfo{person}{Johannes Hofmann}, \bibinfo{person}{Georg Hager},
  {and} \bibinfo{person}{Gerhard Wellein}.} \bibinfo{year}{2018}\natexlab{}.
\newblock \showarticletitle{Automated {{Instruction Stream Throughput
  Prediction}} for {{Intel}} and {{AMD Microarchitectures}}}. In
  \bibinfo{booktitle}{\emph{2018 {{IEEE}}/{{ACM Performance Modeling}},
  {{Benchmarking}} and {{Simulation}} of {{High Performance Computer Systems}}
  ({{PMBS}})}}. \bibinfo{publisher}{{IEEE Press}}, \bibinfo{address}{{Dallas,
  TX, USA}}, \bibinfo{pages}{121--131}.
\newblock
\urldef\tempurl%
\url{https://doi.org/10.1109/PMBS.2018.8641578}
\showDOI{\tempurl}


\bibitem[{LLVM Project}(2022)]%
        {llvm_project_llvm-mca_2022}
\bibfield{author}{\bibinfo{person}{{LLVM Project}}.}
  \bibinfo{year}{2022}\natexlab{}.
\newblock \bibinfo{title}{Llvm-Mca - {{LLVM Machine Code Analyzer}}}.
\newblock
  \bibinfo{howpublished}{\url{https://llvm.org/docs/CommandGuide/llvm-mca.html}}.
\newblock


\bibitem[Loh and Hill(2011)]%
        {loh_efficiently_2011}
\bibfield{author}{\bibinfo{person}{Gabriel~H. Loh} {and}
  \bibinfo{person}{Mark~D. Hill}.} \bibinfo{year}{2011}\natexlab{}.
\newblock \showarticletitle{Efficiently {{Enabling Conventional Block Sizes}}
  for {{Very Large Die-Stacked DRAM Caches}}}. In
  \bibinfo{booktitle}{\emph{Proceedings of the 44th {{Annual IEEE}}/{{ACM
  International Symposium}} on {{Microarchitecture}}}}
  \emph{(\bibinfo{series}{{{MICRO-44}}})}. \bibinfo{publisher}{{Association for
  Computing Machinery}}, \bibinfo{address}{{New York, NY, USA}},
  \bibinfo{pages}{454--464}.
\newblock
\showISBNx{978-1-4503-1053-6}
\urldef\tempurl%
\url{https://doi.org/10.1145/2155620.2155673}
\showDOI{\tempurl}


\bibitem[Loh et~al\mbox{.}(2007)]%
        {loh_processor_2007}
\bibfield{author}{\bibinfo{person}{Gabriel~H. Loh}, \bibinfo{person}{Yuan Xie},
  {and} \bibinfo{person}{Bryan Black}.} \bibinfo{year}{2007}\natexlab{}.
\newblock \showarticletitle{Processor {{Design}} in {{3D Die-Stacking
  Technologies}}}.
\newblock \bibinfo{journal}{\emph{IEEE Micro}} \bibinfo{volume}{27},
  \bibinfo{number}{3} (\bibinfo{date}{May} \bibinfo{year}{2007}),
  \bibinfo{pages}{31--48}.
\newblock
\showISSN{0272-1732}
\urldef\tempurl%
\url{https://doi.org/10.1109/MM.2007.59}
\showDOI{\tempurl}


\bibitem[Ltaief et~al\mbox{.}(2021)]%
        {ltaief_meeting_2021}
\bibfield{author}{\bibinfo{person}{Hatem Ltaief}, \bibinfo{person}{Jesse
  Cranney}, \bibinfo{person}{Damien Gratadour}, \bibinfo{person}{Yuxi Hong},
  \bibinfo{person}{Laurent Gatineau}, {and} \bibinfo{person}{David~E. Keyes}.}
  \bibinfo{year}{2021}\natexlab{}.
\newblock \showarticletitle{Meeting the {{Real-Time Challenges}} of
  {{Ground-Based Telescopes Using Low-Rank Matrix Computations}}}. In
  \bibinfo{booktitle}{\emph{Proceedings of the {{International Conference}} for
  {{High Performance Computing}}, {{Networking}}, {{Storage}} and
  {{Analysis}}}} \emph{(\bibinfo{series}{{{SC}} '21})}.
  \bibinfo{publisher}{{ACM}}, \bibinfo{address}{{New York, NY, USA}},
  \bibinfo{pages}{29:1--29:16}.
\newblock
\urldef\tempurl%
\url{https://doi.org/10.1145/3458817.3476225}
\showDOI{\tempurl}


\bibitem[McCalpin(1995)]%
        {mccalpin_memory_1995}
\bibfield{author}{\bibinfo{person}{J.~D. McCalpin}.}
  \bibinfo{year}{1995}\natexlab{}.
\newblock \showarticletitle{Memory {{Bandwidth}} and {{Machine Balance}} in
  {{Current High Performance Computers}}}.
\newblock \bibinfo{journal}{\emph{IEEE Technical Committee on Computer
  Architecture (TCCA) Newsletter}} \bibinfo{volume}{2},
  \bibinfo{number}{19--25} (\bibinfo{date}{Dec.} \bibinfo{year}{1995}),
  \bibinfo{pages}{1--7}.
\newblock


\bibitem[McKee(2004)]%
        {mckee_reflections_2004}
\bibfield{author}{\bibinfo{person}{Sally~A. McKee}.}
  \bibinfo{year}{2004}\natexlab{}.
\newblock \showarticletitle{Reflections on the {{Memory Wall}}}. In
  \bibinfo{booktitle}{\emph{Proceedings of the 1st {{Conference}} on
  {{Computing Frontiers}}}} \emph{(\bibinfo{series}{{{CF}} '04})}.
  \bibinfo{publisher}{{Association for Computing Machinery}},
  \bibinfo{address}{{New York, NY, USA}}, \bibinfo{pages}{162}.
\newblock
\showISBNx{1-58113-741-9}
\urldef\tempurl%
\url{https://doi.org/10.1145/977091.977115}
\showDOI{\tempurl}


\bibitem[Mendis et~al\mbox{.}(2019)]%
        {mendis_ithemal_2019}
\bibfield{author}{\bibinfo{person}{Charith Mendis}, \bibinfo{person}{Alex
  Renda}, \bibinfo{person}{Dr.Saman Amarasinghe}, {and}
  \bibinfo{person}{Michael Carbin}.} \bibinfo{year}{2019}\natexlab{}.
\newblock \showarticletitle{Ithemal: {{Accurate}}, {{Portable}} and {{Fast
  Basic Block Throughput Estimation}} Using {{Deep Neural Networks}}}. In
  \bibinfo{booktitle}{\emph{Proceedings of the 36th {{International
  Conference}} on {{Machine Learning}}}} \emph{(\bibinfo{series}{Proceedings of
  {{Machine Learning Research}}}, Vol.~\bibinfo{volume}{97})},
  \bibfield{editor}{\bibinfo{person}{Kamalika Chaudhuri} {and}
  \bibinfo{person}{Ruslan Salakhutdinov}} (Eds.). \bibinfo{publisher}{{PMLR}},
  \bibinfo{address}{{Long Beach, California, USA}},
  \bibinfo{pages}{4505--4515}.
\newblock
\newblock
\shownote{\url{https://proceedings.mlr.press/v97/mendis19a.html}}.


\bibitem[Misawa et~al\mbox{.}(2018)]%
        {misawa_mvmc--open-source_2018}
\bibfield{author}{\bibinfo{person}{Takahiro Misawa}, \bibinfo{person}{Satoshi
  Morita}, \bibinfo{person}{Kazuyoshi Yoshimi}, \bibinfo{person}{Mitsuaki
  Kawamura}, \bibinfo{person}{Yuichi Motoyama}, \bibinfo{person}{Kota Ido},
  \bibinfo{person}{Takahiro Ohgoe}, \bibinfo{person}{Masatoshi Imada}, {and}
  \bibinfo{person}{Takeo Kato}.} \bibinfo{year}{2018}\natexlab{}.
\newblock \showarticletitle{{{mVMC--Open-source}} Software for Many-Variable
  Variational {{Monte Carlo}} Method}.
\newblock \bibinfo{journal}{\emph{Computer Physics Communications}}
  \bibinfo{volume}{235}, \bibinfo{number}{Feb. 2019} (\bibinfo{year}{2018}),
  \bibinfo{pages}{447--462}.
\newblock
\showISSN{0010-4655}
\urldef\tempurl%
\url{https://doi.org/10.1016/j.cpc.2018.08.014}
\showDOI{\tempurl}


\bibitem[Mittal and Vetter(2016)]%
        {mittal_survey_2016}
\bibfield{author}{\bibinfo{person}{Sparsh Mittal} {and}
  \bibinfo{person}{Jeffrey~S. Vetter}.} \bibinfo{year}{2016}\natexlab{}.
\newblock \showarticletitle{A {{Survey Of Techniques}} for {{Architecting DRAM
  Caches}}}.
\newblock \bibinfo{journal}{\emph{IEEE Transactions on Parallel and Distributed
  Systems}} \bibinfo{volume}{27}, \bibinfo{number}{6} (\bibinfo{date}{June}
  \bibinfo{year}{2016}), \bibinfo{pages}{1852--1863}.
\newblock
\urldef\tempurl%
\url{https://doi.org/10.1109/TPDS.2015.2461155}
\showDOI{\tempurl}


\bibitem[{Mohd-Yusof} et~al\mbox{.}(2013)]%
        {mohd-yusof_co-design_2013}
\bibfield{author}{\bibinfo{person}{Jamaludin {Mohd-Yusof}},
  \bibinfo{person}{Sriram Swaminarayan}, {and} \bibinfo{person}{Timothy~C.
  Germann}.} \bibinfo{year}{2013}\natexlab{}.
\newblock \bibinfo{booktitle}{\emph{Co-Design for Molecular Dynamics: {{An}}
  Exascale Proxy Application}}.
\newblock \bibinfo{type}{{T}echnical {R}eport} LA-UR 13-20839.
  \bibinfo{institution}{{Los Alamos National Laboratory}}.
\newblock
\newblock
\shownote{\url{http://www.lanl.gov/orgs/adtsc/publications/science_highlights_2013/docs/Pg88_89.pdf}}.


\bibitem[Moore(1975)]%
        {moore_progress_1975}
\bibfield{author}{\bibinfo{person}{Gordon~E. Moore}.}
  \bibinfo{year}{1975}\natexlab{}.
\newblock \showarticletitle{Progress in {{Digital Integrated Electronics}}}.
\newblock \bibinfo{journal}{\emph{International Electron Devices Meeting,
  IEEE}}  \bibinfo{volume}{21} (\bibinfo{year}{1975}), \bibinfo{pages}{11--13}.
\newblock


\bibitem[Morgan(2022)]%
        {morgan_milan-x_2022}
\bibfield{author}{\bibinfo{person}{Timothy~P. Morgan}.}
  \bibinfo{year}{2022}\natexlab{}.
\newblock \bibinfo{title}{"{{Milan-X}}" {{3D Vertical Cache Yields Epyc HPC
  Bang For The Buck Boost}}}.
\newblock
  \bibinfo{howpublished}{\url{https://www.nextplatform.com/2022/03/21/milan-x-3d-vertical-cache-yields-epyc-hpc-bang-for-the-buck-boost/}}.
\newblock


\bibitem[Nakajima et~al\mbox{.}(2014)]%
        {nakajima_ntchem:_2014}
\bibfield{author}{\bibinfo{person}{Takahito Nakajima}, \bibinfo{person}{Michio
  Katouda}, \bibinfo{person}{Muneaki Kamiya}, {and} \bibinfo{person}{Yutaka
  Nakatsuka}.} \bibinfo{year}{2014}\natexlab{}.
\newblock \showarticletitle{{{NTChem}}: {{A High-Performance Software Package}}
  for {{Quantum Molecular Simulation}}}.
\newblock \bibinfo{journal}{\emph{International Journal of Quantum Chemistry}}
  \bibinfo{volume}{115}, \bibinfo{number}{5} (\bibinfo{date}{Dec.}
  \bibinfo{year}{2014}), \bibinfo{pages}{349--359}.
\newblock
\urldef\tempurl%
\url{https://doi.org/10.1002/qua.24860}
\showDOI{\tempurl}


\bibitem[Nickolls and Dally(2010)]%
        {nickolls_gpu_2010}
\bibfield{author}{\bibinfo{person}{John Nickolls} {and}
  \bibinfo{person}{William~J. Dally}.} \bibinfo{year}{2010}\natexlab{}.
\newblock \showarticletitle{The {{GPU Computing Era}}}.
\newblock \bibinfo{journal}{\emph{IEEE Micro}} \bibinfo{volume}{30},
  \bibinfo{number}{2} (\bibinfo{year}{2010}), \bibinfo{pages}{56--69}.
\newblock
\urldef\tempurl%
\url{https://doi.org/10.1109/MM.2010.41}
\showDOI{\tempurl}


\bibitem[Nori et~al\mbox{.}(2018)]%
        {nori_criticality_2018}
\bibfield{author}{\bibinfo{person}{Anant~Vithal Nori}, \bibinfo{person}{Jayesh
  Gaur}, \bibinfo{person}{Siddharth Rai}, \bibinfo{person}{Sreenivas
  Subramoney}, {and} \bibinfo{person}{Hong Wang}.}
  \bibinfo{year}{2018}\natexlab{}.
\newblock \showarticletitle{Criticality {{Aware Tiered Cache Hierarchy}}: {{A
  Fundamental Relook}} at {{Multi-Level Cache Hierarchies}}}. In
  \bibinfo{booktitle}{\emph{Proceedings of the 45th {{Annual International
  Symposium}} on {{Computer Architecture}}}} \emph{(\bibinfo{series}{{{ISCA}}
  '18})}. \bibinfo{publisher}{{IEEE Press}}, \bibinfo{address}{{Los Angeles,
  CA, USA}}, \bibinfo{pages}{96--109}.
\newblock
\showISBNx{978-1-5386-5984-7}
\urldef\tempurl%
\url{https://doi.org/10.1109/ISCA.2018.00019}
\showDOI{\tempurl}


\bibitem[{NVIDIA Corporation}(2022)]%
        {nvidia_corporation_nvidia_2022}
\bibfield{author}{\bibinfo{person}{{NVIDIA Corporation}}.}
  \bibinfo{year}{2022}\natexlab{}.
\newblock \bibinfo{title}{{{NVIDIA H100 Tensor Core GPU}}}.
\newblock
  \bibinfo{howpublished}{\url{https://www.nvidia.com/en-us/data-center/h100/}}.
\newblock


\bibitem[Okazaki et~al\mbox{.}(2020)]%
        {okazaki_supercomputer_2020}
\bibfield{author}{\bibinfo{person}{Ryohei Okazaki}, \bibinfo{person}{Takekazu
  Tabata}, \bibinfo{person}{Sota Sakashita}, \bibinfo{person}{Kenichi
  Kitamura}, \bibinfo{person}{Noriko Takagi}, \bibinfo{person}{Hideki Sakata},
  \bibinfo{person}{Takeshi Ishibashi}, \bibinfo{person}{Takeo Nakamura}, {and}
  \bibinfo{person}{Yuichiro Ajima}.} \bibinfo{year}{2020}\natexlab{}.
\newblock \bibinfo{booktitle}{\emph{Supercomputer {{Fugaku CPU A64FX Realizing
  High Performance}}, {{High-Density Packaging}}, and {{Low Power
  Consumption}}}}.
\newblock \bibinfo{type}{Fujitsu {{Technical Review}}}.
  \bibinfo{institution}{{Fujitsu Limited}}. \bibinfo{pages}{9} pages.
\newblock
\newblock
\shownote{\url{https://www.fujitsu.com/global/documents/about/resources/publications/technicalreview/2020-03/article03.pdf}}.


\bibitem[Oliveira et~al\mbox{.}(2021)]%
        {oliveira_damov_2021}
\bibfield{author}{\bibinfo{person}{Geraldo~F. Oliveira}, \bibinfo{person}{Juan
  {G{\'o}mez-Luna}}, \bibinfo{person}{Lois Orosa}, \bibinfo{person}{Saugata
  Ghose}, \bibinfo{person}{Nandita Vijaykumar}, \bibinfo{person}{Ivan
  Fernandez}, \bibinfo{person}{Mohammad Sadrosadati}, {and}
  \bibinfo{person}{Onur Mutlu}.} \bibinfo{year}{2021}\natexlab{}.
\newblock \showarticletitle{{{DAMOV}}: {{A New Methodology}} and {{Benchmark
  Suite}} for {{Evaluating Data Movement Bottlenecks}}}.
\newblock \bibinfo{journal}{\emph{IEEE Access}}  \bibinfo{volume}{9}
  (\bibinfo{year}{2021}), \bibinfo{pages}{134457--134502}.
\newblock
\urldef\tempurl%
\url{https://doi.org/10.1109/ACCESS.2021.3110993}
\showDOI{\tempurl}


\bibitem[Ono et~al\mbox{.}(2016)]%
        {ono_ffv-c_2016}
\bibfield{author}{\bibinfo{person}{Kenji Ono}, \bibinfo{person}{Masako Iwata},
  \bibinfo{person}{Tsuyoshi Tamaki}, \bibinfo{person}{Yasuhiro Kawashima},
  \bibinfo{person}{Kei Akasaka}, \bibinfo{person}{Soichiro Suzuki},
  \bibinfo{person}{Junya Onishi}, \bibinfo{person}{Ken Uzawa},
  \bibinfo{person}{Kazuhiro Hamaguchi}, \bibinfo{person}{Yohei Miyazaki}, {and}
  \bibinfo{person}{Masashi Imano}.} \bibinfo{year}{2016}\natexlab{}.
\newblock \bibinfo{title}{{{FFV-C}} Package}.
\newblock
  \bibinfo{howpublished}{\url{http://avr-aics-riken.github.io/ffvc_package/}}.
\newblock


\bibitem[{Or-Bach}(2017)]%
        {or-bach_1000x_2017}
\bibfield{author}{\bibinfo{person}{Zvi {Or-Bach}}.}
  \bibinfo{year}{2017}\natexlab{}.
\newblock \showarticletitle{A 1,000x {{Improvement}} in {{Computer Systems}} by
  {{Bridging}} the {{Processor-Memory Gap}}}. In \bibinfo{booktitle}{\emph{2017
  {{IEEE SOI-3D-Subthreshold Microelectronics Technology Unified Conference}}
  ({{S3S}})}}. \bibinfo{publisher}{{IEEE Press}},
  \bibinfo{address}{{Burlingame, CA, USA}}, \bibinfo{pages}{1--4}.
\newblock
\urldef\tempurl%
\url{https://doi.org/10.1109/S3S.2017.8309202}
\showDOI{\tempurl}


\bibitem[Owens et~al\mbox{.}(2008)]%
        {owens_gpu_2008}
\bibfield{author}{\bibinfo{person}{John~D. Owens}, \bibinfo{person}{Mike
  Houston}, \bibinfo{person}{David Luebke}, \bibinfo{person}{Simon Green},
  \bibinfo{person}{John~E. Stone}, {and} \bibinfo{person}{James~C. Phillips}.}
  \bibinfo{year}{2008}\natexlab{}.
\newblock \showarticletitle{{{GPU Computing}}}.
\newblock \bibinfo{journal}{\emph{Proc. IEEE}} \bibinfo{volume}{96},
  \bibinfo{number}{5} (\bibinfo{year}{2008}), \bibinfo{pages}{879--899}.
\newblock
\urldef\tempurl%
\url{https://doi.org/10.1109/JPROC.2008.917757}
\showDOI{\tempurl}


\bibitem[Park et~al\mbox{.}(2015)]%
        {park_high-performance_2015}
\bibfield{author}{\bibinfo{person}{Jongsoo Park}, \bibinfo{person}{Mikhail
  Smelyanskiy}, \bibinfo{person}{Ulrike~Meier Yang}, \bibinfo{person}{Dheevatsa
  Mudigere}, {and} \bibinfo{person}{Pradeep Dubey}.}
  \bibinfo{year}{2015}\natexlab{}.
\newblock \showarticletitle{High-Performance {{Algebraic Multigrid Solver
  Optimized}} for {{Multi-core Based Distributed Parallel Systems}}}. In
  \bibinfo{booktitle}{\emph{Proceedings of the {{International Conference}} for
  {{High Performance Computing}}, {{Networking}}, {{Storage}} and
  {{Analysis}}}} \emph{(\bibinfo{series}{{{SC}} '15})}.
  \bibinfo{publisher}{{ACM}}, \bibinfo{address}{{Austin, TX, USA}},
  \bibinfo{pages}{54:1--54:12}.
\newblock
\showISBNx{978-1-4503-3723-6}
\urldef\tempurl%
\url{https://doi.org/10.1145/2807591.2807603}
\showDOI{\tempurl}


\bibitem[Petersson and Sj{\"o}green(2017)]%
        {petersson_users_2017}
\bibfield{author}{\bibinfo{person}{N.~A. Petersson} {and} \bibinfo{person}{B.
  Sj{\"o}green}.} \bibinfo{year}{2017}\natexlab{}.
\newblock \bibinfo{booktitle}{\emph{User's Guide to {{SW4}}, Version 2.0}}.
\newblock \bibinfo{type}{{T}echnical {R}eport} LLNL-SM-741439.
  \bibinfo{institution}{{Lawrence Livermore National Laboratory}}.
\newblock


\bibitem[Podobas et~al\mbox{.}(2020)]%
        {podobas_survey_2020-1}
\bibfield{author}{\bibinfo{person}{Artur Podobas}, \bibinfo{person}{Kentaro
  Sano}, {and} \bibinfo{person}{Satoshi Matsuoka}.}
  \bibinfo{year}{2020}\natexlab{}.
\newblock \showarticletitle{A {{Survey}} on {{Coarse-Grained Reconfigurable
  Architectures From}} a {{Performance Perspective}}}.
\newblock \bibinfo{journal}{\emph{IEEE Access}}  \bibinfo{volume}{8}
  (\bibinfo{date}{July} \bibinfo{year}{2020}).
\newblock
\urldef\tempurl%
\url{https://doi.org/10.1109/ACCESS.2020.3012084}
\showDOI{\tempurl}


\bibitem[Pouchet and Taylor(2016)]%
        {pouchet_polybenchc_2016}
\bibfield{author}{\bibinfo{person}{Louis-Noel Pouchet} {and}
  \bibinfo{person}{Mark Taylor}.} \bibinfo{year}{2016}\natexlab{}.
\newblock \bibinfo{title}{{{PolyBench}}/{{C}} 4.2.1 (Beta)}.
\newblock
  \bibinfo{howpublished}{\url{https://sourceforge.net/projects/polybench/}}.
\newblock


\bibitem[Renda et~al\mbox{.}(2020)]%
        {renda_difftune_2020}
\bibfield{author}{\bibinfo{person}{Alex Renda}, \bibinfo{person}{Yishen Chen},
  \bibinfo{person}{Charith Mendis}, {and} \bibinfo{person}{Michael Carbin}.}
  \bibinfo{year}{2020}\natexlab{}.
\newblock \showarticletitle{{{DiffTune}}: {{Optimizing CPU Simulator
  Parameters}} with {{Learned Differentiable Surrogates}}}. In
  \bibinfo{booktitle}{\emph{2020 53rd {{Annual IEEE}}/{{ACM International
  Symposium}} on {{Microarchitecture}} ({{MICRO}})}}. \bibinfo{publisher}{{IEEE
  Press}}, \bibinfo{address}{{Athens, Greece}}, \bibinfo{pages}{442--455}.
\newblock
\urldef\tempurl%
\url{https://doi.org/10.1109/MICRO50266.2020.00045}
\showDOI{\tempurl}


\bibitem[{RIKEN AICS}(2015)]%
        {riken_aics_fiber_2015}
\bibfield{author}{\bibinfo{person}{{RIKEN AICS}}.}
  \bibinfo{year}{2015}\natexlab{}.
\newblock \bibinfo{title}{Fiber {{Miniapp Suite}}}.
\newblock \bibinfo{howpublished}{\url{https://fiber-miniapp.github.io/}}.
\newblock


\bibitem[{RIKEN Center for Computational Science}(2021)]%
        {riken_center_for_computational_science_kernel_2021}
\bibfield{author}{\bibinfo{person}{{RIKEN Center for Computational Science}}.}
  \bibinfo{year}{2021}\natexlab{}.
\newblock \bibinfo{title}{The Kernel Codes from {{Priority Issue Target
  Applications}}}.
\newblock
  \bibinfo{howpublished}{\url{https://github.com/RIKEN-RCCS/fs2020-tapp-kernels}}.
\newblock


\bibitem[{RIKEN-RCCS}(2020)]%
        {riken-rccs_riken_simulator_2020}
\bibfield{author}{\bibinfo{person}{{RIKEN-RCCS}}.}
  \bibinfo{year}{2020}\natexlab{}.
\newblock \bibinfo{title}{Riken\_simulator}.
\newblock
  \bibinfo{howpublished}{\url{https://github.com/RIKEN-RCCS/riken_simulator}}.
\newblock


\bibitem[Rodrigues et~al\mbox{.}(2012)]%
        {rodrigues_improvements_2012}
\bibfield{author}{\bibinfo{person}{Arun Rodrigues}, \bibinfo{person}{Elliot
  {Cooper-Balis}}, \bibinfo{person}{Keren Bergman}, \bibinfo{person}{Kurt
  Ferreira}, \bibinfo{person}{David Bunde}, {and} \bibinfo{person}{K.~Scott
  Hemmert}.} \bibinfo{year}{2012}\natexlab{}.
\newblock \showarticletitle{Improvements to the {{Structural Simulation
  Toolkit}}}. In \bibinfo{booktitle}{\emph{Proceedings of the 5th
  {{International ICST Conference}} on {{Simulation Tools}} and
  {{Techniques}}}} \emph{(\bibinfo{series}{{{SIMUTOOLS}} '12})}.
  \bibinfo{publisher}{{ICST (Institute for Computer Sciences,
  Social-Informatics and Telecommunications Engineering)}},
  \bibinfo{address}{{Brussels, BEL}}, \bibinfo{pages}{190--195}.
\newblock
\showISBNx{978-1-4503-1510-4}


\bibitem[Sato et~al\mbox{.}(2020)]%
        {sato_co-design_2020}
\bibfield{author}{\bibinfo{person}{Mitsuhisa Sato}, \bibinfo{person}{Yutaka
  Ishikawa}, \bibinfo{person}{Hirofumi Tomita}, \bibinfo{person}{Yuetsu
  Kodama}, \bibinfo{person}{Tetsuya Odajima}, \bibinfo{person}{Miwako Tsuji},
  \bibinfo{person}{Hisashi Yashiro}, \bibinfo{person}{Masaki Aoki},
  \bibinfo{person}{Naoyuki Shida}, \bibinfo{person}{Ikuo Miyoshi},
  \bibinfo{person}{Kouichi Hirai}, \bibinfo{person}{Atsushi Furuya},
  \bibinfo{person}{Akira Asato}, \bibinfo{person}{Kuniki Morita}, {and}
  \bibinfo{person}{Toshiyuki Shimizu}.} \bibinfo{year}{2020}\natexlab{}.
\newblock \showarticletitle{Co-{{Design}} for {{A64FX Manycore Processor}} and
  "{{Fugaku}}"}. In \bibinfo{booktitle}{\emph{Proceedings of the
  {{International Conference}} for {{High Performance Computing}},
  {{Networking}}, {{Storage}} and {{Analysis}}}} \emph{(\bibinfo{series}{{{SC}}
  '20})}. \bibinfo{publisher}{{IEEE Press}}, \bibinfo{address}{{Atlanta, GA,
  USA}}, \bibinfo{pages}{1--15}.
\newblock
\showISBNx{978-1-72819-998-6}


\bibitem[Shiba et~al\mbox{.}(2022)]%
        {shiba_7-nm_2022}
\bibfield{author}{\bibinfo{person}{Kota Shiba}, \bibinfo{person}{Mitsuji
  Okada}, \bibinfo{person}{Atsutake Kosuge}, \bibinfo{person}{Mototsugu
  Hamada}, {and} \bibinfo{person}{Tadahiro Kuroda}.}
  \bibinfo{year}{2022}\natexlab{}.
\newblock \showarticletitle{A 7-Nm {{FinFET}} 1.2-{{TB}}/s/Mm{$^2$}
  {{3D-Stacked SRAM Module With}} 0.7-{{pJ}}/b {{Inductive Coupling Interface
  Using Over-SRAM Coil}} and {{Manchester-Encoded Synchronous Transceiver}}}.
\newblock \bibinfo{journal}{\emph{IEEE Journal of Solid-State Circuits}}
  (\bibinfo{year}{2022}), \bibinfo{pages}{1--12}.
\newblock
\urldef\tempurl%
\url{https://doi.org/10.1109/JSSC.2022.3224421}
\showDOI{\tempurl}


\bibitem[Shiba et~al\mbox{.}(2021)]%
        {shiba_96-mb_2021}
\bibfield{author}{\bibinfo{person}{Kota Shiba}, \bibinfo{person}{Tatsuo Omori},
  \bibinfo{person}{Kodai Ueyoshi}, \bibinfo{person}{Shinya
  {Takamaeda-Yamazaki}}, \bibinfo{person}{Masato Motomura},
  \bibinfo{person}{Mototsugu Hamada}, {and} \bibinfo{person}{Tadahiro Kuroda}.}
  \bibinfo{year}{2021}\natexlab{}.
\newblock \showarticletitle{A 96-{{MB 3D-Stacked SRAM Using Inductive Coupling
  With}} 0.4-{{V Transmitter}}, {{Termination Scheme}} and 12:1 {{SerDes}} in
  40-Nm {{CMOS}}}.
\newblock \bibinfo{journal}{\emph{IEEE Transactions on Circuits and Systems I:
  Regular Papers (TCAS-I)}} \bibinfo{volume}{68}, \bibinfo{number}{2}
  (\bibinfo{date}{Feb.} \bibinfo{year}{2021}), \bibinfo{pages}{692--703}.
\newblock
\urldef\tempurl%
\url{https://doi.org/10.1109/TCSI.2020.3037892}
\showDOI{\tempurl}


\bibitem[Shilov(2022)]%
        {shilov_tsmc_2022}
\bibfield{author}{\bibinfo{person}{Anton Shilov}.}
  \bibinfo{year}{2022}\natexlab{}.
\newblock \bibinfo{title}{{{TSMC Roadmap Update}}: {{N3E}} in 2024, {{N2}} in
  2026, {{Major Changes Incoming}}}.
\newblock
  \bibinfo{howpublished}{\url{https://www.anandtech.com/show/17356/tsmc-roadmap-update-n3e-in-2024-n2-in-2026-major-changes-incoming}}.
\newblock


\bibitem[Shulaker et~al\mbox{.}(2015)]%
        {shulaker_monolithic_2015}
\bibfield{author}{\bibinfo{person}{Max~M. Shulaker}, \bibinfo{person}{Tony~F.
  Wu}, \bibinfo{person}{Mohamed~M. Sabry}, \bibinfo{person}{Hai Wei},
  \bibinfo{person}{H.-S.~Philip Wong}, {and} \bibinfo{person}{Subhasish
  Mitra}.} \bibinfo{year}{2015}\natexlab{}.
\newblock \showarticletitle{Monolithic {{3D Integration}}: {{A Path}} from
  {{Concept}} to {{Reality}}}. In \bibinfo{booktitle}{\emph{Proceedings of the
  2015 {{Design}}, {{Automation}} \& {{Test}} in {{Europe Conference}} \&
  {{Exhibition}}}} \emph{(\bibinfo{series}{{{DATE}} '15})}.
  \bibinfo{publisher}{{EDA Consortium}}, \bibinfo{address}{{San Jose, CA,
  USA}}, \bibinfo{pages}{1197--1202}.
\newblock
\showISBNx{978-3-9815370-4-8}


\bibitem[Sorby(2017)]%
        {sorby_mpi_2017}
\bibfield{author}{\bibinfo{person}{Hugh Sorby}.}
  \bibinfo{year}{2017}\natexlab{}.
\newblock \bibinfo{title}{{{MPI Stub}}}.
\newblock \bibinfo{howpublished}{\url{https://github.com/hsorby/mpistub}}.
\newblock


\bibitem[{Standard Performance Evaluation Corporation}(2020)]%
        {standard_performance_evaluation_corporation_specs_2020}
\bibfield{author}{\bibinfo{person}{{Standard Performance Evaluation
  Corporation}}.} \bibinfo{year}{2020}\natexlab{}.
\newblock \bibinfo{title}{{{SPEC}}'s {{Benchmarks}}}.
\newblock \bibinfo{howpublished}{\url{https://www.spec.org/benchmarks.html}}.
\newblock


\bibitem[Stephens et~al\mbox{.}(2017)]%
        {stephens_arm_2017}
\bibfield{author}{\bibinfo{person}{Nigel Stephens}, \bibinfo{person}{Stuart
  Biles}, \bibinfo{person}{Matthias Boettcher}, \bibinfo{person}{Jacob Eapen},
  \bibinfo{person}{Mbou Eyole}, \bibinfo{person}{Giacomo Gabrielli},
  \bibinfo{person}{Matt Horsnell}, \bibinfo{person}{Grigorios Magklis},
  \bibinfo{person}{Alejandro Martinez}, \bibinfo{person}{Nathanael Premillieu},
  \bibinfo{person}{Alastair Reid}, \bibinfo{person}{Alejandro Rico}, {and}
  \bibinfo{person}{Paul Walker}.} \bibinfo{year}{2017}\natexlab{}.
\newblock \showarticletitle{The {{ARM Scalable Vector Extension}}}.
\newblock \bibinfo{journal}{\emph{IEEE Micro}} \bibinfo{volume}{37},
  \bibinfo{number}{02} (\bibinfo{date}{March} \bibinfo{year}{2017}),
  \bibinfo{pages}{26--39}.
\newblock
\showISSN{1937-4143}
\urldef\tempurl%
\url{https://doi.org/10.1109/MM.2017.35}
\showDOI{\tempurl}


\bibitem[Strohmaier et~al\mbox{.}(2021)]%
        {strohmaier_top500_2021}
\bibfield{author}{\bibinfo{person}{Erich Strohmaier}, \bibinfo{person}{Jack
  Dongarra}, \bibinfo{person}{Horst Simon}, {and} \bibinfo{person}{Martin
  Meuer}.} \bibinfo{year}{2021}\natexlab{}.
\newblock \bibinfo{title}{{{TOP500}}}.
\newblock \bibinfo{howpublished}{\url{http://www.top500.org/}}.
\newblock


\bibitem[Suggs et~al\mbox{.}(2020)]%
        {suggs_amd_2020}
\bibfield{author}{\bibinfo{person}{David Suggs}, \bibinfo{person}{Mahesh
  Subramony}, {and} \bibinfo{person}{Dan Bouvier}.}
  \bibinfo{year}{2020}\natexlab{}.
\newblock \showarticletitle{The {{AMD}} "{{Zen}} 2" {{Processor}}}.
\newblock \bibinfo{journal}{\emph{IEEE Micro}} \bibinfo{volume}{40},
  \bibinfo{number}{2} (\bibinfo{year}{2020}), \bibinfo{pages}{45--52}.
\newblock
\urldef\tempurl%
\url{https://doi.org/10.1109/MM.2020.2974217}
\showDOI{\tempurl}


\bibitem[Tavakkoli et~al\mbox{.}(2016)]%
        {tavakkoli_analysis_2016}
\bibfield{author}{\bibinfo{person}{Fatemeh Tavakkoli}, \bibinfo{person}{Siavash
  Ebrahimi}, \bibinfo{person}{Shujuan Wang}, {and} \bibinfo{person}{Kambiz
  Vafai}.} \bibinfo{year}{2016}\natexlab{}.
\newblock \showarticletitle{Analysis of Critical Thermal Issues in {{3D}}
  Integrated Circuits}.
\newblock \bibinfo{journal}{\emph{International Journal of Heat and Mass
  Transfer}}  \bibinfo{volume}{97} (\bibinfo{year}{2016}),
  \bibinfo{pages}{337--352}.
\newblock
\showISSN{0017-9310}
\urldef\tempurl%
\url{https://doi.org/10.1016/j.ijheatmasstransfer.2016.02.010}
\showDOI{\tempurl}


\bibitem[Theis and Wong(2017)]%
        {theis_end_2017-1}
\bibfield{author}{\bibinfo{person}{Thomas~N. Theis} {and}
  \bibinfo{person}{H.-S.~Philip Wong}.} \bibinfo{year}{2017}\natexlab{}.
\newblock \showarticletitle{The {{End}} of {{Moore}}'s {{Law}}: {{A New
  Beginning}} for {{Information Technology}}}.
\newblock \bibinfo{journal}{\emph{Computing in Science Engineering}}
  \bibinfo{volume}{19}, \bibinfo{number}{2} (\bibinfo{year}{2017}),
  \bibinfo{pages}{41--50}.
\newblock
\urldef\tempurl%
\url{https://doi.org/10.1109/MCSE.2017.29}
\showDOI{\tempurl}


\bibitem[Tomita and Satoh(2004)]%
        {tomita_new_2004}
\bibfield{author}{\bibinfo{person}{Hirofumi Tomita} {and}
  \bibinfo{person}{Masaki Satoh}.} \bibinfo{year}{2004}\natexlab{}.
\newblock \showarticletitle{A New Dynamical Framework of Nonhydrostatic Global
  Model Using the Icosahedral Grid}.
\newblock \bibinfo{journal}{\emph{Fluid Dynamics Research}}
  \bibinfo{volume}{34}, \bibinfo{number}{6} (\bibinfo{year}{2004}),
  \bibinfo{pages}{357--400}.
\newblock
\newblock
\shownote{\url{http://stacks.iop.org/1873-7005/34/i=6/a=A03}}.


\bibitem[Tramm et~al\mbox{.}(2014)]%
        {tramm_xsbench_2014}
\bibfield{author}{\bibinfo{person}{John~R Tramm}, \bibinfo{person}{Andrew~R
  Siegel}, \bibinfo{person}{Tanzima Islam}, {and} \bibinfo{person}{Martin
  Schulz}.} \bibinfo{year}{2014}\natexlab{}.
\newblock \showarticletitle{{{XSBench}} - {{The Development}} and
  {{Verification}} of a {{Performance Abstraction}} for {{Monte Carlo Reactor
  Analysis}}}. In \bibinfo{booktitle}{\emph{{{PHYSOR}} 2014 - {{The Role}} of
  {{Reactor Physics}} toward a {{Sustainable Future}}}}.
  \bibinfo{publisher}{{JAEA}}, \bibinfo{address}{{Kyoto}},
  \bibinfo{pages}{1--13}.
\newblock
\urldef\tempurl%
\url{https://doi.org/10.11484/jaea-conf-2014-003}
\showDOI{\tempurl}


\bibitem[{Van der Wijngaart}(2002)]%
        {van_der_wijngaart_nas_2002}
\bibfield{author}{\bibinfo{person}{Rob~F. {Van der Wijngaart}}.}
  \bibinfo{year}{2002}\natexlab{}.
\newblock \bibinfo{booktitle}{\emph{The {{NAS Parallel Benchmarks}} 2.4}}.
\newblock \bibinfo{type}{{T}echnical {R}eport} NAS-02-007.
  \bibinfo{institution}{{NASA Ames Research Center}}. \bibinfo{pages}{8} pages.
\newblock
\newblock
\shownote{\url{https://www.nas.nasa.gov/assets/pdf/techreports/2002/nas-02-007.pdf}}.


\bibitem[Vasudevan et~al\mbox{.}(2017)]%
        {vasudevan_parallel_2017}
\bibfield{author}{\bibinfo{person}{Aravind Vasudevan}, \bibinfo{person}{Andrew
  Anderson}, {and} \bibinfo{person}{David Gregg}.}
  \bibinfo{year}{2017}\natexlab{}.
\newblock \showarticletitle{Parallel {{Multi Channel}} Convolution Using
  {{General Matrix Multiplication}}}. In \bibinfo{booktitle}{\emph{2017
  {{IEEE}} 28th {{International Conference}} on {{Application-specific
  Systems}}, {{Architectures}} and {{Processors}} ({{ASAP}})}}.
  \bibinfo{publisher}{{IEEE Press}}, \bibinfo{address}{{Seattle, WA, USA}},
  \bibinfo{pages}{19--24}.
\newblock
\urldef\tempurl%
\url{https://doi.org/10.1109/ASAP.2017.7995254}
\showDOI{\tempurl}


\bibitem[Vetter et~al\mbox{.}(2017)]%
        {vetter_architectures_2017-1}
\bibfield{author}{\bibinfo{person}{Jeffery~S. Vetter}, \bibinfo{person}{Erik~P.
  DeBenedictis}, {and} \bibinfo{person}{Thomas~M. Conte}.}
  \bibinfo{year}{2017}\natexlab{}.
\newblock \showarticletitle{Architectures for the {{Post-Moore Era}}}.
\newblock \bibinfo{journal}{\emph{IEEE Micro}} \bibinfo{volume}{37},
  \bibinfo{number}{04} (\bibinfo{date}{July} \bibinfo{year}{2017}),
  \bibinfo{pages}{6--8}.
\newblock
\showISSN{1937-4143}
\urldef\tempurl%
\url{https://doi.org/10.1109/MM.2017.3211127}
\showDOI{\tempurl}


\bibitem[Voskuilen et~al\mbox{.}(2016)]%
        {voskuilen_analyzing_2016}
\bibfield{author}{\bibinfo{person}{Gwendolyn Voskuilen},
  \bibinfo{person}{Arun~F. Rodrigues}, {and} \bibinfo{person}{Simon~D.
  Hammond}.} \bibinfo{year}{2016}\natexlab{}.
\newblock \showarticletitle{Analyzing {{Allocation Behavior}} for {{Multi-Level
  Memory}}}. In \bibinfo{booktitle}{\emph{Proceedings of the {{Second
  International Symposium}} on {{Memory Systems}}}}
  \emph{(\bibinfo{series}{{{MEMSYS}} '16})}. \bibinfo{publisher}{{Association
  for Computing Machinery}}, \bibinfo{address}{{New York, NY, USA}},
  \bibinfo{pages}{204--207}.
\newblock
\showISBNx{978-1-4503-4305-3}
\urldef\tempurl%
\url{https://doi.org/10.1145/2989081.2989116}
\showDOI{\tempurl}


\bibitem[Wang et~al\mbox{.}(2018)]%
        {wang_3d_2018}
\bibfield{author}{\bibinfo{person}{Shaoxi Wang}, \bibinfo{person}{Yue Yin},
  \bibinfo{person}{Chenxia Hu}, {and} \bibinfo{person}{Pouya Rezai}.}
  \bibinfo{year}{2018}\natexlab{}.
\newblock \showarticletitle{{{3D Integrated Circuit Cooling}} with
  {{Microfluidics}}}.
\newblock \bibinfo{journal}{\emph{Micromachines}} \bibinfo{volume}{9},
  \bibinfo{number}{6} (\bibinfo{year}{2018}), \bibinfo{pages}{1--14}.
\newblock
\showISSN{2072-666X}
\urldef\tempurl%
\url{https://doi.org/10.3390/mi9060287}
\showDOI{\tempurl}


\bibitem[Warnock et~al\mbox{.}(2015)]%
        {warnock_41_2015}
\bibfield{author}{\bibinfo{person}{James Warnock}, \bibinfo{person}{Brian
  Curran}, \bibinfo{person}{John Badar}, \bibinfo{person}{Gregory Fredeman},
  \bibinfo{person}{Donald Plass}, \bibinfo{person}{Yuen Chan},
  \bibinfo{person}{Sean Carey}, \bibinfo{person}{Gerard Salem},
  \bibinfo{person}{Friedrich Schroeder}, \bibinfo{person}{Frank Malgioglio},
  \bibinfo{person}{Guenter Mayer}, \bibinfo{person}{Christopher Berry},
  \bibinfo{person}{Michael Wood}, \bibinfo{person}{Yiu-Hing Chan},
  \bibinfo{person}{Mark Mayo}, \bibinfo{person}{John Isakson},
  \bibinfo{person}{Charudhattan Nagarajan}, \bibinfo{person}{Tobias Werner},
  \bibinfo{person}{Leon Sigal}, \bibinfo{person}{Ricardo Nigaglioni},
  \bibinfo{person}{Mark Cichanowski}, \bibinfo{person}{Jeffrey Zitz},
  \bibinfo{person}{Matthew Ziegler}, \bibinfo{person}{Tim Bronson},
  \bibinfo{person}{Gerald Strevig}, \bibinfo{person}{Daniel Dreps},
  \bibinfo{person}{Ruchir Puri}, \bibinfo{person}{Douglas Malone},
  \bibinfo{person}{Dieter Wendel}, \bibinfo{person}{Pak-Kin Mak}, {and}
  \bibinfo{person}{Michael Blake}.} \bibinfo{year}{2015}\natexlab{}.
\newblock \showarticletitle{4.1 22nm {{Next-generation IBM System}} z
  Microprocessor}. In \bibinfo{booktitle}{\emph{2015 {{IEEE International
  Solid-State Circuits Conference}} - ({{ISSCC}}) {{Digest}} of {{Technical
  Papers}}}}. \bibinfo{publisher}{{IEEE Press}}, \bibinfo{address}{{San
  Francisco, CA, USA}}, \bibinfo{pages}{1--3}.
\newblock
\urldef\tempurl%
\url{https://doi.org/10.1109/ISSCC.2015.7062930}
\showDOI{\tempurl}


\bibitem[Wolf et~al\mbox{.}(2015)]%
        {wolf_task-based_2015}
\bibfield{author}{\bibinfo{person}{M.~M. Wolf}, \bibinfo{person}{J.~W. Berry},
  {and} \bibinfo{person}{D.~T. Stark}.} \bibinfo{year}{2015}\natexlab{}.
\newblock \showarticletitle{A Task-Based Linear Algebra {{Building Blocks}}
  Approach for Scalable Graph Analytics}. In \bibinfo{booktitle}{\emph{2015
  {{IEEE High Performance Extreme Computing Conference}} ({{HPEC}})}}.
  \bibinfo{publisher}{{IEEE Press}}, \bibinfo{address}{{Waltham, MA, USA}},
  \bibinfo{pages}{1--6}.
\newblock
\urldef\tempurl%
\url{https://doi.org/10.1109/HPEC.2015.7322450}
\showDOI{\tempurl}


\bibitem[Yamamura et~al\mbox{.}(2022)]%
        {yamamura_a64fx_2022}
\bibfield{author}{\bibinfo{person}{Shuji Yamamura}, \bibinfo{person}{Yasunobu
  Akizuki}, \bibinfo{person}{Hideyuki Sekiguchi}, \bibinfo{person}{Takumi
  Maruyama}, \bibinfo{person}{Tsutomu Sano}, \bibinfo{person}{Hiroyuki
  Miyazaki}, {and} \bibinfo{person}{Toshio Yoshida}.}
  \bibinfo{year}{2022}\natexlab{}.
\newblock \showarticletitle{{{A64FX}}: 52-{{Core Processor Designed}} for the
  {{442PetaFLOPS Supercomputer Fugaku}}}. In \bibinfo{booktitle}{\emph{{{IEEE
  International Solid-State Circuits Conference}}, {{ISSCC}} 2022, {{San
  Francisco}}, {{CA}}, {{USA}}, {{February}} 20-26, 2022}}.
  \bibinfo{publisher}{{IEEE}}, \bibinfo{address}{{San Francisco, CA, USA}},
  \bibinfo{pages}{352--354}.
\newblock
\urldef\tempurl%
\url{https://doi.org/10.1109/ISSCC42614.2022.9731627}
\showDOI{\tempurl}


\bibitem[Yoshida(2018)]%
        {yoshida_fujitsu_2018}
\bibfield{author}{\bibinfo{person}{Toshio Yoshida}.}
  \bibinfo{year}{2018}\natexlab{}.
\newblock \showarticletitle{Fujitsu {{High Performance CPU}} for the {{Post-K
  Computer}}}. In \bibinfo{booktitle}{\emph{2018 {{IEEE Hot Chips}} 30
  {{Symposium}} ({{HCS}})}}. \bibinfo{publisher}{{IEEE Computer Society}},
  \bibinfo{address}{{California, USA}}, \bibinfo{pages}{22}.
\newblock
\newblock
\shownote{\url{http://www.fujitsu.com/jp/Images/20180821hotchips30.pdf}}.


\bibitem[Young et~al\mbox{.}(2018)]%
        {young_accord_2018}
\bibfield{author}{\bibinfo{person}{Vinson Young}, \bibinfo{person}{Chiachen
  Chou}, \bibinfo{person}{Aamer Jaleel}, {and} \bibinfo{person}{Moinuddin
  Qureshi}.} \bibinfo{year}{2018}\natexlab{}.
\newblock \showarticletitle{{{ACCORD}}: {{Enabling Associativity}} for
  {{Gigascale DRAM Caches}} by {{Coordinating Way-Install}} and
  {{Way-Prediction}}}. In \bibinfo{booktitle}{\emph{2018 {{ACM}}/{{IEEE}} 45th
  {{Annual International Symposium}} on {{Computer Architecture}} ({{ISCA}})}}.
  \bibinfo{publisher}{{IEEE}}, \bibinfo{address}{{Los Angeles, CA, USA}},
  \bibinfo{pages}{328--339}.
\newblock
\urldef\tempurl%
\url{https://doi.org/10.1109/ISCA.2018.00036}
\showDOI{\tempurl}


\bibitem[Zhang et~al\mbox{.}(2014)]%
        {zhang_survey_2014}
\bibfield{author}{\bibinfo{person}{Yuang Zhang}, \bibinfo{person}{Li Li},
  \bibinfo{person}{Zhonghai Lu}, \bibinfo{person}{Axel Jantsch},
  \bibinfo{person}{Minglun Gao}, \bibinfo{person}{Hongbing Pan}, {and}
  \bibinfo{person}{Feng Han}.} \bibinfo{year}{2014}\natexlab{}.
\newblock \showarticletitle{A Survey of Memory Architecture for {{3D}} Chip
  Multi-Processors}.
\newblock \bibinfo{journal}{\emph{Microprocessors and Microsystems}}
  \bibinfo{volume}{38}, \bibinfo{number}{5} (\bibinfo{year}{2014}),
  \bibinfo{pages}{415--430}.
\newblock
\showISSN{0141-9331}
\urldef\tempurl%
\url{https://doi.org/10.1016/j.micpro.2014.03.007}
\showDOI{\tempurl}


\end{thebibliography}


\iftoggle{includeAppendix}{
    \appendixpage\appendix
          
}{}

\iftoggle{includeResponse}{
    \newpage
    \input{999_taco_response_v2}
}{}

\end{document}